\documentclass[twocolumn,final,prb,superscriptaddress,eqsecnum,amssymb,amsmath,amsfonts,showpacs,floatfix,citeautoscript]{revtex4-1}
\usepackage{graphicx,epstopdf}
\usepackage{subfigure}
\usepackage{mathrsfs}
\usepackage{bm}
\usepackage[latin1]{inputenc}
\usepackage{wasysym}    
\usepackage{color}

\newcommand{\beq}{\begin{equation}}
\newcommand{\eeq}{\end{equation}}
\newcommand{\nn}{\nonumber}

\newcommand{\al}{\alpha}

\newcommand{\be}{\beta}

\newcommand{\mathbbm}{}

\definecolor{mygreen}{rgb}{0,0,0}

\def\be{\begin{equation}}
\def\ee{\end{equation}}
\def\bc{\begin{center}}
\def\ec{\end{center}}
\newcommand{\bea}{\begin{eqnarray}}
\newcommand{\eea}{\end{eqnarray}}

\newcommand{\dagga}{{\phantom{\dagger}}}

\usepackage[colorlinks,bookmarks=true,citecolor=blue,linkcolor=blue,urlcolor=blue]{hyperref}

\begin{document}


\title{Mixtures of ultra-cold atoms in 1D disordered potentials}
\author{Fran\c cois Cr\'epin}
\email[E-mail: ]{francois.crepin@physik.uni-wuerzburg.de}
\affiliation{Laboratoire de Physique des Solides, CNRS UMR-8502 Universit\'e Paris Sud, 91405 Orsay Cedex, France}
\altaffiliation[Now at: ]{Institute for Theoretical Physics and Astrophysics, University of Würzburg.}
\author{ Gergely Zar\'and}
\affiliation{Freie Universit\"at Berlin, Fachbereich Physik, Arnimallee 14, D-14195 Berlin, Germany}
\affiliation{MTA-BME Quantum Phases Lendulet Research Group, Budapest University of Technology and Economics, Budafoki ut 8, H-1521, Hungary}
\author{Pascal Simon}
\affiliation{Laboratoire de Physique des Solides, CNRS UMR-8502 Universit\'e Paris Sud, 91405 Orsay Cedex, France}
\date{\today}

\begin{abstract}
We study interacting 1D two-component mixtures of cold atoms in
a random potential, and extend the results reported 
earlier [{\it Phys. Rev. Lett.} {\bf 105}, 115301
(2010)]. 
We construct the phase diagram of a disordered Bose-Fermi mixture 
as a function of  the strength of the Bose-Bose and Bose-Fermi 
interactions, and  the ratio of the bosonic sound velocity 
and the Fermi velocity.
Performing renormalization group and  variational calculations,  three
phases are identified: (i) a fully delocalized  two-component
Luttinger liquid with superfluid bosons and fermions 
(ii) a fully localized phase with both components 
pinned by disorder, and (iii) an intermediate phase where fermions are
localized but bosons are superfluid.  Within the
variational approach, each phase corresponds to
a different level of replica symmetry breaking. 
In the fully localized phase we find that the bosonic and fermionic  
localization lengths can largely differ. 
We also compute the momentum distribution as well as
the structure factor  of the atoms (both experimentally  accessible), 
and discuss how the three phases can be  experimentally
distinguished.   
\end{abstract}


\maketitle

\section{Introduction}\label{sec:intro}

Since the original work of Anderson~\cite{Anderson58} on the
conductivity of electrons in a disordered crystal, the topic of
localization has been of major importance in the field of condensed
matter. Recent experiments on ultra-cold atomic gases have shed a
different light on the subject~\cite{Palencia10} as they strive to
systematically study such important factors as dimensionality and
interactions~\cite{Billy08, Roati08, Fallani07, DeMarco09, Deissler10,
  RDSV10, DeMarco11, Aspect11}. The interest in the interplay between
interactions and disorder dates back to Anderson's paper and it was
later understood, thanks to the work of Mott on the metal-insulator
transition~\cite{Mott68}, that, { at least at zero temperature}, the repulsion between electrons could 
actually favor localization by disorder, instead of preventing it. It
was recently proposed that interactions can be responsible for a
metal-insulator transition { at finite temperature} by allowing for a many-body mobility edge
in systems where all single-particle states should be
localized~\cite{Altshuler06}.

Interacting disordered bosonic systems have been  first thoroughly 
studied in the context of  dirty high temperature
superconductors, where Cooper pairs 
were thought to behave 
as bosons  in random media. The Bose-Hubbard model with random
on-site chemical potentials is one of the most famous 
models studied in this context~\cite{Fisher89}. This particular model
sustains a gapless but compressible  disordered insulating phase, 
the so-called {\it Bose glass}. This phase is surrounded 
by incompressible Mott phases -- associated with various commensurate
fillings --,  and a compressible superfluid phase. 
Its existence has been confirmed by several numerical studies~\cite{Batrouni91,
  Krauth91, Pollet09, Pollet10}. Bosons are peculiar when it
comes to disorder since, in the absence of interactions, they should
condense 
in a single lowest energy localized state at $T=0$
temperature.\cite{Giamarchi88} A tiny interaction, however, destroys this
state and  drives the system to a glassy insulating phase, 
the aforementioned 
Bose glass phase. 
However, increasing the interaction further, a transition from 
this localized Bose glass phase to a
superfluid phase takes place,
as interactions eventually favor the overlap between the localized
wave-functions and hence restore the long-range phase coherence. This
particular transition was intensively studied in the past few
years~\cite{Giamarchi88, Lugan07, Lugan07b, Lugan07c, Lugan09,
  Nattermann09, Altman11, Muller11} since it should be very relevant to
current experiments on cold atoms~\cite{Billy08, Roati08, DeMarco09, Deissler10}. 

The case of strong interactions is well described in 1D,
where one can use the harmonic fluid approach~\cite{Haldane81} to
treat interactions and disorder on the same
footing.
Using the renormalization group (RG), Giamarchi
and Schulz\cite{Giamarchi88}  showed that there exists a Kosterlitz-Thouless 
transition from a superfluid phase to a localized phase, 
corresponding to the pinning of the density wave by 
a weak random potential. In the
harmonic fluid approach the transition occurs, both for bosons and
for spinless fermions at a Luttinger parameter $K=3/2$, that is,
repulsive interactions for bosons and attractive interactions for
spinless fermions. The localized phase lies in the region $K<3/2$,
the Luttinger parameter $K=1$ corresponding to 
free fermions or hardcore bosons, respectively.  The harmonic
fluid approach was successfully tested in a recent cold atom
experiment \cite{Haller10} probing the superfluid to Mott insulator
transition, in a clean 1D lattice. This system was indeed well
described by the sine-Gordon model, that predicts the pinning of the
density wave for strong enough interactions (in the clean case,
$K=2$).~\cite{Zwerger03}

In the present paper, we focus on  Bose-Fermi(BF)
mixtures, but our results carry over to 
Bose-Bose or Fermi-Fermi cold atomic mixtures 
of incommensurate  (imbalanced) densities in a 1D random potential. 
Three-dimensional two-component mixtures have been 
recently realized experimentally in various cold atomic systems,~\cite{Schreck01, Ketterle02,
  Esslinger06, Bongs06, Bongs06b, Ferlaino04, Bloch09, Sugawa11},
where the densities, but also mass ratios, and the 
sign and magnitude of interactions can
be tuned. This versatility has fueled many 
analytical~\cite{Cazalilla03,
  Mathey04, Lewenstein04, Demler06, Demler06b, Mathey07, Mathey07b,
  Refael08, Burovski09, Jolicoeur10, Orignac10, Crepin10b, Roux11}
and numerical studies on clean Bose Fermi mixtures~\cite{Pollet06, Sengupta07, Hebert07, Hebert08,
  Pollet08b, Burovski09, Jolicoeur10, Roux11, Roscilde11}. Most of these works, 
especially the analytical ones, focussed on the 1D case where
 the harmonic fluid approaches~\cite{Haldane81, Cazalilla04, GiamarchiBook} proved very fruitful.
 Though quantum statistics
in 1D is somewhat less relevant than in higher
dimensions (see however Ref. \onlinecite{Crepin11}), these mixtures  already present very rich phase diagrams
and  many possible instabilities have been found.~\cite{Cazalilla03, Mathey04, Mathey07,Mathey07b, Burovski09, Orignac10, Roux11} 

In this work, we primarily focus on the role of a weak disorder. More
specifically,  
we analyze the effects of an external random potential on generic 1D
two-component   
mixtures using the harmonic fluid (bosonization) method.
We establish the phase diagram by combining a renormalization group 
approach with the so-called Gaussian variational method. 
This latter approach allows us to capture the glassy phases, 
which appear as saddle point solutions with broken replica symmetry.

From our perspective,   the work of Giamarchi and
Schulz on electrons in a random potential~\cite{Giamarchi88}  
thus focussed on the non-generic case of a balanced (commensurate)
Fermi-Fermi mixture with equal Fermi velocities. In this special 
case,  a perfect spin-charge separation occurs, back scattering plays
also an important role, and  several instabilities (pairing, charge
density waves, spin density waves) 
compete with the disorder, leading
to the rich phase diagram of Ref.~\onlinecite{Giamarchi88}.

Our approach -- and thus our results -- should { also} be contrasted to   
those of  Ref. \onlinecite{SanchezP05} { on disordered Bose-Fermi mixtures}: there, the system is placed on a
random lattice, and the limit of very strong interactions is taken, so
that various composite particles are created. This results in an
effective Hamiltonian with random couplings for the composite
particles, allowing for localized, metallic and Mott-insulating
phases of the latter. In contrast,  we do not have an
underlying lattice (or the densities are incommensurate with it), 
and  we are not a priori in a situation where composite particles 
(such as pairs of bosons and fermions) are likely to form in the 
clean system. 

We find that 
disorder-induced localization of one species (say fermions)
can influence the localization of the other species through
interactions. This is one of the main lessons of our analysis.
We find typically three distinct phases: a  delocalized phase
described by a 2-component Luttinger liquid, a hybrid phase where only one species is localized
and a fully localized phase where both components of the mixture are localized.
The latter phase turns out to be the most interesting one since it is characterized by two
interlaced localization length scales. The larger localization
length depends on the smaller one through the interaction between both species
and the ratio between the velocities of the density waves. 
In other words, though the two components of the mixture  are
localized, 
interactions between them still play an important role. 
%
This can be revealed through
the dynamical structure factor of one of the species, which exhibits two peaks
whose width are proportional to the inverse of the localization length
of each species (see Figs \ref{Fig:structure_factor} and \ref{sketch2}).
From a more technical point of view, it is worth emphasizing that the 
fully localized phase is characterized
 by  2-step replica symmetry breaking, which can be seen as a 
mathematical consequence of the interlacing localization
length scales. 

The plan of the paper is as follows:
In Sec. \ref{sec:model}, we describe the specific model under
consideration, justify 
its derivation and write its low energy bosonized form. Then, in
Sec.\ref{sec:loc}, we proceed to study localization by disorder in a
Bose-Fermi mixture, using RG and the variational method in replica
space. Finally in Sec. \ref{sec:observables}, we compute correlations
functions using the replica formalism. We analyze possible signatures
of these phases in various observables such as the structure factors
(related to Bragg scattering experiments) and the momentum
distributions observed in time--of-flight experiments.
The section \ref{sec:conclusion} contains a  non-technical detailed summary
of the results. The reader interested only in a snapshot of the results derived in this paper
 can jump directly to this section. Finally,
details of the calculations are given in the appendices.

\section{Model}
\label{sec:model}
\subsection{Low-energy theory}
In this section we present a phenomenological approach to the problem
of localization in 1D Bose-Fermi (BF) mixtures, where we start from the
low-energy hydrodynamic theory -- a two-component Luttinger liquid --
and then perturb it with a random chemical potential. 

We start from a microscopic 1D Hamiltonian 
\beq \label{eq:hgeneral}
H = H_f + H_b + H_{bf} +H_{ext},
\eeq
 with 
\bea
&H_f =&\hspace{-0.2cm} \int dx \  \psi_f^\dagger(x) \left[ -\frac{1}{2M_f} \frac{d^2}{dx^2} 
 \right] \psi^\dagga_f(x), \\
&H_b =&\hspace{-0.2cm} \int dx \  \psi_b^\dagger(x) \left[ -\frac{1}{2M_b} \frac{d^2}{dx^2}  \right] \psi^\dagga_b(x) \nn \\
&&+ \frac{U_b}{2} \int dx \ \psi_b^\dagger(x) \psi_b^\dagger(x) \psi_b^\dagga(x) \psi_b^\dagga(x),  \\
&H_{bf} =& U_{bf} \int dx \ \rho_b(x) \rho_f(x), \\
&H_{\text{ext}} =&\hspace{-0.2cm} \int dx \  \left[ V_f(x)\rho_f(x) + V_b(x)\rho_b(x)\right]. 
\eea
{Notice that we have set $\hbar=1$ in the whole paper.}
A discussion on the derivation of this Hamiltonian from a real, 3D,
experimental system can be found in Ref. \onlinecite{Crepin10b} and \onlinecite{Mathey07b}. Here,
$\psi_f^\dagger(x)$, $ \psi^\dagga_f(x)$ (resp. $\psi_b^\dagger(x)$, $
\psi^\dagga_b(x)$) are creation and annihilation operators for
spinless fermions (resp. bosons) while $\rho_f(x)$=$\psi_f^\dagger(x)
\psi^\dagga_f(x)$ and $\rho_b(x)$=$\psi_b^\dagger(x)
\psi^\dagga_b(x)$ are the density operators. $H_{\text{ext}}$
represents a random chemical potential shift. The overall bosonic and
fermionic chemical potentials do not appear in 
the Hamiltonian, since we rather take the bosonic and fermionic
densities as fixed. Furthermore, in the spirit of local density approximation,
we do not include a harmonic trapping potential either, 
certainly present in real cold atom experiments. 
We chose not to work with an underlying lattice and therefore do not include umklapp
scattering processes that could lead to gapped phases. 

This being said, we now
define the distributions and correlation functions of the random
potentials $V_f$ and $V_b$. In most experimental setups the same external potential
will couple to both bosons and fermions, and it is safe to assume that
$V_f$ and $V_b$ are indeed proportional. Let us define an optical
potential $V$ such that, $V_f = \alpha_f V$ and $V_b = \alpha_b V$. We
will take for $V$ a Gaussian distribution with zero mean and no
spatial correlation, such that \footnote{Overlining a quantity will
  indicate averaging over all possible realizations of the disorder.}  
\beq
\overline{V(x)V(x')} = D\delta(x-x'). \label{eq:corr1}
\eeq
In order to write the low-energy form of the Hamiltonian we follow the work of Haldane \cite{Haldane81} and introduce two quantum fields $\phi_{\alpha}$ and $\theta_{\alpha}$ for each species $\alpha=f,b$. The field $\phi_\alpha$ encodes density fluctuations through 
\beq 
\label{eq:density_bosonization}
\rho_\alpha(x) = \left[\rho_\alpha -\frac{1}{\pi}\nabla \phi_\alpha(x) \right] \sum_{p } e^{i2p[\pi\rho_\alpha x - \phi_\alpha(x)]}.
\eeq
One can understand this formula by considering a
 hypothetical classical configuration where atoms of a 1D gas are at a
 distance $\rho_\alpha^{-1}$ apart from each other. A good starting
 point is the density wave of wave vector $q=2\pi \rho_\alpha$, so
 that, $\rho_\alpha(x)$=$\rho_\alpha \cos[2\pi\rho_\alpha x
   -2\phi_\alpha]$. Density fluctuations are then allowed by letting
 the phase $\phi_\alpha$ of the density wave vary in space. The true
 density operator,  $\rho_\alpha(x) = \sum_{i=1}^N \delta(x-x_i)$, is
 reconstructed by summing over all even harmonics of $2\pi
 \rho_\alpha$. In (\ref{eq:density_bosonization}), the $\nabla \phi$
 term describes long wavelength fluctuations of the density. This
 exact formula is complemented by the expression of the creation
 operators  
\beq
\psi_\alpha^\dagger(x) = \sqrt{\rho_\alpha(x)}e^{-i\theta_\alpha(x)},
\eeq
where $\theta_\alpha(x)$ is the quantum phase operator. The fields $\phi_\alpha$ and $\theta_\alpha$ obey the following commutation relations
\beq \label{eq:quantum_fluctuations}
\left[ \phi_{\al}(x), \nabla \theta_{\beta}(x') \right] = i\pi\delta_{\alpha\beta}\delta (x-x'),
\eeq
and quantum statistics impose that \cite{Haldane81,GiamarchiBook}
\bea
&\hspace{-0.8cm}\psi_b^\dagger(x) =&\hspace{-0.1cm} \sqrt{\rho_b}\  e^{-i\theta_b(x)} \sum_{p} e^{i2p[\pi\rho_b x - \phi_b(x)]},\\
&\hspace{-0.8cm}\psi_f^\dagger(x) =&\hspace{-0.1cm} \sqrt{\rho_f}\  e^{-i\theta_f(x)} \sum_{p} e^{i(2p+1)[\pi\rho_f x - \phi_f(x)]}.
\eea
Haldane's theory also states that the universal low-energy Hamiltonian of a fermionic or bosonic 1D interacting system is of the form
\beq \label{eq:Luttinger_fixed_point}
H_\alpha = \frac{v_\alpha}{2\pi} \int dx \left[ K_\alpha (\nabla \theta_\alpha)^2 + \frac{1}{K_\alpha} (\nabla \phi_\alpha)^2\right],
\eeq
where $v_\alpha$ and $K_\alpha$ are two non-universal parameters
depending on the exact details of the microscopic model
considered. For $U_{bf}=0$, $K_f=1$ and $v_f = \pi \rho_f/M_f$ is
the Fermi velocity, while $K_b$ and $v_b$ can be extracted from the
solution of the Lieb-Liniger model\cite{Cazalilla04}. They depend on a
single dimensionless parameter $\gamma = M_b U_b/\rho_b$,
characterizing the strength of bosonic interactions. $K_b$ is a
monotonously  decreasing function of $\gamma$. For all values of
$\gamma$, $K_b \geq 1$ and $K_b=1$ for hard-core bosons, that is
$\gamma \rightarrow \infty$. The velocity $v_b$ can be identified
to the sound velocity in the quasi-BEC, is an increasing function of
$\gamma$ and saturates to $ \pi \rho_b/M_b$. 

We then add  interactions between the two species perturbatively. 
The lowest order term is 
\beq
H_{bf} = \frac{U_{bf}}{\pi^2} \int dx \ \nabla \phi_f \nabla \phi_b,
\eeq
a term that couples density fluctuations of each species.  It encodes
forward scattering processes for fermions, i.e. low momentum
scattering events that leave fermions on the same branch of the Fermi
surface. Backscattering processes, that transform right-moving
fermions into left-moving fermions and vice versa, would arise from a
term such as 
\beq
H_{bf} = g_{bf}  \int dx \ \cos\left[2\phi_f(x) - 2\phi_b(x)\right],
\label{back-scattering}
\eeq
as can be seen from (\ref{eq:density_bosonization}), when $\rho_f =
\rho_b$. For the rest of the paper, however, we will assume that $\rho_f \neq
\rho_b$, and drop Eq.~\eqref{back-scattering}. Also, since there is no 
underlying lattice,  dangerous umklapp processes do not appear either.
 Therefore, to next order in
perturbation theory only a current-current interaction term appears
\cite{Orignac10}, and renormalizes slightly  the Luttinger parameters. 
We thus neglect all these effects and retain only the following
quadratic  Hamiltonian 
\bea
\label{eq:H0}
H_0 = \hspace{-0.3cm} \sum_{\alpha=f,b} \frac{v_\alpha}{2\pi} &&\int dx \left[ K_\alpha (\nabla \theta_\alpha)^2 + \frac{1}{K_\alpha} (\nabla \phi_\alpha)^2\right] \nn \\
+&& \ \frac{U_{bf}}{\pi^2} \int dx \ \nabla \phi_f \nabla \phi_b\;.
\eea

We now proceed to couple the system to the external random
potential. Being interested in the low-energy sector of the theory,
only certain Fourier components of the random potential will couple to the
BF mixture. The low-momentum Fourier components couple to the density
fluctuations fields $\nabla \phi_f$ and $\nabla \phi_b$, whereas the
Fourier components around $2\pi \rho_f$ and $2\pi\rho_b$ couple
directly to the density waves. Therefore,  following Ref.
\onlinecite{Giamarchi88}, we decompose $V_{\alpha}$ as 
\bea 
V_\alpha(x) &\approx & \gamma_\alpha(x) + \xi_\alpha(x) e^{i 2\pi
  \rho_\alpha x} + {h.c.}+\dots
\\
\gamma_{f(b)}(x) &=& \frac{1}{L}\sum_{q \sim 0} e^{iqx}V_{f(b),q},  \\
\xi_{f(b)}(x) &=& \frac{1}{L}\sum_{q \sim 0} e^{iqx}V_{f(b),q-2\pi\rho_{f(b)}}, 
\eea
with $V_{f(b),q}$ the Fourier transform of $V_{f(b)}(x)$. Notice that
$\xi_f$ and $\xi_b$ are uncorrelated for $\rho_f \neq \rho_b$. 
 Indeed from equation \eqref{eq:corr1} we have
$\overline{V_q V_{q'}} = D\delta_{qq'}$. Therefore
$\overline{\xi_\alpha(x)\xi^*_\beta(x')}=D\delta_{\alpha\beta}\delta(x-x')$
as well as $\overline{\xi_\alpha(x)\xi_\beta(x')}=0$ and
$\overline{\xi_\alpha^*(x)\xi_\beta^*(x')}=0$. On the contrary, since
$V_f$ and $V_b$ are proportional, so are $\gamma_f$ and
$\gamma_b$. The resulting hydrodynamic  Hamiltonian reads 
\bea
H_{\text{ext}} = \sum_{\alpha=f,b}\ &&\int dx \left[-\frac{\gamma_\alpha(x) }{\pi} \nabla \phi_\alpha  \right. \nn \\
&&\left.+ \ \rho_\alpha \xi_\alpha(x)e^{- i2\phi_\alpha(x)} + h.c. \hspace{-0.2cm}\phantom{\frac{1}{1}}\right]. 
\eea
It appears that $\gamma_f$ and $\gamma_b$ act as random chemical
potentials. However, they just describe forward scattering, 
have no effect on the pinning of the density waves,
and can indeed be eliminated  through a gauge
transformation\cite{Giamarchi88}, 
\bea
\widetilde{\phi}_\alpha(x) &= &\phi_\alpha(x) - \int^x dy
\ \lambda_\alpha(y) \label{eq:gauge}\;,
\\
\widetilde{\xi}_\alpha(x) &= &\xi_\alpha(x) e^{-i2\int^x dy \ \lambda_\alpha(y)}, \nn 
\eea
with the static fields $\lambda_\alpha(x) $ defined as 
\bea
\lambda_f(x) &=& \frac{K_f/v_f}{1-g^2} \left[ \gamma_f(x)
  -g\sqrt{\frac{v_f}{v_b}\frac{K_b}{K_f}}\gamma_b(x) \right],  \\ 
\lambda_b(x) &=& \frac{K_b/v_b}{1-g^2} \left[ \gamma_b(x)
  -g\sqrt{\frac{v_b}{v_f}\frac{K_f}{K_b}}\gamma_f(x) \right], 
\eea 
and the fields $\theta_f(x)$ and $\theta_b(x)$ remaining unchanged. 
Here we have
introduced  the dimensionless Bose-Fermi coupling, 
\beq
\label{eq:g}
g=\frac{U_{bf}}{\pi}\sqrt{\frac{K_f K_b}{v_f v_b}}\;,
\eeq
an essential parameter in our future analysis.
After this gauge transformation our  Hamiltonian reads
\bea
\label{eq:Hgauged}
H =&&  \sum_{\alpha=f,b} \frac{ v_\alpha}{2\pi} \int dx \left[
  K_\alpha (\nabla \theta_\alpha)^2 + \frac{1}{K_\alpha} (\nabla
  \widetilde{\phi}_\alpha)^2\right]  \nn \\
+&& \ \frac{U_{bf}}{\pi^2} \int dx \ \nabla \widetilde{\phi}_f \nabla \widetilde{\phi}_b \nn \\
+&& \sum_{\alpha=f,b}\ \int dx \left[\rho_\alpha
  \widetilde{\xi}_\alpha(x)e^{- i2\widetilde{\phi}_\alpha(x)} +
  h.c. \hspace{-0.2cm}\phantom{\frac{1}{1}}\right].  
\eea

We remark that the gauge transformation above does not affect the fermion pair
correlation function nor current-current correlation function, nor
does it affect the bosonic propagator. However, it does effect 
the density operator, and results in an exponential decay of its
correlation function (see Section \ref{sec:density_corr}). 
Forward scattering on disorder thus does not compete with superfluid 
or normal currents and does not lead to  localization, though it
generates an exponential decay in certain correlation functions.
Pinning of the density waves occurs because of back-scattering  with transfered
momenta $2\pi \rho_f$ or $2\pi \rho_b$,  as decribed by the fields 
$\widetilde{\xi}_f(x)$ and $\widetilde{\xi}_b(x)$ in
\eqref{eq:Hgauged}.  Notice that since $\gamma_{f,b}$ and
$\xi_{f,b} $ are uncorrelated, $\widetilde{\xi}_{f,b}$ are also
independent Gaussian random variables with the same correlation functions as
$\xi_{f,b} $. Although the gauge transformation has consequences when
computing quantities depending directly on the density, for
simplicity, we shall omit  the tildes in our
subsequent analysis.

\subsection{Correlations in the homogeneous system}
\label{sec:LL}
In this paragraph, we consider the quadratic Hamiltonian $H_0$ of
equation \eqref{eq:H0} in the absence of a random potential. It is
natural to introduce the two normal
modes $\phi_\pm$ that diagonalize
$H_0$, \cite{Cazalilla03,Mathey04,Orignac10}
\bea
&\phi_f &= \  f_+ \phi_+ + f_- \phi_-\;, \\
&\phi_b &= \ b_+ \phi_+ + b_- \phi_-\;.
\eea
The coefficients $f_\pm$ and $b_\pm$ can be found in Appendix~\ref{app:coeff}. 
Similarly, the corresponding transformations for the $\theta$ fields read 
\bea 
&\theta_f &= \ \bar{f}_+ \theta_+ + \bar{f}_- \theta_-,\\
&\theta_b &=\  \bar{b}_+ \theta_+ + \bar{b}_- \theta_-.
\eea
The sound velocities of these normal modes are 
\beq
v_{\pm}^2=\frac{1}{2}(v_f^2 + v_b^2) \pm \frac{1}{2}\sqrt{(v_f^2 - v_b^2)^2 + 4g^2v_f^2 v_b^2}
\eeq
with $g$ defined in equation \eqref{eq:g}. It appears that when
$|g|>1$ the theory is unstable as $v_-$ becomes imaginary. As pointed
out in \onlinecite{Cazalilla03} this dynamical instability is a signal
of phase separation ($U_{bf}>0$) or collapse ($U_{bf}<0$) of the BF
mixture. Using the above decomposition one can compute correlation
functions for several instabilities and deduce a phase diagram. As
usual in one dimension, the nature of a given phase is determined by
the slowest decaying correlation function, since in a Luttinger liquid
only quasi long-range ordering can occur. As pointed out in
\onlinecite{Orignac10}, four instabilities compete in the
two-component Luttinger liquid, a charge density wave of fermions
($CDW_f$), $p$-wave pairing or fermions ($FP$), a charge density wave
of bosons ($CDW_b$) and superfluidity of bosons ($SF$). The
corresponding order parameters are $O_{CDW_f}(x) =
\psi_{f,R}^\dagger(x)\psi_{f,L}(x) \approx \rho_f e^{2i\phi_f(x)}$,  $O_{FP}(x) =
\psi_{f,R}(x)\psi_{f,L}(x) \approx \rho_f e^{2i\theta_f(x)}$, where we have used
the operators for right and left-moving fermions \cite{GiamarchiBook},
and $O_{CDW_b}(x)= \rho_b e^{i2\phi_b(x)}$, $O_{SF}(x)= \psi_b(x) \approx
\sqrt{\rho_b} \,e^{i\theta_b(x)}$. For a simple 1D Fermi gas we would have 
\bea
\langle O^\dagga_{CDW_f}(x) O_{CDW_f}^\dagger(0)  \rangle_{g=0} &\sim&
\left(\frac{\alpha}{|x|}\right)^{2K_f}, \\ 
\langle O^\dagga_{FP}(x) O_{FP}^\dagger(0) \rangle_{g=0} &\sim&
\left(\frac{\alpha}{|x|}\right)^{2/K_f}, 
\eea
with $\alpha$ a short distance cutoff, of the order of the
inter-particle distance. It turns out that our case of $K_f=1$ is a transition point
between a phase dominated by charge density wave fluctuations
($K_f<1$, repulsive interactions), with wave-vector $2k_F$, and a
phase dominated by pairing fluctuations ($K_f>1$, attractive
interactions).\cite{GiamarchiBook} Turning now to the Bose-Fermi
mixture, we find 
\bea
\langle O^\dagga_{CDW_f}(x) O_{CDW_f}^\dagger(0) \rangle_{g\ne0}
&\sim& \left(\frac{\alpha}{|x|}\right)^{2f_+^2 + 2f_-^2}, \\ 
\langle O^\dagga_{FP}(x) O_{FP}^\dagger(0) \rangle_{g\ne0} &\sim&
\left(\frac{\alpha}{|x|}\right)^{2\bar{f}_+^2 +
  2\bar{f}_-^2}, \label{eq:pair}  
\eea
with:
\bea
f_+^2 + f_-^2 &=& \frac{K_f}{\sqrt{1-g^2}}\frac{1+t\sqrt{1-g^2}}{\sqrt{1+2t\sqrt{1-g^2}+t^2}}, \\
\bar{f}_+^2 + \bar{f}_-^2 &=& \frac{1}{K_f}\frac{t+\sqrt{1-g^2}}{\sqrt{1+2t\sqrt{1-g^2}+t^2}},
\eea
where $t=v_f/v_b$. For any ratio of velocities $t$ and $|g|<1$, $f_+^2
+ f_-^2 >K_f$ and $\bar{f}_+^2 + \bar{f}_-^2<1/K_f$. Therefore,
starting from non-interacting fermions and $K_f=1$, because of the
Bose-Fermi interactions the pairing fluctuations will always dominate
over the charge density wave fluctuations. The effect of the
Bose-Fermi interaction is thus to create an effective attractive
Fermi-Fermi interaction. The situation is very similar to the one of
interacting electrons in a metal, coupled to phonons, where an
effective attractive interaction arises from the integration of the
phonon degrees of freedom. 

 A similar analysis can be carried out for the bosons, where charge
 density wave fluctuations, with wave-vector $2\pi \rho_b$, compete
 with superfluid fluctuations. Indeed, 
\bea
\langle O^\dagga_{CDW_b}(x) O_{CDW_b}^\dagger(0) \rangle_{g\ne0} &\sim& \left(\frac{\alpha}{|x|}\right)^{2b_+^2 + 2b_-^2}, \\
\langle O^\dagga_{SF}(x) O_{SF}^\dagger(0) \rangle_{g\ne0} &\sim& \left(\frac{\alpha}{|x|}\right)^{\frac{1}{2}\bar{b}_+^2 + \frac{1}{2}\bar{b}_-^2} \label{eq:boson}, 
\eea
with
\bea
b_+^2 + b_-^2 &=& \frac{K_b}{\sqrt{1-g^2}}\frac{t+\sqrt{1-g^2}}{\sqrt{1+2t\sqrt{1-g^2}+t^2}}, \\
\bar{b}_+^2 + \bar{b}_-^2 &=& \frac{1}{K_b}\frac{1+t\sqrt{1-g^2}}{\sqrt{1+2t\sqrt{1-g^2}+t^2}}.
\eea
Similar to the fermionic sector,  $b_+^2 + b_-^2 >K_b$ and
$\bar{b}_+^2 + \bar{b}_-^2<1/K_b$. Superfluidity is thus enhanced by
the Bose-Fermi interactions, which create an effective attractive
bosonic interaction as well, that reduces the original repulsive
Bose-Bose interactions and therefore favors superfluidity.  
\begin{figure}[h!] 
\includegraphics[width=8cm,clip]{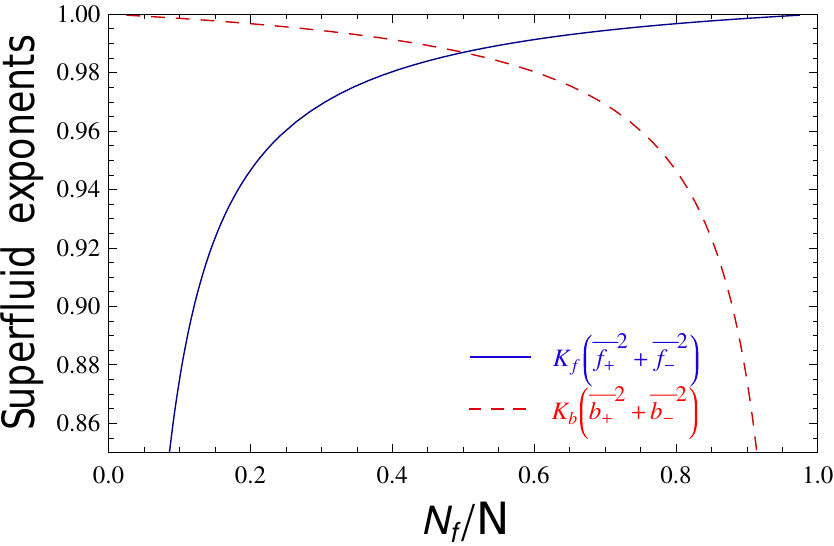}
\caption{Superfluid exponents for $K_b=1$ and $U_{bf}/v_0 = 1$, with $v_0 = (v_f + v_b)/2$, as a function of the fraction of fermions. We take equal masses for both species, so that $U_{bf}/v_0$ is kept a constant for all fillings.  Solid blue line: fermions. Red dashed line: bosons.}
\label{fig:superfluid_exp}
\end{figure}

\begin{figure}[h!] 
\includegraphics[width=8cm,clip]{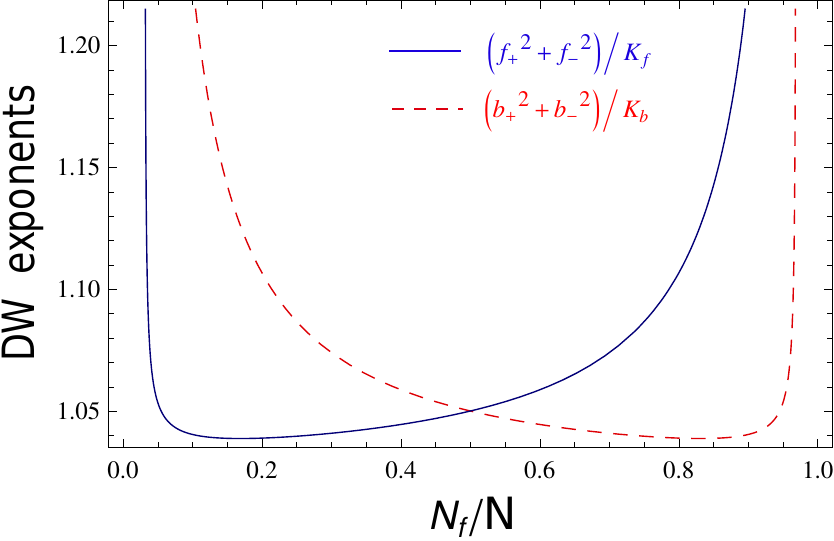}
\caption{Density wave exponents for $K_b=1$ and $U_{bf}/v_0 = 1$, with $v_0 = (v_f + v_b)/2$, as a function of the fraction of fermions. We take equal masses for both species, so that $U_{bf}/v_0$ is kept a constant for all fillings.  Solid blue line: fermions. Red dashed line: bosons}
\label{fig:CDW_exp}
\end{figure}

In the system under consideration there are three independent
parameters, the ratio of velocities $t = v_f/v_b$, the Luttinger
parameter of bosons $K_b$ ($K_f$ is fixed and equal to 1), and
Bose-Fermi interactions, through $U_{bf}$. As an example we plot the
superfluid and density wave exponents in Fig.~\ref{fig:superfluid_exp}
and Fig.~\ref{fig:CDW_exp} respectively, taking $K_f = K_b = 1$,
corresponding to hard core bosons and noninteracting fermions. Also we
have taken equal masses ($M_f = M_b = M$) for both species. In that
case (and since we are working here in the continuum), $v_f = 
\pi \rho_f / M$ and  $v_b =  \pi \rho_b / M$ and the ratio of
velocities can be expressed as a function of the fraction of fermions
only $N_f/N$ ($N$ the total number of particles being fixed). For
Bose-Fermi interactions, we use the following dimensionless parameter
$U_{bf}/v_0$ with $v_0 = (v_f + v_b)/2$ the mean velocity, which is a
constant independent of $N_f/N$. We wish to illustrate here that the
ratio of velocity is a crucial parameter that is ultimately related to
clear parameters of an experimental system, such as the number of
particles. In Fig.~\ref{fig:superfluid_exp} and Fig.~\ref{fig:CDW_exp}
one notices that density wave correlations are mostly suppressed for the
slowest species of the two, while on the contrary, its superfluid
correlations are enhanced. One has to keep this feature in mind which will be 
crucial for the
analysis of localization.

\section{Localization in 1D Bose-Fermi mixtures}
\label{sec:loc}

\subsection{A preliminary variational argument}
\label{sec:var}

\begin{figure}[h!] 
\includegraphics[width=8cm,clip]{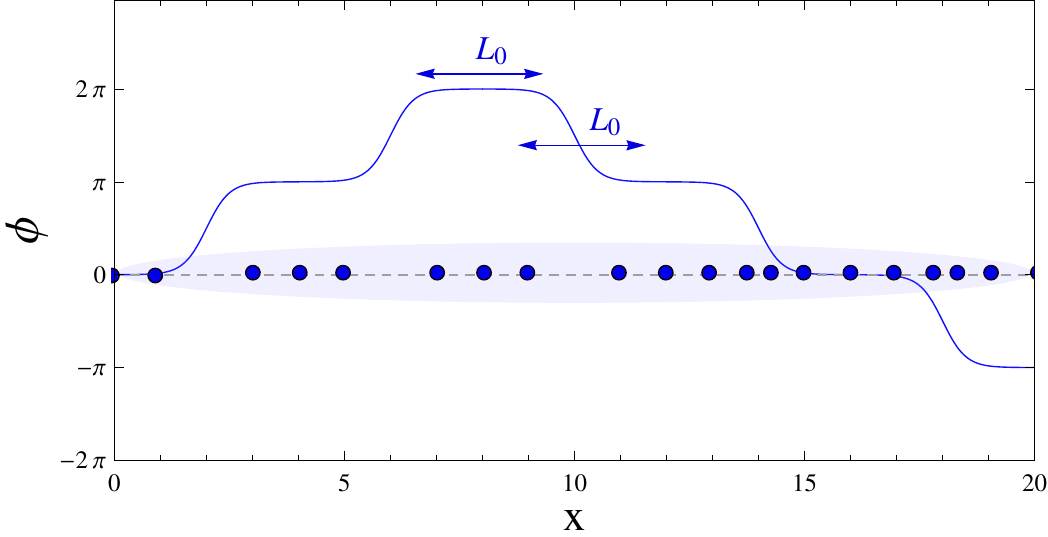}
\caption{Schematic representation of the phase $\phi(x)$ from the variational solution for the pinning of a density wave by a random potential. The system breaks into domains of order $L_0$ separated by domain walls of the same average size. A schematic description of the classical density also appears. $x$ is in unit of the inverse density. On a domain with constant $\phi$ a regulare DW appears while on domain walls, an extra particle ($\nabla \phi <0$) or a hole ($\nabla \phi >0$) is pinned by disorder.}
\label{Fig:density1}
\end{figure}

Consider the case of a single species of interacting particles with backscattering on a random external potential. We recast ist low-energy Hamiltonian (see \eqref{eq:Hgauged}) into 
\bea
H &=& \frac{  v}{2\pi}\int dx \left[ K( \nabla \theta)^2 + \frac{1}{K} ( \nabla \phi)^2 \right]\nn \\
 &+&\rho \int dx \left[ \xi(x)e^{-i2\phi(x)}+H.c. \right], \label{eq:H_Fukuyama}
\eea
where again $\xi(x)$ is the $2\pi\rho$ Fourier component of the random potential. We briefly review the variational argument proposed by Fukuyama and Suzumura in Ref. \onlinecite{Suzumura83}. It starts by looking for a classical configuration $\phi_0(x)$ satisfying
\beq
\left.\frac{\delta H}{\delta \phi}\right|_{\phi=\phi_0} = 0,
\eeq
a differential equation with random coefficients. A variational
solution is found by assuming that the charge density wave breaks into
domains of size $L_0$ on which $\phi_0(x)$ is a constant. By doing so
it takes advantage from  the random potential ($\xi$ and $\xi^*$) as
much as possible.  A typical energy of order
$-(L/L_0)\sqrt{D L_0}$ is gained in this way. However, the optimal
value $\phi_0$ varies randomly from domain to domain with
differences of order $\pi$. This costs elastic energy, through the
interaction term $(\nabla \phi)^2$, of order $L/L_0$ if domain walls
are of the same typical size $L_0$. According to this Imry-Ma-like
analysis\cite{Imry75}, there is always a finite value $L_0$ for which
the energy is minimum and negative (as compared to the value of zero
one would obtain for $\phi_0$ constant over the whole system). As a
next step, quantum fluctuations are  added self-consistently. These 
tend to reduce the
amount of potential energy gained from the random potential. Expanding
$\phi(x)$ around the classical 
solution $\phi_0(x)$, $\phi(x) = \phi_0(x) + \psi(x)$ with $\psi$ a 
quantum field, the effective Hamiltonian per unit length and up to a
constant is: 
\bea
H/L &=& E_{\text{el}} + \frac{  v}{2\pi}\int dx \left[ K( \nabla \theta)^2 + \frac{1}{K} :(\nabla \psi)^2: \right] \nn \\
&-& \left(\frac{D}{L_0}\right)^{1/2}\left(\frac{m}{\Lambda}\right)^{K}\int dx :\cos[2\psi(x)]:,
\eea
where we have normal-ordered the effective sine-Gordon
Hamiltonian\cite{Coleman75} and $E_{\text{el}}$ is the elastic energy
(of order $1/L_0$) coming from the classical configuration. $\Lambda$
is a UV cut-off and the mass $m$ can be obtained using a
self-consistent harmonic approximation\cite{Suzumura83},
$m^2=\mathcal{B}\sqrt{D/L_0}(m/\Lambda)^{K}$, with $\mathcal{B}$ some
unimportant prefactor. Minimizing the energy with respect to $L_0$ one
then finds 
\beq
\label{eq:loc_length1} 
L_0 \propto \left( \frac{1}{D} \right)^{\frac{1}{3-2K}}, \ \text{for} \ K<3/2.
\eeq
In the context of 1D interacting particles this length can be understood as the localization length of the system. The CDW is pinned by the random potential and correlations in the phase of the CDW are lost above the localization length. It appears from \eqref{eq:loc_length1} that $L_0$ diverges as $K$ approaches $3/2$. Therefore a gas of 1D fermions should undergo a transition from a localized to superfluid phase as attractive interactions are increased. Similarly for bosons, a transition to superfluidity would occur as interactions are decreased from the hard-core limit. A schematic description of the pinning as understood from the variational solution is given in Fig.~\ref{Fig:density1}.

Turning now to the case of an interacting BF mixture, similar arguments can be put forward. Consider the classical solution that minimizes the energy
\bea
E =&&   \frac{  v_f}{2\pi K_f} \int dx (\nabla \phi_{f})^2 + \frac{  v_b}{2\pi K_b} \int dx (\nabla \phi_{b})^2 \nonumber \\
+&& \frac{U_{bf}}{\pi^2} \int dx \ \nabla \phi_{f} \nabla \phi_{b} \nn \\
+&& \sum_{\alpha=f,b} \rho_\alpha \int dx \left[ \xi_\alpha(x)e^{-i2\phi_{\alpha}(x)}+H.c. \right]. 
\label{eq:classical_energy}
\eea
In the simpler case where the random potential couples only to bosons,
the situation is very similar to the one exposed in the single species
problem. The boson gas breaks up into domains of size $L_b$, on which
the classical phase $\phi_{b,0}$ adjusts to the random phase. The
fermionic density wave then deforms its phase so as to minimize the
energy of the system. There are two elastic contributions  for
fermions: $(\nabla \phi_{f,0})^2$ that tries to keep the fermionic
phase constant and $U_{bf}/\pi^2 \nabla \phi_{f,0} \nabla \phi_{b,0}$
that tries to keep both density waves in phase ($U_{bf}<0$) or out of
phase ($U_{bf}>0$). The optimal configuration is readily exhibited by
recasting the Hamiltonian in the following form: 
\bea
\label{eq:classical_energy_2}
E =&&   \frac{  v_f}{2\pi K_f} \int dx (\nabla \widetilde{\phi}_f)^2 \nn \\
+&& \frac{  v_b}{2\pi K_b}\left( 1-\frac{U_{bf}}{\pi^2}\frac{K_b K_f}{ ^2 v_f v_b}\right)
 \int dx (\nabla \phi_{b})^2\nn \\
+&& \rho_b \int dx \left[ \xi_b(x)e^{-i2\phi_{b}(x)}+H.c. \right],
\eea
where we have made the following change of variables, $\widetilde{\phi}_f = \phi_f + \frac{U_{bf}}{\pi}\frac{K_f}{  v_f}\phi_b$. The elastic energy cost of deforming the bosonic phase is reduced by a factor $1-g^2$ if $\tilde{\phi}_f$ is kept constant. Then, the bosonic gas breaks into domains of size $L_b$,
\beq
\label{eq:Lb}
L_b \propto \left( \frac{1}{D_b} \right)^{\frac{1}{3-2\tilde{K}_b}}, \ \text{for} \ \tilde{K}_b<3/2,
\eeq
with $\tilde{K}_b = K_b/\sqrt{1-g^2}$, while the fermionic density adjusts accordingly in a way given by the classical solution:
\beq
\label{eq:fermion_phase_classical}
\phi_{f,0}(x) = -\frac{U_{bf}}{\pi}\frac{K_f}{  v_f}\phi_{b,0}(x) + \text{cste}.
\eeq
We emphasize two important aspects at this point. Note that the
instability toward phase separation or collapse when $g^2=1$ is
apparent already at the classical level. The coefficient before
$(\nabla \phi_b)^2$ becomes negative as $g^2$ exceeds $1$, favoring
maximum distortion of the bosonic density wave. Note also that the
present transformation for the fields \eqref{eq:classical_energy_2}
does not diagonalize the full quantum Hamiltonian, and would not allow
for a correct treatment of quantum fluctuations.  

 Finally, we consider the situation where the disorder couples to both
 components of the gas. Although a full self-consistent treatment is
 needed, here we only give a few qualitative arguments; we postpone
 the self-consistent calculation to section \ref{sec:RSB} where we use
 replica-symmetry breaking in order to describe the fully localized
 phase. We are now in a situation where the disorder tries to pin both
 components of the gas independently (as $\xi_f$ and $\xi_b$ are
 uncorrelated) while elastic deformations are coupled. We still look
 for a solution where both components break into domains of size $L_b$
 and $L_f$. A way to disentangle the problem is to consider the case
 where one of the localization lengths is (much) larger than the other,
 say $L_b>L_f$. Building on equation
 \eqref{eq:fermion_phase_classical}, it seems reasonable to assume
 that the fermionic phase is of the form: 
\beq
\label{eq:fermion_phase_classical_2}
\phi_{f,0}(x) = -\frac{U_{bf}}{\pi}\frac{K_f}{  v_f}\phi_{b,0}(x) + \lambda_f.
\eeq
Here, $\lambda_f$ which is related to the random phase of the
disordered potential $\xi_f$ has replaced the constant in equation
\eqref{eq:fermion_phase_classical}. On domains where $\phi_b$ is a
constant, $\phi_f$ makes random jumps of order $\pi$ to accommodate
the random potential, very much as in the single species case. On the
contrary, when $\phi_b$ deforms between two domains, it has the effect
of a chemical potential -- much like in a Mott-$\delta$ transition --
and imposes a finite slope on $\phi_f$. In that case the variations of
$\phi_f$ follow a nested pattern, in the sense that the coupling to
bosons imposes variations on a length scale of the order of $L_b$,
while each domain of size $L_b$ breaks down into smaller domains of
size $L_f$ in order to accommodate the random phase of the
disorder. Of course at this level one should take into account quantum
fluctuations. It is the object of the next two sections. First we
perform a renormalization group calculation in order to identify the
regions of parameter space where disorder is relevant and likely to
pin one or both components of the gas. Then we use the concept of
replica symmetry breaking to confirm the findings of the RG
calculation and the intuitions we got from the classical approach.

 
\subsection{Renormalization Group calculation and a tentative phase diagram}
\label{sec:RG}
The RG approach is especially powerful to treat both the effects of
interactions and disorder in 1D systems. Following the approach
introduced by Giamarchi and Schulz in Ref.~\onlinecite{Giamarchi88},
we treat disorder as a perturbation of the Luttinger liquid fixed
point. Our starting point is the Hamiltonian of equation
\eqref{eq:Hgauged}. The low-energy fixed point is the two-component
Luttinger liquid described in section \ref{sec:LL} and, once again,
the random potential tries to pin each component independently. The RG
transformation is constructed by integrating out high energy degrees
of freedom -- here, short distance density fluctuations -- at the
level of the partition function, through a rescaling of the UV
cut-off. A detailed calculation is presented in appendix
\ref{app:RG}. A complication arises in a system with quenched
disorder. There, the thermodynamic quantity of interest is the average
free energy $F$ defined as: 
\beq
-\beta F =  \overline{\log Z}
\eeq
where $Z$ is the partition function for a given realization of the
random potential, and $\overline{\phantom i \ldots \phantom i}$
denotes averaging over all possible realizations of the disorder. The
average free energy is very difficult to compute and one way of action
is to use the so-called replica trick \cite{MezardBook}. It rests upon
the following observation, 
\beq
\lim\limits_{n\rightarrow 0} \frac{1}{n}\log \overline{Z^n} = \overline{\log Z}.
\eeq
The trick consists in introducing $n$ identical copies of the system,
average over the disorder realizations and in the end take the limit
$n\rightarrow 0$. Practically we will work with the quantity
$\overline{Z^n}$ to perform the RG calculation. Using the path
integral formulation, the partition function $Z$ for a given
realization of the disorder is 
\beq 
Z= \int D\phi_f D\phi_b \ e^{-S[\phi_f,\phi_b]} 
\;,
\eeq
with $S$ the action derived from the Hamiltonian \eqref{eq:Hgauged},
that is, $S = S_0 + S_{\text{dis}}$, with: 
\bea
S_0 =  \sum_{\alpha=f,b}&&\hspace{-0.01cm}\frac{1}{2\pi K_\alpha}\int dx d\tau \left[\frac{1}{v_\alpha}\left( \partial_\tau \phi_\alpha\right)^2 + v_\alpha \left( \partial_x \phi_\alpha\right)^2 \right] \nn \\
+&& \hspace{-0.01cm} \frac{U_{bf}}{\pi^2}\int dx d\tau
\ \partial_x\phi_f \partial_x \phi_b, 
\label{S_0}
\\
S_{\text{dis}} = \sum_{\alpha=f,b}&&\hspace{-0.01cm} \rho_\alpha \int
dx d\tau \left[ \xi_\alpha(x) e^{-2\phi_\alpha(x,\tau)} +
  H.c. \right].
\label{S_dis}
\eea
Assuming that $\xi_f$ and $\xi_b$ have Gaussian distributions, we compute the replicated action defined through:
\beq
\overline{Z^n} = \int \prod_{a=1}^n D\phi_f^a D\phi_b^a \  e^{-S_{\text{rep}} },
\eeq
and find $S_{\text{rep}} = S_0^{\text{rep}} + S_{\text{dis}}^{\text{rep}}$ with
\bea
S_0^{\text{rep}} &&\hspace{0.cm}=  \sum_{a=1}^n
\sum_{\alpha=f,b}\frac{1}{2\pi K_\alpha}\int dx d\tau
\left[\frac{1}{v_\alpha}\left( \partial_\tau \phi^a_\alpha\right)^2 +
  v_\alpha \left( \partial_x \phi^a_\alpha\right)^2 \right] \nn \\ 
&&+\frac{U_{bf}}{\pi^2}\int dx d\tau \ \partial_x\phi^a_f \partial_x
\phi^a_b, 
\label{S_0^rep} 
\\
S_{\text{dis}}^{\text{rep}} && =-{D_f \rho_f^2} \sum_{a,b}\int dx d\tau d\tau ' \cos[2\phi_f^a(x,\tau) - 2\phi_f^b(x,\tau ')]\nn \\
&&\hspace{-0.3cm}-{D_b \rho_b^2} \sum_{a,b}\int dx d\tau
d\tau ' \cos[2\phi_b^a(x,\tau) - 2\phi_b^b(x,\tau ')]\;.
\label{S_dis^rep} 
\eea
Using the following parametrization for the UV cut-off, $\Lambda(l) =
\Lambda_0e^{-l}$, $\Lambda_0$ being the bare cut-off, we find the
following RG flow equations (see Appendix~\ref{app:RG}) 
\bea
\frac{d \tilde{D}_f}{dl} &=& (3 -X_f)\tilde{D}_f(l) 
\label{eq:Df}, 
\\
\frac{d \tilde{D}_b}{dl} &=& (3 -X_b)\tilde{D}_b(l),
\label{eq:Db}
\eea
where we have defined the dimensionless couplings $\tilde{D}_f = \frac{D_f \rho_f^2}{ v_f^2 \Lambda^3} $ and $\tilde{D}_b = \frac{D_b \rho_b^2}{ v_b^2 \Lambda^3}$. $K_f, K_b, v_f$ and $v_b$ are also renormalized. Their flow equations are written in appendix \ref{app:RG}. The anomalous dimensions, $X_f$ and $X_b$, of the disorder operators are obtained from the diagonalization of $S_0$. They are $X_f = 2f_+^2 + 2f_-^2$ and  $X_b = 2b_+^2 + 2b_-^2$. We recall their analytical expressions as a function of $t=v_f/v_b$ and $g$:
\bea
X_f &=&  \frac{2K_f}{\sqrt{1-g^2}}\frac{1+t\sqrt{1-g^2}}{\sqrt{1+2t\sqrt{1-g^2}+t^2}},\\
X_b &=&  \frac{2K_b}{\sqrt{1-g^2}}\frac{t+\sqrt{1-g^2}} {\sqrt{1+2t\sqrt{1-g^2}+t^2}}.
\eea

\begin{figure}
\includegraphics[width=8cm]{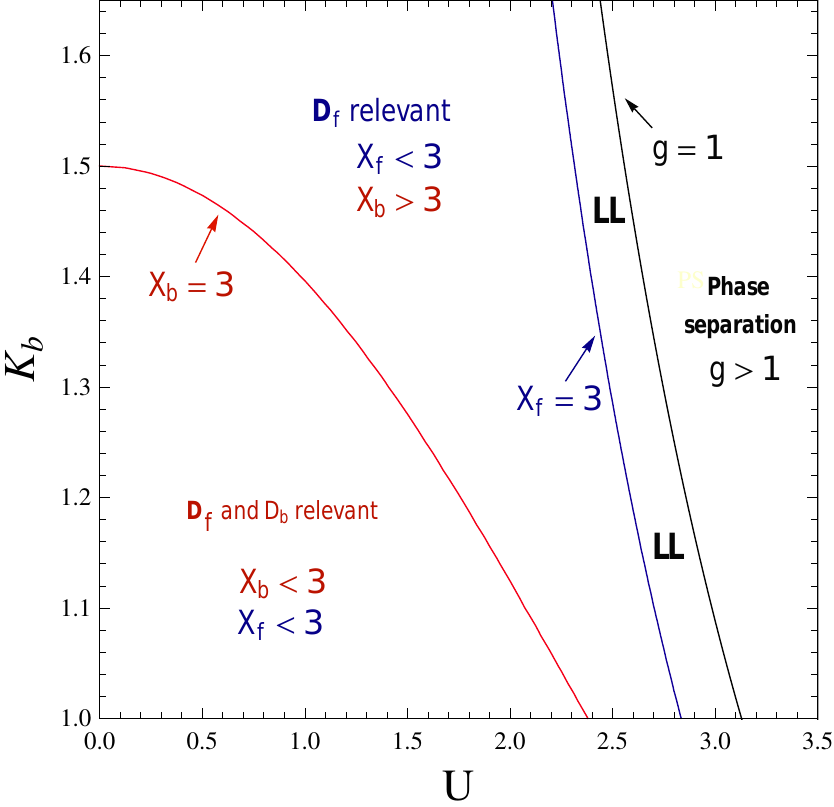}
\caption{Critical lines as obtained from the RG flow \eqref{eq:Df} and \eqref{eq:Db}, for a ratio of velocities $t=3$, and $K_f=1$. Here we have chosen to parametrize interactions as $U=U_{bf}/\sqrt{v_f v_b}$.}
\label{fig:phase_diagram_RG_1}
\end{figure}

\begin{figure}
\includegraphics[width=8cm]{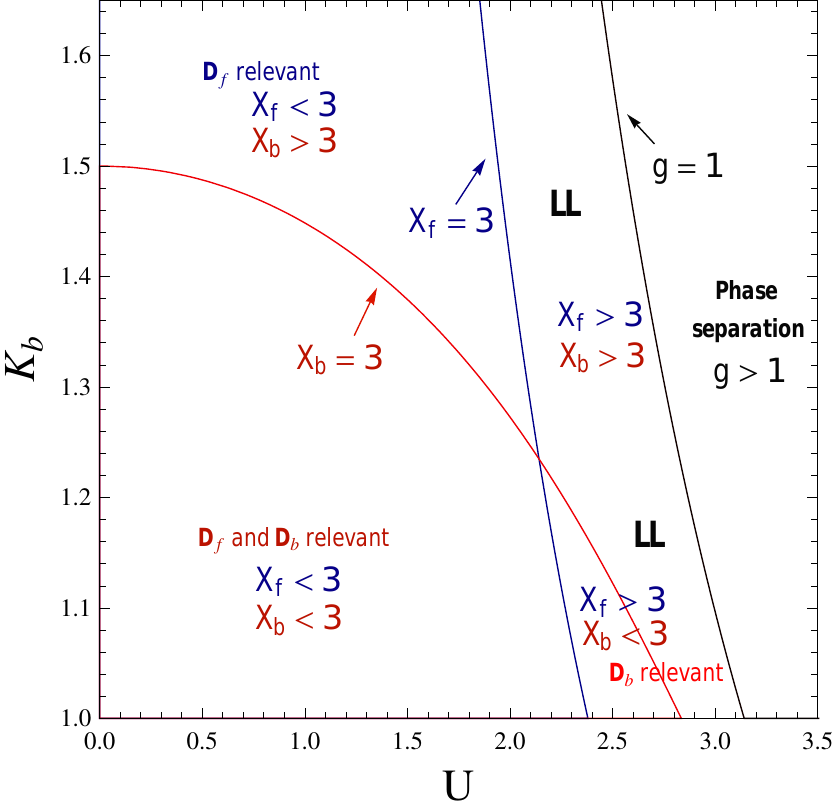}
\caption{Critical lines as obtained from the RG flow \eqref{eq:Df} and \eqref{eq:Db}, for a ratio of velocities $t=1/3$, and $K_f=1$. We have defined $U=U_{bf}/\sqrt{v_f v_b}$.}
\label{fig:phase_diagram_RG_2}
\end{figure}

For uncoupled species ($g=0$), $X_f=2K_f$ and $X_b=2K_b$. Thus
spinless fermions (bosons) are localized when $K_f<3/2$
($K_b<3/2$).\cite{Giamarchi88} As Bose-Fermi interactions are turned
on, new phases appear. As explained in section \ref{sec:LL},
Bose-Fermi interactions tend to enhance superfluid correlations and
impair the formation of density waves. Formally, $X_f>2K_f$ and
$X_b>2K_b$ and there exist regions of parameters for which disorder is
an irrelevant perturbation in the RG sense although single species
would be localized. In the variational language of section
\ref{sec:var}, it means that quantum fluctuations are enhanced by the
Bose-Fermi interactions and tend to reduce the localization length.

 In
Figs.~\ref{fig:phase_diagram_RG_1} and \ref{fig:phase_diagram_RG_2} we
show two examples of the critical lines for two ratios of velocities,
$v_f/v_b=3$ and $v_f/v_b=1/3$. Although the mechanism by which
superfluidity is enhanced seems clear enough one should be careful in
drawing conclusions about the actual phase diagram from the positions
of the critical lines given by Eqs.~\eqref{eq:Df} and
\eqref{eq:Db}. Indeed, when one or both perturbations are relevant,
disorder parameters flow to a strong coupling phase, out of reach of
the perturbative RG we have used so far. This is of special importance
in some regions of the phase diagram. For instance in
Fig. \ref{fig:phase_diagram_RG_1}, there is a large portion of the
diagram for which $X_b>3$ and $X_f<3$.  Here $D_f$
appears to be relevant while $D_b$ is irrelevant. If $K_b>3/2$ the
nature of this phase is quite clear: fermions are localized (they are
in the so-called Anderson glass phase) while bosons remain
superfluid. Indeed, if $U_{bf}=0$, spinless fermions are localized, as
they should be, and bosons are in a superfluid phase ($K_b>3/2$). The
effects of non-zero BF interactions are two-fold. The phonons of the
bosonic gas mediate an effective attractive interactions that
eventually leads to a transition to a phase where fermions are
superfluid (and pair correlation are dominant). Similarly the phonons
of the fermionic gas tend to reduce the repulsion between bosons and
enhance superfluidity. Note that although fermions are localized this
mechanism is possible since their localization length $L_f$ is quite
large in the limit of weak disorder and phonons do exist below
$L_f$. Now if $K_b <3/2$ the interpretation of the RG flow is more
delicate. Phonons in the fermionic gas do renormalize the bosonic
interactions in such a way that $X_b>3$ and $D_b$ is
irrelevant. However as soon as the UV cut-off is rescaled down to the
inverse fermionic localization length, fluctuations in the fermionic
density are pinned by disorder and 'gapped', and thus no longer affect
the bosons. Below this 
cut-off, bosons interact with their bare interactions and, as
$K_b<3/2$, disorder is relevant again and bosons are
localized. Therefore above a certain value $l = l_f$ for which
$\tilde{D}_f(l_f)=1$ and $L_f^{-1} = \Lambda(l_f)$, the flow of
$\tilde{D}_b$ should be modified as follows: 
\beq
\frac{d \textrm{log} \tilde{D}_b}{dl} =
\left\{
\begin{array}{ll}
3 - X_b     & \mbox{ if $\Lambda \gg L_f^{-1}$}\;\\
3-2K_b     & \mbox{ if $\Lambda \ll L_f^{-1}$}.
\end{array}
\label{eq:Db_scaling}
\right.
\eeq
This should hold for any value of $K_b$. The important point here is
that when $X_b>3$ and $K_b<3/2$ bosons are still localized once the
fermions become localized. However, the
structure of the flow indicates that the bosonic localization length
(defined as $L_b = \Lambda(l_b)^{-1}$ with $\tilde{D}_b(l_b) =1$) will
be extremely large. 

\begin{figure}
\includegraphics[width=8cm]{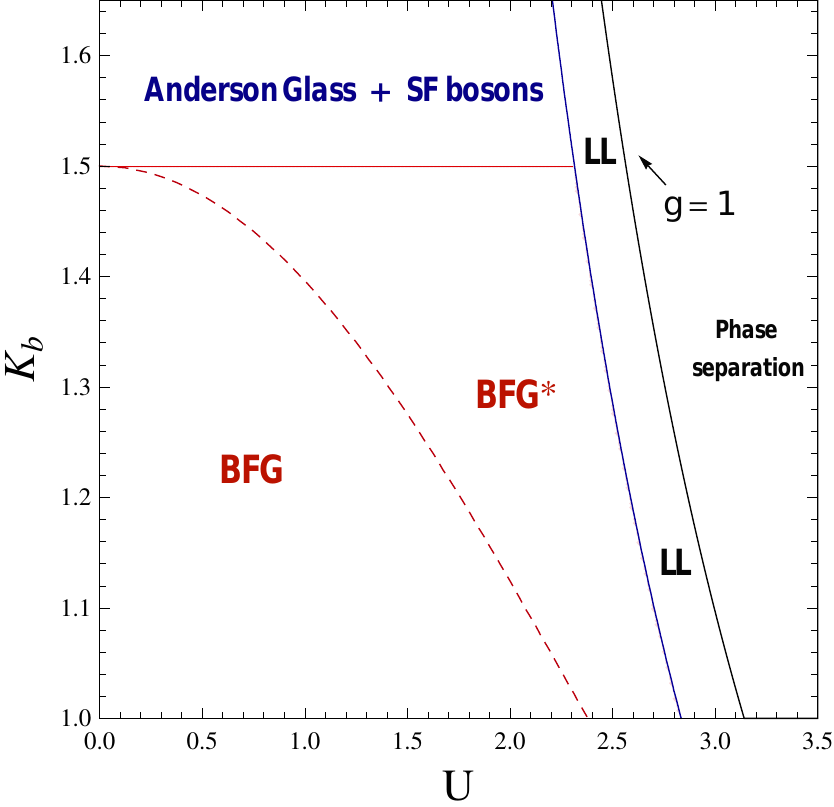}
\caption{Phase diagram of the Bose-Fermi mixture in a random
  potential as a function of $K_b$ and $U=U_{bf}/\sqrt{v_bv_f}$ 
for $t=3$. BFG stands for Bose-Fermi glass. BFG$^*$ is the same
  phase, however we identify a crossover regime for which the
  localization length of bosons is much larger than the one of
  fermions. } 
\label{fig:phase_diagram_RG_3}
\end{figure}

Therefore, based on this RG approach we propose the
following tentative phase diagram, summarized in
Fig.~\ref{fig:phase_diagram_RG_3} for $v_f/v_b >1$. We identify three
phases: the usual two-component Luttinger liquid (LL) as disorder is
irrelevant for both species, a phase where fermions are localized
while bosons remain superfluid and a phase where both species are
localized but still coupled. We call the latter a Bose-Fermi glass, by
analogy with the Bose glass phase. In this phase, despite
localization, interactions have very important effects. Notably, the
localization length of bosons varies greatly with interactions. We
highlight a crossover regime where the boson localization length
becomes much larger than the fermion localization length as indicated
by \eqref{eq:Db_scaling}. 
To confirm our findings we look for a variational solution in replica
space. This is the subject of the next section.

\subsection{Variational calculation in replica space}
\label{sec:RSB}

\subsubsection{Self-consistent equations}
In the study of one dimensional systems, a variational calculation if
often a complementary tool with respect to, e.g., a renormalization
group calculation. We find that, in our case,  even if the RG can provide
some information on the structure of the phase diagram, it fails to
describe properly {\it strong disorder} phases. An example was given
in the previous subsection, in which once the fermionic disorder has
flown to strong coupling -- beyond the reach of perturbation theory --
the behavior of bosons became unclear, regardless of what we should
conclude from the dimension of the disorder operator. 

The variational
method aims at finding the best Gaussian approximation to the
complicated action $S_{\text{rep}}$. It builds on the knowledge that
there exists a phase where it is energetically favorable to lock the
field $\phi$ to a certain value, and considers only quadratic
fluctuations around this minimum value. Within the replica formalism
we actually look for a distribution of these optimal values, very much
like the solution in Fig.~\ref{Fig:density1}. 
Complications arise, however, since  replicas are coupled and one
needs to take the limit $n\rightarrow 0$ in the end. We have seen that 
the RG to first order is replica symmetric. However, 
 to describe the localized phases properly, it is necessary to study
 solutions which break  replica symmetry in the limit
$n\rightarrow 0$. 

The concepts we will use in this section have been
introduced by Parisi and M\a'ezard in  
Ref.~[\onlinecite{Mezard91}] and further developed by Le Doussal and
Giamarchi in Ref.~[\onlinecite{Ledoussal96}] to study the problem of
interacting electrons in a disordered potential. Here we generalize
the method to the case of two coupled species. Let us fix a few
notations before turning to the main points of the calculation, 
details of which can be found in Appendix~\ref{app:RSB}. We rewrite the
action $S_0$ in Fourier space as 
\begin{equation}
S^0 = \frac{1}{2}\frac{1}{\beta L} \sum_{q,i\omega_n}\phi_\alpha^a(q,i\omega_n)(G_0^{-1})_{\alpha \beta}^{ab}(q,i\omega_n)\phi_\beta^b(-q,-i\omega_n),
\end{equation}
where $\alpha,\beta$=$f,b$ while Latin indices $a,b$ run from 1 to $n$,
the number of replicas. There are two implicit summations over $\alpha,\beta$ and $a,b$. $(G_0^{-1})_{\alpha \beta}^{ab}$ is a $2n\times 2n$ matrix whose structure is:
\beq
G_0^{-1}=\left(
\begin{array}{cc}
\frac{v_f}{\pi K_f}\left[ \frac{\omega_n^2}{v_f^2}+q^2\right]\mathbbm{1}_n & \frac{U_{bf}}{\pi^2}q^2\mathbbm{1}_n \\\\
\frac{U_{bf}}{\pi^2}q^2\mathbbm{1}_n  & \frac{v_b}{\pi K_b}\left[ \frac{\omega_n^2}{v_b^2}+q^2\right]\mathbbm{1}_n 
\end{array}
\right)
\eeq
with $\mathbbm{1}_n$ the $n\times n$ unit matrix. As stated earlier,
we want to replace $S_{rep}$ by its best Gaussian approximation,
$S_G$,  with 
\beq
(G^{-1})_{\alpha \beta}^{ab} = (G_0^{-1})_{\alpha
  \beta}^{ab}-\sigma_{\alpha \beta}^{ab}, 
\eeq  
  
and $\sigma_{\alpha\beta}^{ab}$ the self-energy. The best $G$ is obtained by minimizing
the variational free energy, 
$F_{\rm var}$=$F_G  +  \langle S-S_G\rangle_G/\beta$ with respect to $G_{\alpha\beta}^{ab}$. We find the following expression for $F_{var}$:
\bea
F_{var} &=&  - \frac{1}{2\beta}\sum_{q,i\omega_n} \textrm{Tr} \log[G(q,i\omega_n)]\nn \\ &+&\frac{1}{2}\sum_{\alpha,\beta}\sum_{q,i\omega_n}\left(G_0^{-1}\right)_{\alpha\beta}(q,i\omega_n)\textrm{Tr}[G_{\alpha\beta}(q,i\omega_n)] 
\nn \\ &+& \frac{1}{2} \sum_{a,b}  L \int d\tau \left[ V_F[F^{ab}(\tau)] + V_B[B^{ab}(\tau)] \right]
\eea
with:
\bea
F^{ab}(\tau) &=& \langle \left[ \phi_f^a(x,\tau) - \phi_f^b(x,0) \right]^2 \rangle_G,
\\B^{ab}(\tau) &=& \langle \left[ \phi_b^a(x,\tau) - \phi_b^b(x,0) \right]^2 \rangle_G,
\eea
and $V_F(x) = -2{\rho_f^2 D_f}e^{-2x}$ and $V_B(x) =
-2{\rho_b^2 D_b}e^{-2x}$. In the case of static disorder,
off-diagonal quantities (say, $F^{ab}$ or $B^{ab}$ with $a\neq b$) do
not depend on time\cite{Ledoussal96}. This is because
off-diagonal elements describe correlations between replicas locked 
to different minima, but experiencing the same disorder.  
The experienced random potential being static, these correlations 
are also time-independent. 
Bearing this in mind we then derive the following saddle-point equations: 
\bea
\sigma_{ff}^{aa}(q,\omega_n) &=& 2\int_0^\beta d\tau \left(1-\cos[\omega_n\tau]\right) V'_F(F^{aa}(\tau))\nn \\ &+& 2\int_0^\beta d\tau \sum_{b\neq a} V_F'[F^{ab}], \\
\sigma_{bb}^{aa}(q,\omega_n) &=& 2\int_0^\beta d\tau \left(1-\cos[\omega_n\tau]\right) V'_B(B^{aa}(\tau))\nn \\ &+& 2\int_0^\beta d\tau \sum_{b\neq a} V_B'[B^{ab}], \\
\sigma_{ff}^{ab}(q,\omega_n) &=& -2\beta \delta_{n,0} V'_F(F^{ab}) \; \; \; (a\neq b), \\
\sigma_{bb}^{ab}(q,\omega_n) &=& -2\beta \delta_{n,0} V'_B(B^{ab}) \; \; \; (a\neq b), \\
\sigma_{fb}^{ab}(q,\omega_n) &=& \sigma_{bf}^{ab}(q,\omega_n) = 0 \; \; \; \; \; \; \;  (\forall \ a,b).
\eea
The next step is to take the limit $n \rightarrow 0$. We follow
Parisi's parameterization of $0\times 0$ matrices.\cite{Mezard91} 
If $A$ is a matrix in replica space, taking $n$ to $0$, it can be parameterized by a couple
$(\tilde{a},a(u))$, with $\tilde{a}$ corresponding to the (equal)
replica-diagonal elements and $a(u)$ a function of $u\in [0,1]$, 
parameterizing the off-diagonal elements. 
Then the self-energy matrix is expressed as 
\\ 
\beq
\sigma(q,\omega_n=0) = 
\left( 
\begin{array}{cc}
[\tilde{\sigma}_f,\sigma_f(u)] & 0 \\\\
0 & [\tilde{\sigma}_b,\sigma_b(u)]
\end{array}
\right)\;.
\eeq
Then we proceed to invert $G^{-1}$ in order to solve the saddle-point
equations. To do so we are led to make assumptions on the {\it
  off-diagonal} functions $\sigma_f(u)$ and $\sigma_b(u)$. Either we
look for replica-symmetric (RS) solutions, with constant $\sigma_f(u)$
and/or $\sigma_b(u)$, or replica symmetry breaking (RSB) solutions,
with non-constant off-diagonal functions. First shall we focus on the
phase with localized fermions and superfluid bosons then on the phase
in which both species are pinned. In both cases we find a consistent
solution by first making an  intelligent guess for  the structure of 
the replica symmetry-breaking solution, and then verifying its stability.

\subsubsection{Phase with localized fermions and superfluid bosons}

As was shown in Ref.~[\onlinecite{Ledoussal96}], the localized phase
of fermions in a disordered potential is well-described by a RSB
solution. More precisely, a level 1 symmetry breaking is required to
describe the localized phase. It means that $\sigma_f(u)$ is a step
function, $\sigma_f(u<u_f) = 0$ and $\sigma_f(u>u_f) = 1$, with $u_f$
a breaking point that needs to be fixed. To describe the phase with
localized fermions and free bosons that is predicted by the RG, we
look for a solution with level 1 RSB in the fermionic sector and
replica symmetry in the bosonic sector. The details of the calculation
are presented in Appendix~\ref{app:RSB}. 

{ We introduce the inverse connected
Green function\cite{Ledoussal96},
\beq
(G^{-1})^c_{\alpha \beta} 
\equiv \lim_{n \rightarrow 0} \sum\limits_{b}(G^{-1})^{ab}_{\alpha \beta},
\eeq
which, using Parisi's notation becomes
\be 
(G^{-1})^c_{\alpha \beta} = \widetilde{G^{-1}}_{\alpha
    \beta} - \int_0^1 du \ G^{-1}_{\alpha \beta}(u)\;.
\ee
For this choice of replica symmetry breaking it can be cast into
\bea
(G^{-1})^c_{ff} &=& (G_0^{-1})_{ff}(q,i\omega_n) + I_F(\omega_n) + \Sigma_F(1-\delta_{n,0}), \nn \\
(G^{-1})^c_{bb} &=& (G_0^{-1})_{bb}(q,i\omega_n) + I_B(\omega_n), \nn \\
(G^{-1})^c_{fb} &=& (G^{-1})^c_{bf} = (G_0^{-1})_{fb}(q,i\omega_n),
\eea
}
where
\bea
I_f(\omega_n) &=& 2  \int_0^\beta d\tau \left(1-\cos[\omega_n\tau]\right) \times(V'_F[\widetilde{F}(\tau)]- V_F'[F]),\nn \\  \\
\label{Ib}I_b(\omega_n) &=& 2 \int_0^\beta d\tau \left(1-\cos[\omega_n\tau]\right) V'_B[\widetilde{B}(\tau)], \nn \\
\eea
and
\beq
\label{eq:sigmaf_first}
\Sigma_F = 2\beta  u_f V'_F[F].
\eeq
{ The structure of the connected propagator allows to identify several features of the RSB solutions. A mass term $\Sigma_F$ is here 
generated and we will confirm later in this section that it indeed controls the localization length. However $(G^{-1})^c_{ff}(q=0,\omega_n = 0)$ is still $0$, as it should for a system that is, after averaging on disorder, translationaly invariant. Finally, RSB endows the Green's functions with a new dynamical content through the functions $I_F(\omega_n)$ and $I_F(\omega_n)$.\\

The system of equations is effectively closed by writing an equation for the breaking point $u_f$. This is done, by inspecting the stability of such a solution and explicitely requiring the marginality of the so-called \emph{replicon mode}. This choice is made on physical grounds -- as it gives sensible results for dynamical quantities, such as the conductivity -- following the path set in Ref. \onlinecite{Ledoussal96} (details about the calculation are given in appendix \ref{app:RSB}). Finally, equation \eqref{eq:sigmaf_first} is replaced by
}

\bea
\Sigma_f^{3/2} = \frac{8}{\sqrt{1-g^2}} {\rho_f^2 D_f}\left( \frac{\pi K_f}{v_f}\right)^{1/2} e^{-2F} 
\eea

{ Note that $F$, $\widetilde{F}$ and $\widetilde{B}$ are obtained through inversion of $G^{-1}$. They read}
\begin{widetext}
\bea
F = \frac{2}{\beta L}\sum_{q,i\omega_n} \frac{\pi K_f}{v_f} \frac{ \left(\frac{\omega_n^2}{v_b^2}+q^2+\hat{I}_b(\omega_n)\right)}{\left(\frac{\omega_n^2}{v_b^2}+q^2+\hat{I}_b(\omega_n)\right)\left(\frac{\omega_n^2}{v_f^2}+q^2+\hat{I}_f(\omega_n)+\hat{\Sigma}_f\right)-g^2q^4} \\
\widetilde{F}(\tau) = \frac{2}{\beta L} \sum_{q,i\omega_n} \frac{\pi K_f}{v_f}\left(1-\cos[\omega_n\tau]\right) \frac{\frac{\omega_n^2}{v_b^2}+q^2+\hat{I}_b(\omega_n)}{\left(\frac{\omega_n^2}{v_b^2}+q^2+\hat{I}_b(\omega_n)\right)\left(\frac{\omega_n^2}{v_f^2}+q^2+\hat{I}_f(\omega_n)+\hat{\Sigma}_f\right)-g^2 q^4} \\
\widetilde{B}(\tau) = \frac{2}{\beta L} \sum_{q,i\omega_n} \frac{\pi K_b}{v_b}\left(1-\cos[\omega_n\tau]\right) \frac{\frac{\omega_n^2}{v_f^2}+q^2+\hat{I}_f(\omega_n)+\hat{\Sigma}_f}{\left(\frac{\omega_n^2}{v_b^2}+q^2+\hat{I}_b(\omega_n)\right)\left(\frac{\omega_n^2}{v_f^2}+q^2+\hat{I}_f(\omega_n)+\hat{\Sigma}_f\right)-g^2 q^4} 
\eea
\end{widetext}
One can  recognize in $F$ and $\widetilde{F}$ the fermionic
propagator and in $\widetilde{B}$ the bosonic propagator. We have introduced $\hat{I}_f = \frac{\pi K_f}{v_f} I_F$,  $\hat{\Sigma}_F = \frac{\pi
  K_f}{v_f} \Sigma_F$, and used similar notations for bosons. The
complete numerical solution of this self-consistent set of functional
equations is beyond the scope of the present paper, and the presence
of the functions $\hat{I}_f(\omega_n)$ and $\hat{I}_b(\omega_n)$
indeed makes the situation complicated. We proceed in several steps in
order to analyze the equations. 

First we take $\hat{I}_f(\omega_n)$ and $\hat{I}_b(\omega_n)$ to be
zero. In this case, our  variational approach is analogous to the
theory of
Fukuyama and Suzumura, summarized in Sec.~\ref{sec:var}, however with a
more accurate treatment of quantum fluctuations that does not require
a detailed knowledge of the underlying classical solution. We show
that within this approximation we obtain sensible results in
good agreement with the RG results. 

First let us
refine the RG analysis of Sec. \ref{sec:RG} by looking at the flow of
$\tilde{D}_b$ when bosons are coupled to localized fermions. To do so,
we perturb the Gaussian action $S_G$ with a disorder term coupling
only to bosons: 
\beq
S = S_G-\frac{D_b \rho_b^2}{\Lambda^3} \sum_{a,b} \Lambda^3 \int dx d\tau d\tau ' \cos[2\phi_b^a(x,\tau) - 2\phi_b^b(x,\tau ')]\;.
\eeq
The quadratic propagator $G$ is replica symmetric in the bosonic
sector and has level-1 RSB  in the fermionic sector. To obtain the
flow of $\tilde{D}_b$, we proceed as explained in 
Appendix~\ref{app:RG}, 
and integrate out high-momentum degrees of freedom,
between $\Lambda'$ and the original cut-off $\Lambda$. To first order
in $D_b$, the flow equation is obtained by requiring that: 
\beq
\tilde{D}_b(\Lambda') = \tilde{D}_b(\Lambda)\left(
\frac{\Lambda'}{\Lambda}\right)^{-3} \langle
e^{i2\phi_b^a(x,\tau)}\rangle_>^2. 
\eeq
$\langle e^{i2\phi_b^a(x,\tau)}\rangle_>^2$ is only related to the
diagonal part of $G_{bb}$, that is
$\widetilde{G}_{bb}(q,\omega_n)$. We find: 
\beq
\langle e^{i2\phi_b^a(x,\tau)}\rangle_>^2 = \exp\left[-\int_{\Lambda'}^{\Lambda} dq \ \mathcal{J}_b(q)\right],
\eeq
with
\begin{widetext}
\beq
\label{eq:Jb}
\mathcal{J}_b(q) = \frac{2 K_b \left[ t(q^2 + \hat{\Sigma}_f)+
    q\sqrt{q^2(1-g^2)+\hat{\Sigma}_f}\right]}{q\sqrt{q^2(1-g^2)+\hat{\Sigma}_f}\sqrt{q^2(1+t^2)+t^2
    \hat{\Sigma}_f + 2tq\sqrt{q^2(1-g^2)+\hat{\Sigma}_f}}}.  
\eeq
\end{widetext}
Finally, by taking $\Lambda' = \Lambda(1+dl)$,the flow equation reads
\beq
\frac{d \log D_b}{dl} = 3 - \Lambda(l) \mathcal{J}_b(\Lambda(l)).
\label{good_RG}
\eeq
$\mathcal{J}_b(q)$ has a power law decay at small and large $\Lambda$
but with different prefactors. Indeed: 
\bea
\Lambda \mathcal{J}_b(\Lambda) &= 2K_b ~&\textrm{when}~ \Lambda
\ll \left(\hat{\Sigma}_f/\sqrt{1-g^2}\right)^{1/2}, \\ 
\Lambda \mathcal{J}_b(\Lambda) &= X_b  ~&\textrm{when}~ \Lambda
\gg \left(\hat{\Sigma}_f/\sqrt{1-g^2}\right)^{1/2}. 
\eea

\begin{figure}[t]
\includegraphics[width=8cm]{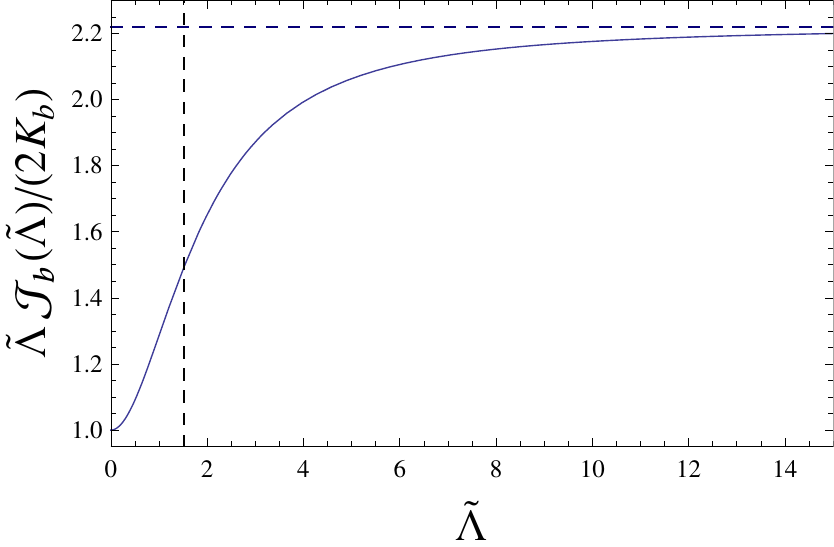}
\caption{Crossover behavior of the function $\mathcal{J}_b$ as defined
  in Eq.~\eqref{eq:Jb}, for $t=3$ and $g=0.9$. Here we have plotted
  $\tilde{\Lambda}\mathcal{J}_b(\tilde{\Lambda})/(2K_b)$, with
  $\tilde{\Lambda}=\Lambda/\sqrt{\hat{\Sigma}_f}$. It goes to $1$ at
  low momenta and saturates at $X_b/(2K_b)$ as momentum is
  increased. We identify a crossover region around $\Lambda =
  (\hat{\Sigma}_f/\sqrt{1-g^2})^{1/2}$ (vertical dashed line). } 
\label{Fig:Jb}
\end{figure}
\begin{figure}[b]
\includegraphics[width=8cm]{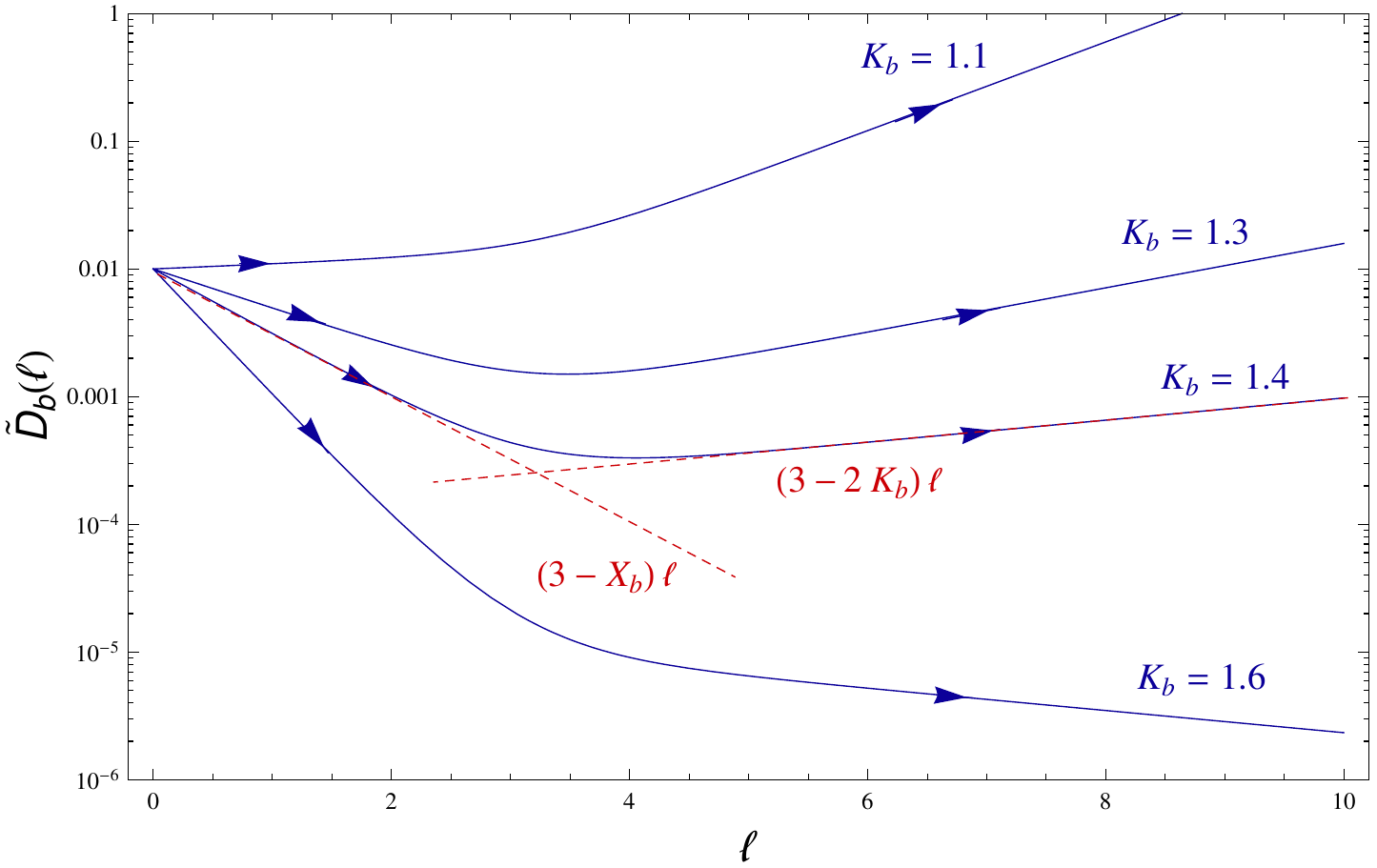}
\caption{Modified RG flow for bosons using the Gaussian variational solution.}
\label{Fig:flow_bosons1}
\end{figure}

The cross-over of the anomalous dimension appearing in Eq.~\eqref{good_RG}
and the corresponding flow of the disorder are illustrated in
Figs.~\ref{Fig:Jb} and \ref{Fig:flow_bosons1}. 
The flows in Fig.~\ref{Fig:flow_bosons1}
confirm entirely  our intuitive arguments in Section~\ref{sec:RG}. The
 fermionic mass $\hat{\Sigma}_f$ sets a length scale below which 
bosons interact with their bare interactions. The corresponding length scale can thus be identified as the fermionic localization length, 
$L_f$, 
\beq
L_f \equiv \left(\frac{\sqrt{1-g^2}}{\hat{\Sigma}_f}\right)^{1/2}\;.
\eeq
We shall return to this equation in the next section when we
compute correlation functions for fermions. This RG analysis 
also confirms that below $K_b=3/2$ bosonic disorder is 
relevant, and the variational solution with
RSB only in the fermionic sector is insufficient. To 
proceed and describe the phase where disorder is relevant 
for both species we shall need
a variational solution with replica symmetry breaking in both the
fermionic and the bosonic sector.

\subsubsection{Phase with both species pinned by disorder, and
  complete phase diagram}

{\em Case of Faster fermions, $v_f>v_b$.} In this case 
we found that in the regime with both species localized 
it is impossible to obtain a 
self-consistent solution
with only level 1 RSB in both sectors, and one needs to allow 
for level 2 RSB in at least one of the sectors. It turns out, that 
with level 2 RSB the marginal stability (marginalitly) of
 the  saddle point solution can be satisfied, and that 
physically meaningful results are thus obtained. Level 2 RSB is 
thus suffucient to describe this phase.  
The structure of the solution, and the derivation of the corresponding
integral equations
are detailed in Appendix~\ref{app:RSB}. The resulting 
(rather complicated) integral equations were solved numerically.

\begin{figure}[h]
\includegraphics[width=8cm]{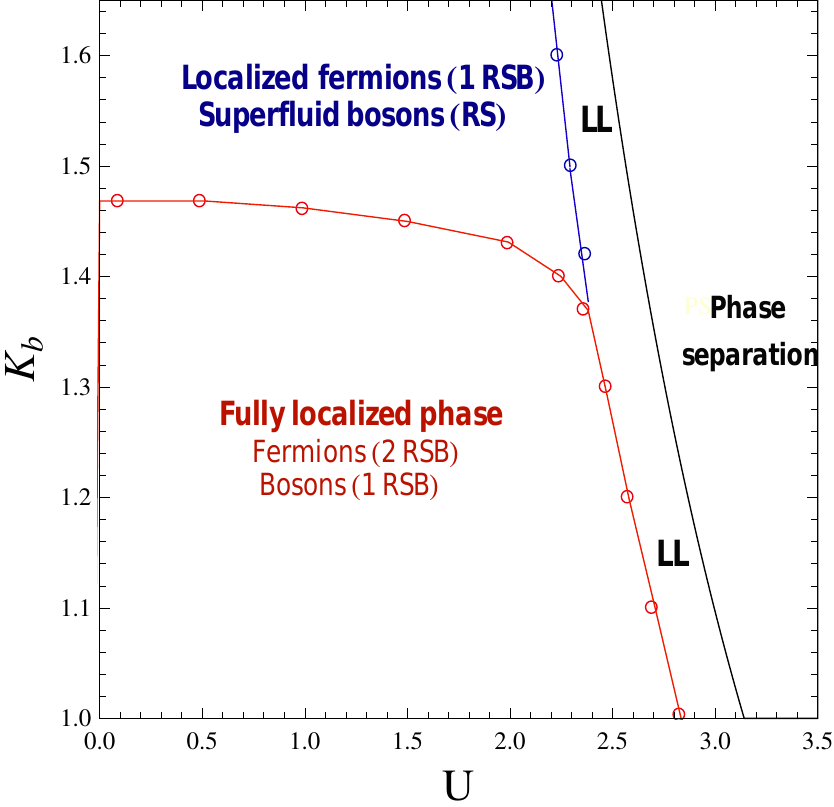}
\caption{Phase diagram obtained by the Gaussian variational method in
  replica space as a function of $K_b$ and $U=U_{bf}/\sqrt{v_b v_f}$. 
We took $\tilde{D}_f = \tilde{D}_b = 0.005$ and  $t=v_f/v_b=3$.}  
\label{Fig:phase_diagram_RSB_1}
\end{figure}

\begin{figure}[t]
\includegraphics[width=8cm]{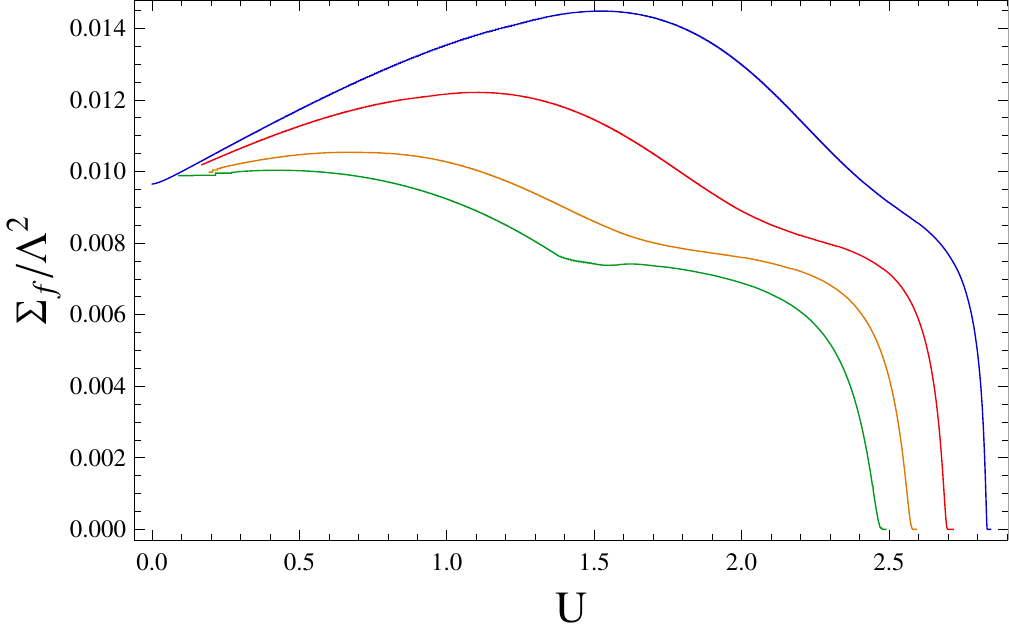}
\includegraphics[width=8cm]{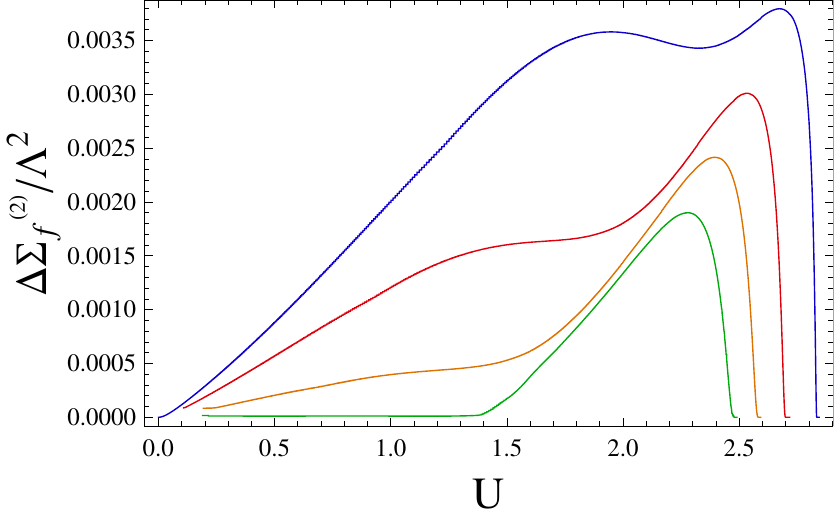}
\caption{Fermion mass $\hat{\Sigma}_f$ and
  $\Delta\hat{\Sigma}_f^{(2)}$ 
as a function of $U=U_{bf}/\sqrt{v_f v_b}$, for $t=v_f/v_b$
and $K_b=1,1.1,1.2,1.3$ (top to bottom). 
 Note that  $\Delta\hat{\Sigma}_f^{(2)}$ goes to zero simultaneously with $\hat{\Sigma}_f$ as $U$ is increased.} 
\label{Fig:sigma_f}
\end{figure}

\begin{figure}[b]
\includegraphics[width=8cm]{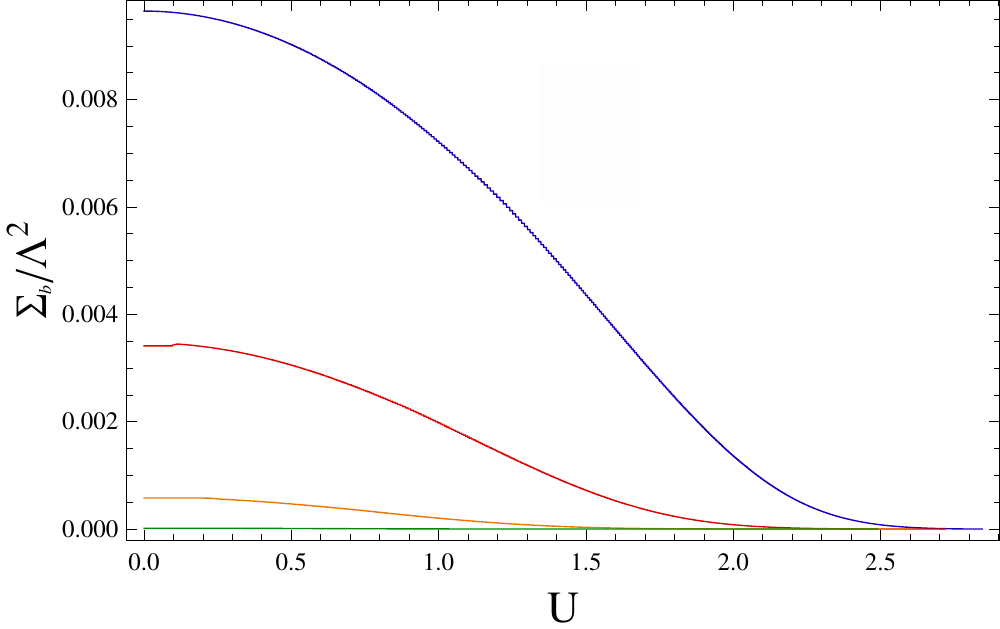}
\caption{The boson mass $\hat{\Sigma}_b$ as a function of
  $U=U_{bf}/\sqrt{v_f v_b}$, 
for $t=v_f/v_b$ and $K_b=1,1.1,1.2,1.3$ (top to bottom). 
  Note that $\hat{\Sigma}_b$ goes to zero, simultaneously with
  $\hat{\Sigma}_f$ } 
\label{Fig:sigma_b}
\end{figure}

For $v_f > v_b$ we always 
find a stable numerical solution with 2RSB for fermions and 1RSB for
bosons with the following self-energy structure:  
$\sigma_f(u)$ is a 2-step function, $\sigma_f(u<u_1) = 0$,
$\sigma_f(u_1<u<u_2) = \sigma_f^{(1)}$ and $\sigma_f(u_2<u<1) =
\sigma_f^{(2)}$, while $\sigma_b(u)$ is a 1-step function,
$\sigma_b(u<u_2)=0$, $\sigma_f(u_2<u<u_1) = \sigma_b^{(2)}$.  Note
that the structure of the solution is reminiscent of the physical 
arguments we
developed in section \ref{sec:var} for the classical solution. We
argued there that in the situation where $L_b > L_f$ -- which is the case in
the phase diagram of Fig. \ref{Fig:phase_diagram_RSB_1}, and apparent
on Figs. \ref{Fig:sigma_f} and \ref{Fig:sigma_b} -- the fermion
density wave should have a nested structure as it breaks into domains
to accommodate the random phase and the bosonic density. 

In Figs.~\ref{Fig:sigma_f} and \ref{Fig:sigma_b}
we show examples of the solution for various
values of $K_b$ as $U_{bf}$ is increased. Note that the fermion 
mass is related to the 2-step self-energy as 
\beq
\Sigma_f =  \Sigma_f^{(1)} + \Delta \Sigma_f^{(2)}, 
\eeq
with  $\Sigma_f^{(1)} = u_1 \sigma_f^{(1)}$, and $\Delta
\Sigma_f^{(2)} = u_2 \left(\sigma_f^{(2)} -
\sigma_f^{(1)}\right)$. For the bosonic mass we have $\Sigma_b = u_2
\sigma_b^{(2)}$ and $L_b = \hat{\Sigma}_b^{-1/2}$. The boundary of the
region where the level-2 RSB solution exists is obtained by the
condition,
 $\Delta \Sigma_f^{(2)}=0$. This condition is fulfilled
 either when  $\Sigma_f^{(1)} \neq 0$,
$\Sigma_b =0$ and the system is in the region with 1 level RSB in the
fermionic sector only, or when $\Sigma_f^{(1)} = 0$, $\Sigma_b =0$ and the
system is in the replica symmetric phase.

Let us notice two important points here. For any  $U_{bf}\neq 0$
we obtain
$\hat{\Sigma_b} \leq \hat{\Sigma_f}$, impying that the fermionic
localization length is smaller than the bosonic localization
length. There are two reasons for that. First, if $K_b > 1$, quantum
fluctuations are more important for bosons than for fermions (for
which $K_f = 1$), and tend to increase the localization length of
bosons with respect to that of fermions (disorder pins the fermion
density wave more efficiently). Second, Bose-Fermi interactions
enhance superfluid correlations of both components of the
mixture. Nevertheless, as we pointed out in Sec.~\ref{sec:model},
superfluid correlations of the slower species are more strongly
enhanced.  This behavior appears in Figs.~\ref{Fig:sigma_f} and
\ref{Fig:sigma_b}, where one can see that the mass of the slow species
(here bosons) decreases to extremely small values, long before the
true transition to the Luttinger liquid phase takes place, 
whereas the mass of the
fast species (here fermions) is weakly renormalized, excepting the close
proximity of the transition. This is a crucial point for the possible
observation of the fully localized phase: in a finite-size system,
bosons could appear as superfluid simply because their localization
length exceeds the size of the trap, even though they should be
localized in the thermodynamic limit.\\

\begin{figure}[h]
\includegraphics[width=8cm]{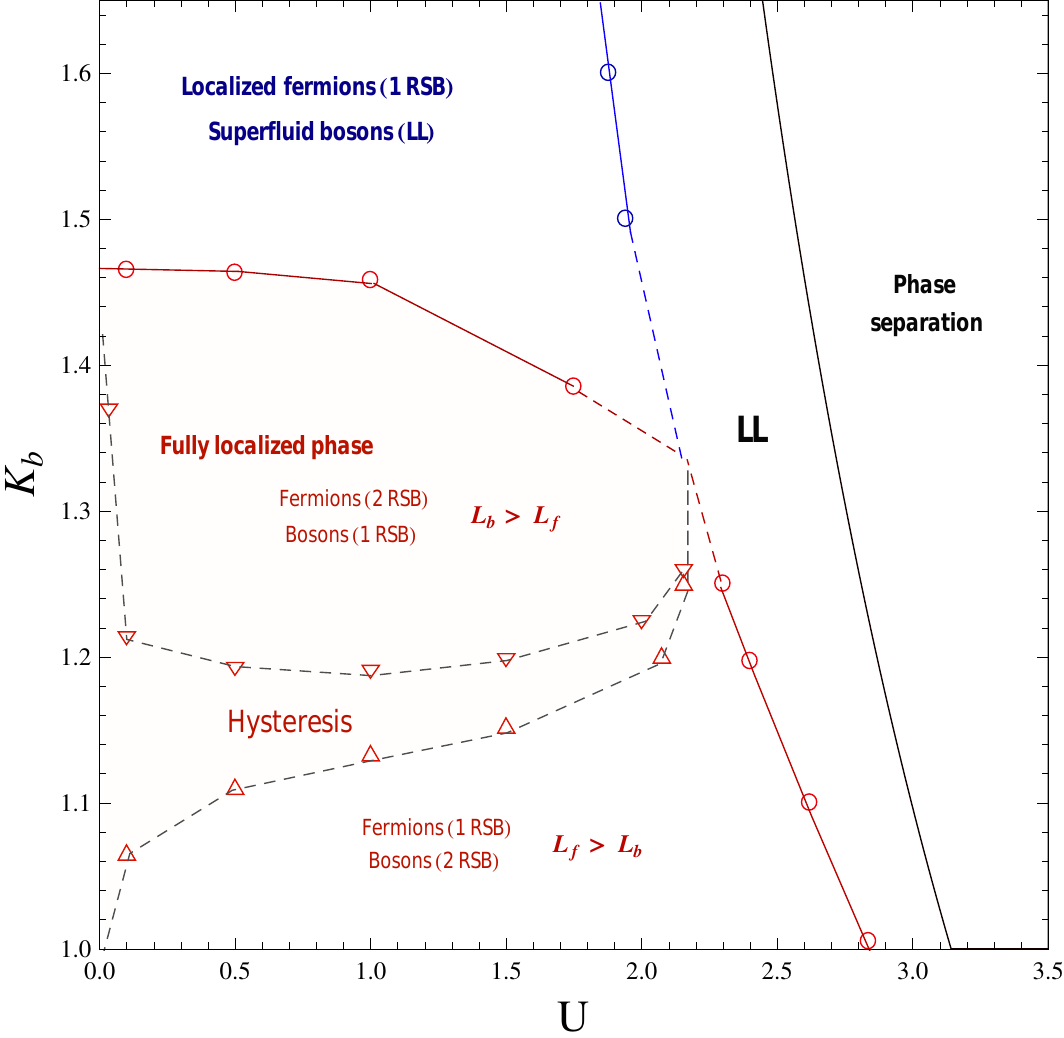}
\caption{Phase diagram obtained by the Gaussian variational method in
  replica space for $\tilde{D}_f = \tilde{D}_b = 0.005$ and $v_f/v_b
  =1/3$. In the fully localized phase there is crossover between two
  regions where the order of replica symmetry breaking is
  reverted.  } 
\label{Fig:phase_diagram_RSB_2}
\end{figure}

{\em Case of faster bosons, $v_f<v_b$}
Now let us turn to the more delicate case of $v_b > v_f$. In the
RG-based tentative phase diagram of Fig.~\ref{fig:phase_diagram_RG_2} 
two intermediate regions appear, where only one of the species appears
to be localized. However, for the same reasons as in the previous case
$v_f > v_b$, 
these two phases are just artifacts of the  RG procedure, and 
in each of them we need to repeat our two step localization
argument. Correspondingly, a 2-step RSB shall appear  in the  variational
solution,  too. However, before discussing
the phase diagram of Fig.~\ref{Fig:phase_diagram_RSB_2}, let us make a
few remarks to gain a reasonable intuition for the results. 

In the  $v_b < v_f$ case we had established that the bosonic localization
length was always larger than the fermionic one, because of (i) larger
quantum fluctuations ($K_b \geq K_f$) and (ii) because of the
enhancement of superluid correlations of the slow species because of
interactions. In the present case the situation is somewhat
different. Along the line $K_b = K_f = 1$, bosons are now the faster
species, and their localization length is now smaller than that of
fermions. This translates into an inversion of the levels of symmetry
breaking in the fermionic and bosonic sectors. 
However as soon as $K_b > 1$, larger quantum
fluctuations for bosons counteract this effect, and 
if $K_b$ is large enough, then the bosonic localization
length exceeds the fermionic one again, and the order of replica
symmetry breaking is reversed . 

This can be clearly seen in Fig.  \ref{Fig:phase_diagram_RSB_2},
where we show the phase diagram emerging from a full numerical
solution of the integral equations. We indeed find an inversion of the levels 
of symmetry breaking. We have to point out here  that the numerical
solution has a {\em hysteresis}: we obtain a different phase 
boundary by increasing $K_b$ at fixed $U_{bf}$ instead of decreasing
it.  This can be explained by the asymmetric structure of the equations for the
2RSB+1RSB solutions (see Appendix~\ref{app:RSB}, $K_b$ playing a singular role). 
The appearing hysteresis could be a signature of a first order
transition, too, however, on physical grounds we tend to believe that
there is simply a cross-over between the 2RSB+1RSB and 
1RSB+2RSB regimes. We emphasize again that the difference between the two cases, $v_f >
v_b$ and $v_b > v_f$, is related to
the inversion of length scales that can only take place for $v_b >
v_f$.

\section{Experimental consequences 
}
\label{sec:observables}

In this section we present a few observables that we believe would
help to characterize the various phases in an experiment on cold atoms.  
One possibility is provided by 
time-of-flight (TOF) experiments, measuring the momentum
distribution of the gas inside the trap. For bosons, TOF provides 
an indirect measurement of the superfluid correlations (and the single-particle
Green's function), the behavior of which varies significantly from the
localized to the superfluid phase. Another usual probe is  Bragg
scattering, giving access to structure factors, that is the
Fourier Transform of the density-density correlation functions. The
latter quantity we already computed in
Ref.~\onlinecite{Crepin10b}. Here we briefly present how one can
compute both density correlations (to compute the structure factor for
instance) and superfluid correlations from the variational method. 
Then we apply these results to study the relevant physical quantities.    


\subsection{Density correlations }

\label{sec:density_corr}

Density correlations are of the form $\overline{\langle \rho(x,t)
  \rho(0,0) \rangle}$, where, as usual, brackets stand for the quantum
averaging and overlining denotes averaging over
disorder realizations. In practice we will use the variational 
solution in replica
space, and more precisely the diagonal Green's functions. 
As experimentalists can address each
species separately, we will focus on  $\overline{\langle \rho_f(x,t)
  \rho_f(0,0) \rangle}$ and $\overline{\langle \rho_b(x,t) \rho_b(0,0)
  \rangle}$ and shall not consider cross-terms for now.  
It is instructive to look at the two particular terms
\bea
C_f(x) &=& \overline{\langle e^{i2\phi_f(x)} e^{-i2\phi_f(0)}\rangle}, \\
C_b(x) &=& \overline{\langle e^{i2\phi_b(x)} e^{-i2\phi_b(0)}\rangle},
\eea
which characterize the $q\approx2\pi \rho_{b/f}$-momentum density-density
correlations (see Eq.~\ref{eq:density_bosonization}). 
In order ot compute $C_f$ and $C_B$, one first needs to recall the
gauge transformation that we performed to get rid of the forward
scattering processes. Once included they lead to an
exponential decay of $C_f$ and $C_b$ in every phase, localized or
not. The general form of $C_\alpha(x)$ is therefore 
\bea
C_f(x) &=& \exp\left[ -x/L_{f,{\rm FW}}\right] \times \nn \\ &\times&
\exp \left[ -2 \frac{1}{\beta L} \sum_{q,\omega_n}
  (1-\cos[qx])\widetilde{G_{ff}}(q,\omega_n)\right] \\ 
C_b(x) &=& \exp\left[ -x/L_{b,{\rm FW}}\right] \times \nn \\ &\times&
\exp \left[ -2 \frac{1}{\beta L} \sum_{q,\omega_n}
  (1-\cos[qx])\widetilde{G_{bb}}(q,\omega_n)\right] 
\eea
where 
\bea
L_{f,{\rm FW}} &=& \frac{(1-g^2)^2}{K_f^2/v_f^2} \left[ \alpha_f -
  \alpha_b g\sqrt{\frac{v_f}{v_b}\frac{K_b}{K_f}}
  \right]^{-2} \hspace{-0.2cm}D_f^{-1}, \\ 
L_{b,{\rm FW}} &=& \frac{(1-g^2)^2}{K_b^2/v_b^2} \left[ \alpha_b - \alpha_f g\sqrt{\frac{v_b}{v_f}\frac{K_f}{K_b}} \right]^{-2} \hspace{-0.2cm}D_b^{-1},
\eea
are length scales related to disorder forward scattering.
In the localized phases, backscattering also leads to an exponential
decay of these correlations functions, and it  might therefore be
difficult to disentangle contributions from forward and backward
scattering. Therefore, experimentally, one should rather focus on 
correlation functions, which are not influenced by the 
forward scattering contribution (see Sec.~\ref{sec:structure} and \ref{sec:SF}). 
It is nevertheless instructive
to write down the explicit form of $C_f(x)$ and $C_b(x)$.  

Let us start with the Luttinger liquid phase. There, the mixture is
not pinned by disorder and the self-energies $\sigma_f$ and $\sigma_b$
are zero. The inversion of $G^{-1}$ leads to 
\bea
\widetilde{G_{ff}}(q,\omega_n) &=& \frac{\pi K_f}{v_f} \frac{q^2 +
  b(\omega_n)}{[q^2 + b(\omega_n)][q^2 + f(\omega_n)]-g^2q^4},\nn
\\ \\ 
\widetilde{G_{bb}}(q,\omega_n) &=& \frac{\pi K_b}{v_b} \frac{q^2 +
  b(\omega_n)}{[q^2 + f(\omega_n)][q^2 + f(\omega_n)]-g^2q^4}. \nn \\ 
\eea
Here we have introduced the following general notation 
\bea
b(\omega_n) &=&
\omega_n^2/v_b^2 + \hat{I}_b(\omega_n) + \hat{\Sigma}_b\;,
\\
f(\omega_n) &= &\omega_n^2/v_f^2 + \hat{I}_f(\omega_n) +
\hat{\Sigma}_f\;.
\eea
In the present case $ \hat{\Sigma}_f = \hat{\Sigma}_b = 0$, and 
therefore we recover the propagators of the
Luttinger liquid, with simply a renormalization of the frequency
behavior by the functions $\hat{I}_b(\omega_n)$ and
$\hat{I}_f(\omega_n)$. Note that these functions are directly
proportional to $\tilde{D}_f$ et $\tilde{D}_b$. Although we have not
solved the self-consistency equations for
$\hat{I}_f(\omega_n)$ and $\hat{I}_b(\omega_n)$,
we expect that in the weak disorder limit they do not modify
drastically the propagators. In the Luttinger liquid phase, $C_f$ and
$C_b$ thus decay algebraically at short distances, however, 
this algebraic decay is cut at long distances by the
exponential decay due to disorder-induced forward scattering processes.

Now let us turn to the phase where fermions are localized and bosons
superfluid. Here we found a variational solution with 1RSB in
the fermionic sector. The inversion of $G^{-1}$ now leads to 
\bea
\widetilde{G_{ff}}(q,\omega_n) &=& \frac{\pi K_f}{v_f} 
\Biggl(
\frac{q^2 +
  b(\omega_n)}{[q^2 + b(\omega_n)][q^2 + f(\omega_n)]-g^2q^4} \nn \\ 
&+&  \delta_{n,0} 
\frac{1}{1-g^2}\frac{\sigma_f}{q^2[q^2(1-g^2) + \hat{\Sigma}_f]}\Biggr)\;,
\\   
\widetilde{G_{bb}}(q,\omega_n) &=& \frac{\pi K_b}{v_b}
\Biggl(
 \frac{q^2 +
  f(\omega_n)}{[q^2 + b(\omega_n)][q^2 + f(\omega_n)]-g^2q^4} \nn 
\\ 
&+&  \delta_{n,0} 
 \frac{g^2}{1-g^2}
\frac{\sigma_f}{q^2[q^2(1-g^2) + \hat{\Sigma}_f]}
\Biggr)\;
.   
\eea
In each propagator, the first term controls the algebraic short distance decay of correlations. The second term, once the sum over momenta is done,  leads to a long distance exponential decay. Indeed,
\bea
\frac{1}{L} \sum_q \frac{1-\cos[qx]}{q^2[q^2(1-g^2)+\hat{\Sigma}_f]} &=& \nn \\
 && \hspace{-3.7cm}\frac{1}{2\hat{\Sigma}_f} \left[ x + \sqrt{\frac{1-g^2}{\hat{\Sigma}_f}}\left(e^{-x\sqrt{\frac{\hat{\Sigma}_f}{1-g^2}}} -1 \right)\right].
\eea
Furthermore $\sigma_f = \beta \frac{v_f}{\pi K_f} \sqrt{1-g^2}\Sigma_f^{3/2}$, and one can isolate in $C_f$, a decaying exponential of the form
\beq
C_f(x) \sim e^{-x/L_f},
\eeq
with
\beq
L_f = \left( \frac{1-g^2}{\hat{\Sigma}_f}\right)^{1/2},
\eeq
which we are tempted to identify with the localization
length. Remarkably, although only the fermions are localized, 
 bosonic correlations also display an extra
exponential decay, with an exponent smaller by a factor $g^2$. This contribution comes directly from the 
Bose-Fermi interaction term $\nabla \phi_f \nabla \phi_b$ 
where $\nabla \phi_f$ acts as a random potential inducing 
forward scattering. The expressions for the propagators in the fully localized phase are given in appendix \ref{app:structure}. At this
stage we think that probing the dynamics of the system, through the dynamical structure factor, would be a better way to test the level of replica
symmetry breaking.

\subsection{Structure factors}
\label{sec:structure}
The dynamical structure factors $S_{b/f}(q,\omega)$ are response 
functions which can be probed through Bragg scattering
experiments.\cite{Inguscio07, Fabbri11}  
They are defined by
\beq\label{eq:sq}
S_{b/f}(q,\omega)=\int\int\, {\rm d}t\,{\rm d}x \, e^{iqx-i\omega t}
\;\overline{\langle \rho_{b/f}(x,t) \rho_{b/f}(0,0) \rangle}.
\eeq
Using Eq. (\ref{eq:density_bosonization}), we can see 
that several Fourier components contribute to the structure factors.
However, at small momenta, $q\ll 2\pi\rho_{b/f}$, it is essentially given by

\beq
S_{b/f}(q,\omega) \approx \int\int\frac{{\rm d}t\,{\rm d}x}{\pi^2}
e^{iqx-i\omega t}\overline{\langle \partial_x\phi_{b/f}(x,t) 
\partial_x\phi_{b/f}(0,0) \rangle}
\eeq
which, in turn, can be computed with the variationnal solution. Note that we ignore here the static contribution from the forward scattering on disorder.\cite{Crepin10b} For fermions it reads: 
\beq
\label{eq:sf_1}
S_{f}({q},\omega) = - \textrm{Im} \left[ q^2 \widetilde{G}_{ff}(q,i\omega_n \to \omega + i\epsilon) \right],
\eeq
and we have a similar expression for bosons. Here, $\widetilde{G}_{ff}$ is the replica-diagonal contribution for the fermion propagator. The complete analytical expression of $S_f$ is given in appendix \ref{app:structure}. To perform the analytical continuation it is necessary to elaborate on the expression of functions $I_f$ et $I_b$. The only thing that seems analytically feasible is to adapt the argument of Ref.~\onlinecite{Ledoussal96} to the case of a mixture. In the fully localized phase, $I_b$ and $I_f$ are given by equations \ref{eq:A_Ib} and \ref{eq:A_If} of appendix \ref{app:RSB} and one can obtain two simple self-consistent equations by assuming that $K_f \ll 1$ and $K_b \ll 1$ (thus taking a sort of classical limit) and expanding the functions $V_f$ and $V_b$ to leading order. In addition, by taking the limit $\Sigma_b \ll \Sigma_f$ we arrive to the following approximate expressions, at low frequency, $I_f(\omega) \sim \alpha_f \omega$ and  $I_b(\omega) \sim \alpha_b \omega$ with
\bea
\label{eq:alphaf}
\alpha_f &=& \sqrt{\Sigma_f} \frac{2}{\sqrt{3}}(1+t^2g^2), \\
\label{eq:alphab}
\alpha_b &=& \sqrt{\Sigma_b} \frac{2}{\sqrt{3}}.
\eea

\begin{figure}[t]
\centering
\includegraphics[width=\columnwidth,clip]{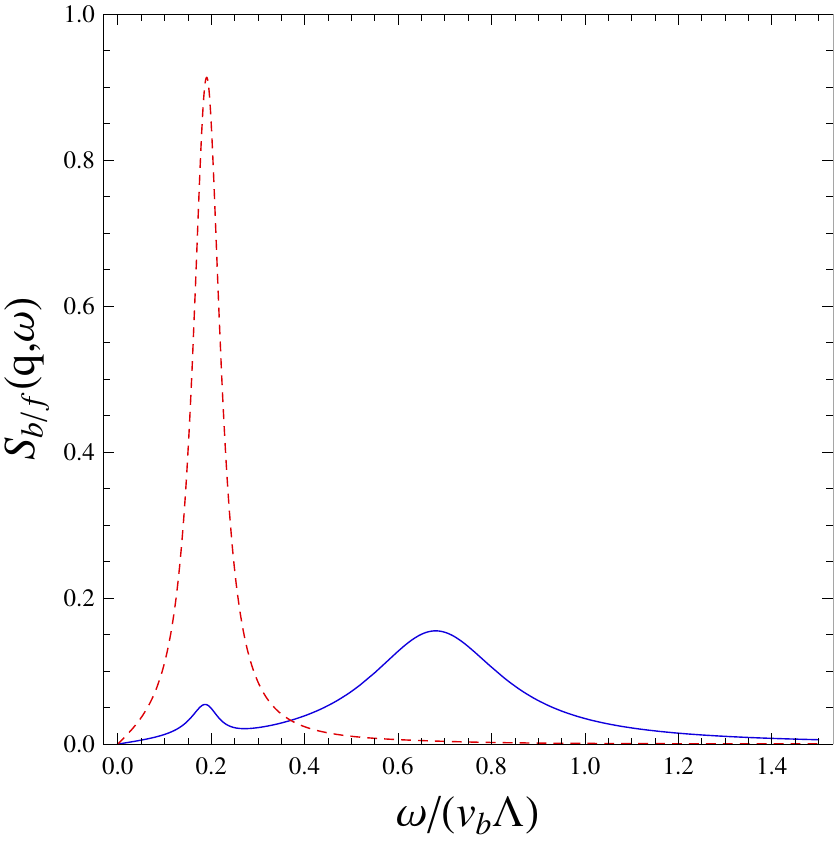}
\caption{\label{Fig:structure_factor} Frequency dependence of the structure factor for bosons (red dashed line) and fermions (blue solid line) in the fully localized phase. Momentum is fixed to $q=0.2 \Lambda$ with $\Lambda$ the UV cut-off of the theory. } 
\end{figure}

If $g=0$ we get back the expression of Ref. \onlinecite{Ledoussal96}. It appears, that the relevant parameters for our model fall off the domain of validity of this approximation. Nevertheless, in order to get at least a qualitative view of the structure factor, we perform the analytical continuation on these expressions for $I_f$ and $I_b$.\\

In the fully localized phase, the structure factor has the profile depicted in Fig. \ref{Fig:structure_factor}. There we took the following parameters: $K_b = K_f = 1$, $v_f/v_b = 3$, $U_{bf}/\sqrt{v_f v_b} = 1.5 \; (g=0.47)$. From the results of Fig. \ref{Fig:sigma_f} and \ref{Fig:sigma_b}, we also took $\hat{\Sigma}_f/\Lambda^2 = 0.014$ and $\hat{\Sigma}_b/\Lambda^2 = 0.0045$.  Several interesting features appear at this level. The fermion structure factor shows a two-peak profile, due to the strongly coupled nature of the fully localized phase. The main peak is at a frequency $\omega/v_b \approx t\sqrt{q^2 + \hat{\Sigma}_f}$ and has a width controlled by $\sqrt{\hat{\Sigma}_f}$ while the "bosonic" peak is at $\omega/v_b \approx \sqrt{q^2 + \hat{\Sigma}_b}$ and has a width controlled by $\sqrt{\hat{\Sigma}_b}$. A a consequence, the "bosonic" peak is much sharper than its fermionic counterpart, a sign of the the enhancement of the bosonic localization length, by interactions, in the fully localized phase. The bosonic structure factor shows only one peak at $\omega/v_b \approx \sqrt{q^2 + \hat{\Sigma}_b}$. In that particular case, the extra peak is obscured by the vicinity of the main peak.

\subsection{Superfluid correlations} \label{sec:SF}

By adapting the variational method, it is possible to 
proceed and try and compute superfluid correlations. 
This calculation is, however,  somewhat delicate: while equal 
time  correlation functions are found to behave in a meaningful way, 
unequal-time correlation functions are apparently pathological 
\cite{Giamarchi03}. The reason is that the bosonic field 
operator  $\psi_b(x)\sim e^{i\theta_b(x)}$ creates a soliton in the
field configuration (shifts the density wave), 
which costs infinite energy in the Gaussian approximation of pinning by disorder. Therefore, unequal time correlation functions 
turn out to vanish identically. 

We therefore restrict ourself to equal time correlation functions. 
As we shall see, the variational method seems to give physically 
reasonable  results for these quantities, though the results of this
subsection  should be taken with some caution.

The quantity we  want to compute is
\beq
\label{eq:corr_sf}
A_b(x) = \overline{\langle e^{i\theta_b(x)} e^{-i\theta_b(0)}\rangle}\;.
\eeq
Note that the original action depends on both $\theta$ and $\phi$. It is only by integrating out $\theta$ fields that we were able to write an effective action depending only on $\phi$ fields and then proceed to the variational calculation. By doing so one can easily compute the correlation functions depending only on $\phi$. However the original action is really of the form (after introducing replicas and averaging over disorder)
\begin{widetext}
\bea
S_0^{\text{rep}} &=& \sum_{a=1}^n \sum_{\alpha=f,b} \int dx d\tau
\left[i\partial_x \theta_\alpha^a \partial_\tau \phi_\alpha^a +
  \frac{v_\alpha}{2\pi K_\alpha} \left( \partial_x
  \theta^a_\alpha\right)^2 + \frac{v_\alpha K_\alpha}{2\pi} \left(
  \partial_x \phi^a_\alpha\right)^2 \right]  +
\frac{U_{bf}}{\pi^2}\int dx  d\tau \ \partial_x\phi^a_f \partial_x
\phi^a_b,  \\ \nn \\ 
S_{\text{dis}}^{\text{rep}} &=& -{D_f \rho_f^2}
\sum_{a,b}\int dx d\tau d\tau ' \cos[2\phi_f^a(x,\tau) -
  2\phi_f^b(x,\tau ')] -\frac{D_b \rho_b^2}{\hbar} \sum_{a,b}\int dx
d\tau d\tau ' \cos[2\phi_b^a(x,\tau) - 2\phi_b^b(x,\tau ')]. 
\eea
\end{widetext}

On can then replace $S_{\text{dis}}^{\text{rep}}$ by the self-energy
$\sigma^{ab}$ obtained from the variational calculation. With this 
approximation, one ends
up with a quadratic action and is able to compute the propagator
$\langle \theta_b^a(q,\omega_n) \theta_b^a(-q,-\omega_n) \rangle$
(using inversion formulas for hierarchical matrices~\cite{Mezard91}). 
In the end, we  find the following form, irrespective of  the
level of replica symmetry breaking 
\begin{widetext}
\beq
\label{eq:theta_generale}
\langle \theta_b^a(q,\omega_n) \theta_b^a(-q,-\omega_n) \rangle = \frac{\pi}{v_b K_b} \frac{[q^2 + \hat{I}_b(\omega_n) + \hat{\Sigma}_b][q^2 + f(\omega_n)]-g^2q^4}{q^2\left[[q^2 + f(\omega_n)][q^2 + b(\omega_n)]-g^2q^4 \right] }.
\eeq
\end{widetext}
The function $\hat{I}_b(\omega_n)$ depends on the state of bosons (superfluid or localized) and $\hat{\Sigma}_b = 0$ in the superfluid phase. Also $\hat{I}_f(\omega_n)$ and $\hat{\Sigma}_f$ vary from phase to phase. We study this function in more details in the next section.

\subsection{Time of flight}

TOF experiments aim at measuring the momentum distribution inside the
trap. To do so one releases the trap, then, after a given time $t$ of
free expansion, one images the density of the atomic cloud. In the case of
a quasi-1D tube, and for long enough times $t$, the average density at
point $\mathbf{r}$ is approximately $\langle \psi_b^\dagger
(\mathbf{r}) \psi_b(\mathbf{r}) \rangle_t \simeq  W(y,z) \langle
n_{Q(x)} \rangle $ with $W(y,z)$ a Gaussian envelope (resulting from
the transverse confinement in directions $y$ and $z$, in a given
tube), $\langle n_{Q(x)} \rangle$ the momentum distribution in the
longitudinal direction,  and $Q(x) =  M_b x/t$
\cite{Altman04}. A detailed calculation actually  leads to 
\bea
\label{eq:n_tof}
\hspace{-1cm}\langle \psi_b^\dagger (\mathbf{r}) \psi_b(\mathbf{r})
\rangle_t &\propto& \nn \\ 
 && \hspace{-3cm}\int_0^L dx_1  \int_0^L dx_2 e^{-i Q(x) (x_1 - x_2)}
\langle \psi_b^\dagger (x_1) \psi_b(x_2) \rangle,  
\eea
where we have introduced here a finite size $L$ for each tube. In our
case, $\langle \psi_b^\dagger (x_1) \psi_b(x_2) \rangle \simeq
\rho_b\; A_b(x_1-x_2)$, with $A_b(x)$ given in \eqref{eq:corr_sf}.  
Typically, the
right hand side of \eqref{eq:n_tof} is the convolution of the Fourier
transform of $A_b(x)$ -- that is, the momentum distribution -- and a
function similar to a rectangle of width $1/L$, imposing an infra-red
cut-off. In an infinitely long tube, without disorder, and at zero
temperature, $A_b(x) \sim x^{-1/(2K_b)}$, for $x \gg \Lambda^{-1}$,
and $\Lambda$ the UV cutoff. Correspondingly, its Fourier transform, 
$n_b(q)$ is typically a power law, too, $n_b(q)\sim q^{1/(2K_b)-1}$ 
for $q \ll \Lambda$. At large $q$ it is known to  decay as $q^{-4}$, 
for the Lieb-Liniger model
\cite{Olshanii03}. For a finite size system the power-law behavior is
cut, and for $q<1/L$ one finds $n_b(q=0) \sim
L^{2-1/(2K_b)}$. These regimes were indeed observed experimentally in
[\onlinecite{Paredes04}]. Note that at finite temperature, the infrared
cutoff is given by $q_0 = \textrm{max}\{ 1/L, 1/v_b \beta\}$ since the
quasi long range order is destroyed beyond the thermal length $v_b
\beta$. 

\begin{figure}
\centering
\includegraphics[width=\columnwidth,clip]{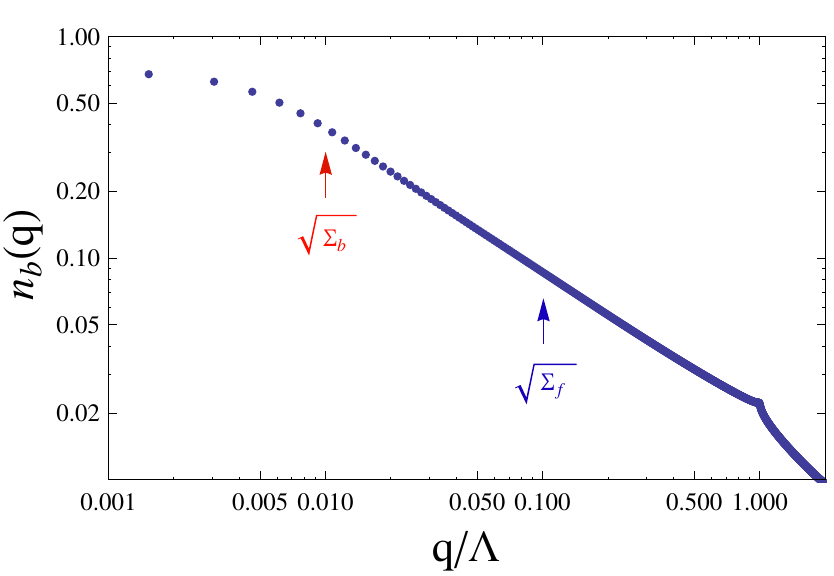}
\caption{\label{Fig:n_of_q_loc} Momentum distribution $n_b(q)$ (Fourier transform of $A_b(x)$) in the fully localized phase. We have taken,  $v_f/v_b = 9$, $K_b = 1.3$ and $g = 0.4$. Also, $\hat{\Sigma}_f = 0.01 \Lambda$ and $\hat{\Sigma}_b = 0.0001 \Lambda$. Related momentum scales, $\sqrt{\hat{\Sigma}_f}$ and $\sqrt{\hat{\Sigma}_b}$
 are pinpointed on the plot. Momentum $q$ is in unit of the UV cut-off  $\Lambda$. } 
\end{figure}

For localized bosons, the localization length $L_b$ plays a role 
similar  to that of the size of the system or the thermal length. We 
cannot compute $A_b(x)$ for a finite size system at finite
temperature. Therefore, in Fig.~\ref{Fig:n_of_q_loc} 
we just plot  $n_b(q)$, for an infinite system 
at zero temperature in the fully localized phase.  
In this case $A_b(x)$ can simply be expressed as
$A_b(x) = \exp[-1/2\langle[\theta^a_b(x)-\theta^a_b(0)]^2\rangle]$,
with 
\begin{widetext}
\beq
\langle[\theta^a_b(x)-\theta^a_b(0)]^2\rangle = \frac{1}{\beta L} \sum_{|q|<\Lambda,\omega_n} 2[1-\cos(qx)]
\langle \theta_b^a(q,\omega_n) \theta_b^a(-q,-\omega_n) \rangle\;.
\eeq
\end{widetext}
Here the sum over momenta is limited by the UV cut-off $\Lambda$, which
explains the cusp for $q=\Lambda$. For $q<\Lambda$ we find the power
law associated with Luttinger liquid physics, but of yourse, 
 the true behavior for
$q>\Lambda$ is cut-off dependent, and is not captured by 
our simple cut-off scheme. For $q<\sqrt{\hat{\Sigma}_b
}$ the distribution bends away from the algebraic law. This is a
signature of the localization of the bosonic gas on a typical length
scale $L_b \sim 1/\sqrt{\hat{\Sigma}_b}$, and the exponential decay of
$A_b(x)$ at large distances. We also indicated the position of
$\sqrt{\hat{\Sigma}_f}$. Indeed, boson interactions are renormalized
on length scales smaller than the localization length of fermions. We
therefore expect a crossover between two power law behaviors with
different exponents. Here for the values of the parameter we have
chosen, the renormalization of the exponent is rather small ($\sim
0.85$ of its initial value). The most prominent effect is thus that of
the infrared cut-off introduced  by the localization length. 
The small-momentum saturation of $n_b(q)$
 induced by the localization  should be an observable effect as long
as $L_b < L, v_b \beta$.


\section{Conclusion}
\label{sec:conclusion}

\begin{figure}
\centering
\includegraphics[width=\columnwidth,clip]{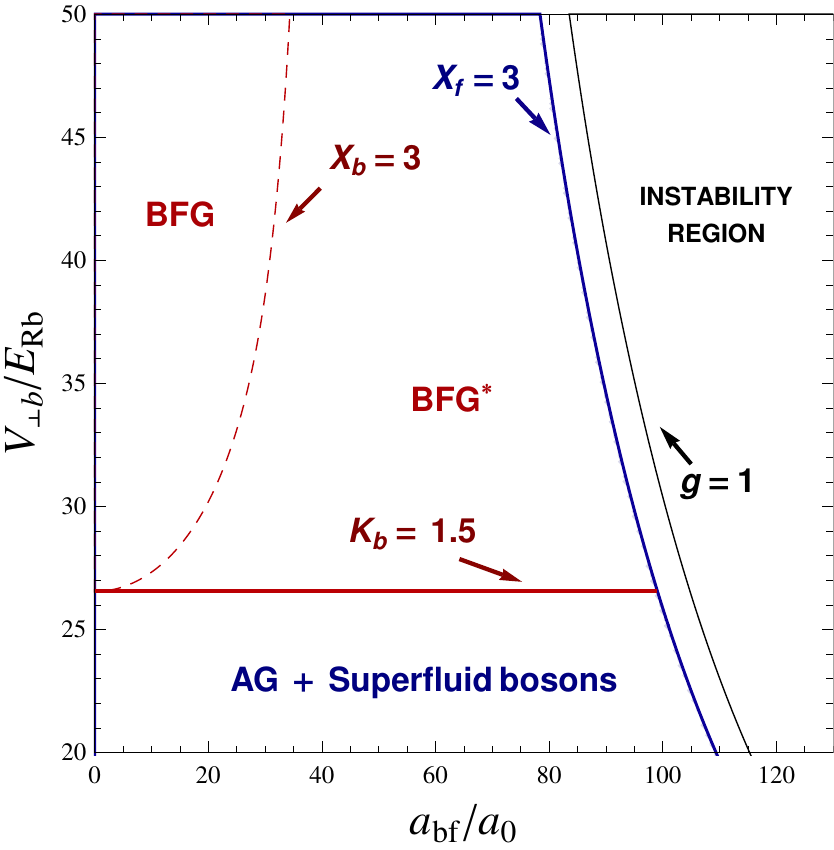}
\caption{\label{Fig:diagramme_exp} Phase diagram of a 1D $^{87}$Rb-$^{40}$K Bose-Fermi mixture, in the
weak disorder limit. We consider an array of tubes,
created with  lasers of wavelength  $\lambda$ = 755 nm, which corresponds
to a (2D) lattice constant $d=\lambda/2$.
We take  $a_{bb}=100 a_0$ as 
Bose-Bose scattering length, while  the Bose-Fermi scattering length, $a_{bf}$, is
tuned using a Feshbach resonance. The one.dimensional densities are
chosen to be $\rho_f d = 0.3 $ and $\rho_b d= 0.2 $.   
The recoil energy is $E_{R,b}=h^2 \lambda^{-2}/(2m_{Rb})$, while
$V_{\perp b}$ is the transverse confining potential creating the 1D
tubes. { The Bose-Bose interaction $U_b$ increases with $V_{\perp b}$}. Note that despite a superficial similarity between the present figure and, for instance, Fig. \ref{fig:phase_diagram_RG_3}, the vertical axis is reverted, since $K_b$ decreases as $U_b$ increases. Four phases and the region { of instability of the Luttinger liquid theory}  are shown.
BFG: Bose-Fermi glass,  BFG$^*$ (BFG
with an extremely large bosonic localization length), AG+SFB: Anderson
Glass + Superfluid Bosons, LL: Luttinger  liquid. } 
\end{figure}

In this article, we have studied in detail a 1D mixture of bosons and
fermions in a random potential. More precisely we have considered the
localization of the gas by analyzing the pinning of density waves by
weak disorder. 
In the case of incommensurate densities, to which we focused
throughout this paper, the two
components of the gas are coupled to  Fourier components of the random
potential that are effectively uncorrelated.  The two density 
waves are, however,
coupled through the Bose-Fermi interactions. Using 
renormalization group methods as well as a self-consistent 
harmonic approximation
in replica space, we arrived at the following general conclusions: 

- For weak disorder, the phase diagram can be plotted adequately as a
function of two parameters, the Luttinger parameter $K_b$ for bosons,
and the Bose-Fermi interaction parameter $U_{bf}$. The structure of
the phase diagram and the properties of the phases depend on a
third parameter, the ratio of sound velocities, $v_f/v_b$. Whatever
the value   of this ratio, we can identify three distinct phases, (i) a
two-component Luttinger liquid, dominated by superfluid correlations
for bosons and pair correlations for fermions, (ii) a fully localized
phase where both components of the gas are pinned by disorder and
(iii) an intermediate phase where fermions are localized and bosons
are superfluid.  In Fig.~\ref{Fig:diagramme_exp} we propose a translation of
the diagram of Fig.~\ref{fig:phase_diagram_RG_3} to microscopic parameters 
relevant for an experiment using a mixture of $^{87}$Rb and $^{40}$K \cite{Bloch09}. This translation 
is done along the lines detailed in Ref. \onlinecite{Crepin10b}.

\begin{figure}
\centering
\includegraphics[width=\columnwidth,clip]{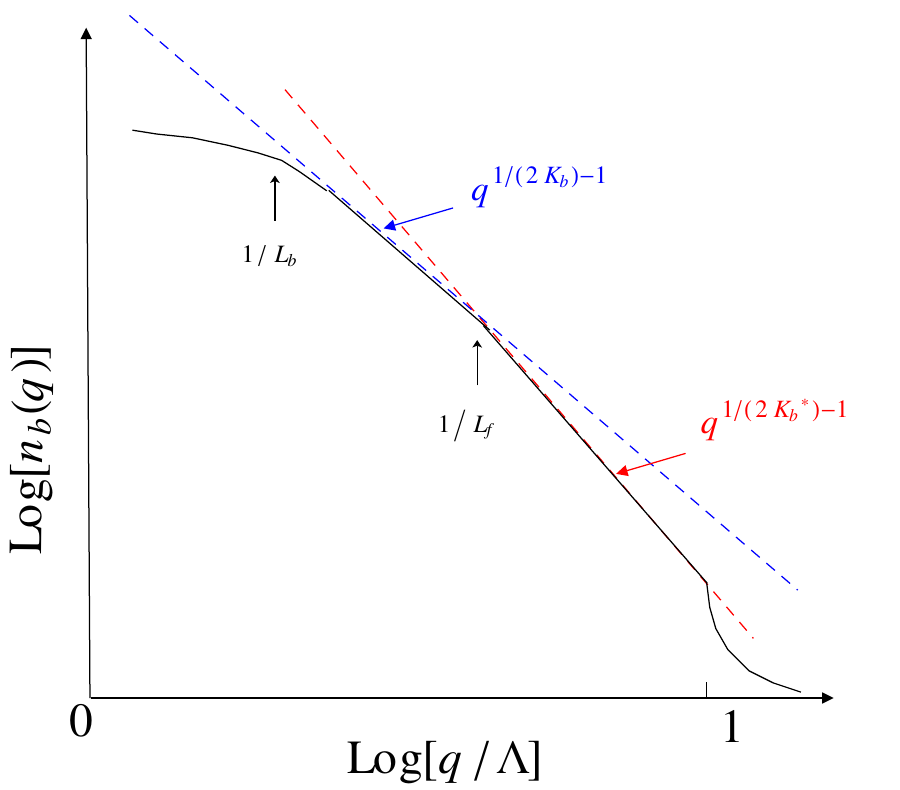}
\caption{\label{sketch1} The momentum distribution of bosons in the fully localized phase 
exhibits a saturation for momenta below $1/L_b$. Above this threshold,
we identify a crossover between two algebraic regimes. Bosonic
interactions are renormalized by Bose-Fermi interactions on length
scales smaller than the fermionic localization length $L_f$. Indeed
for $q > 1/L_f$ the exponent of the power law is renormalized by BF
interactions, while for $1/L_b < q < 1/L_f$ the exponent is controlled
by bare interactions.
}  
\end{figure}

\begin{figure}
\centering
\includegraphics[width=\columnwidth,clip]{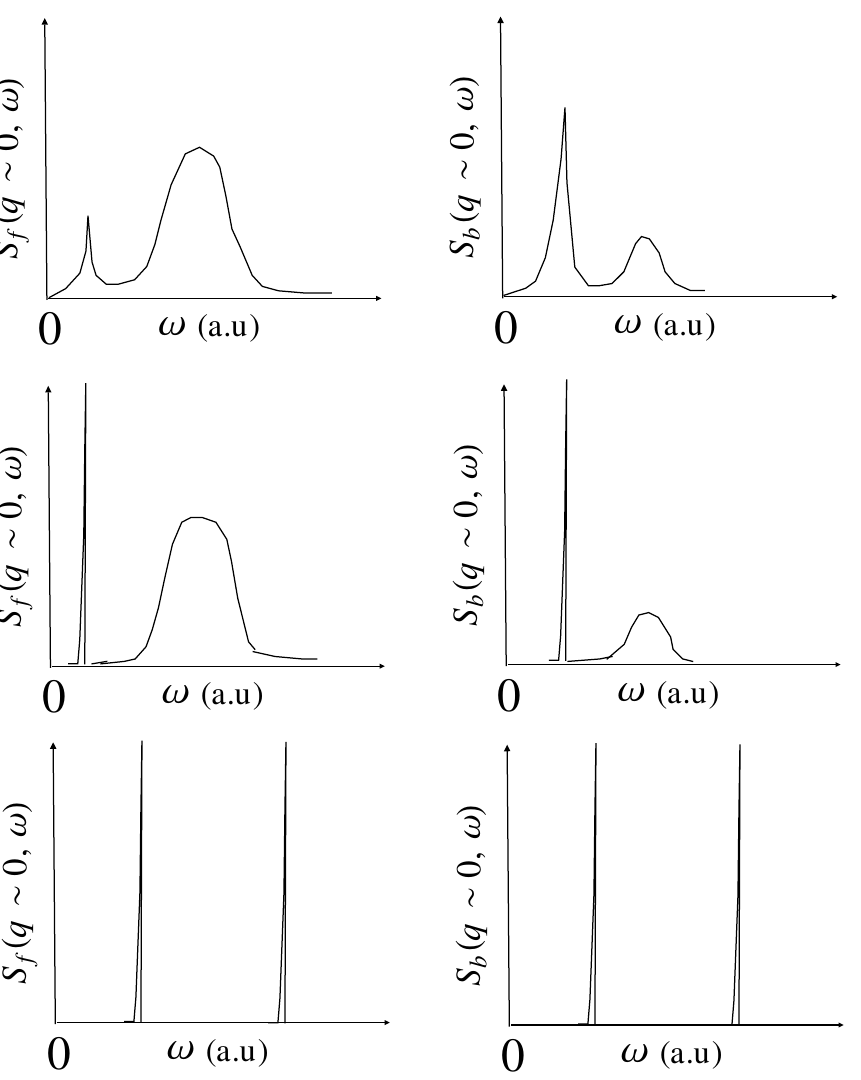}
\caption{\label{sketch2} Sketch of the structure factor -- for bosons
  and fermions -- in the three phases identified in
  Fig.\ref{Fig:diagramme_exp}. Top pannel: fully localizd phase,
  middel pannel: intermediate phase with localized fermions and
  superfluid bosons, bottom pannel: Luttinger liquid phase.  The
  presence of two peaks in each phase is characteristic of the strong
  coupling between the two components of the gas, even in the
  localized phases. In the latter the widths of the peaks are
  inversely proportional to the localization lengths.}  
\end{figure}

- The properties of the fully localized phase depend strongly on the
strength of Bose-Fermi interactions as well as on the ratio of
velocities. Both from the RG and from the variational calculations 
we conclude that  this phase is
characterized by two length scales, $L_f$ and $L_b$, which can be
identified as the fermionic and bosonic localization lengths.  
Beyond these length scales  the phase  correlations of
the density waves are lost. For strong Bose-Fermi
interactions, these two length scales can be very 
different. In the case where $v_f > v_b$ (and for similar amplitudes of
the disorder), $L_b$ is larger than $L_f$ for two reasons: first, 
because $K_b > K_f$
and therefore quantum fluctuations tend to suppress more strongly the
pinning of the bosonic density wave, and second, because despite
localization, fast fermionic phonons screen repulsive  bosonic
interactions and increase $L_b$ further. In the case  $v_b > v_f$
on the other hand, the order of $L_f$ and $L_b$ can be reverted, because,
quantum fluctuations through $K_b$ and the effective attractive
interactions for the fermions have competing effects 
on localization. 

- In any case, for a finite size system, it is likely that one of the
localization length exceeds the size of the system. One of the species
would then appear as  delocalized. Also, finite temperature can
overshadow the effects of disorder if the thermal length is comparable
to one of the localization lengths.

Experimentally, the localization phase transition can be most 
easily observed in correlation functions, which uniquely depend on the 
phase. For the bosons, such a quantity is provided by the momentum
distribution $n_b(q)$ of the quasi-condensate, which is directly 
measurable through
time of flight (TOF) experiments. As shown in Fig.~\ref{Fig:n_of_q_loc} and
sketched in Fig.~\ref{sketch1}, there bosonic localization 
should simply be observed as a saturation of    $n_b(q)$ at momenta smaller 
than $\sim 1/L_b$, provided temperature is low enough (if the thermal length $v_b/(k_B T)$ is smaller than $L_b$ then
localization is obscured by thermal fluctuations).

Unfortunately, pure fermionic phase 
correlations are much more difficult to access.  They appear as 
p-wave superconducting fluctuations, and would probably be only  measurable 
through rather difficult noise correlation measurments. However, the
fermionic localization does have an impact on $n_b(q)$ and 
should also be visible in the dynamical structure factor. Indeed we computed the latter quantity
using our variational solution in replica space. As sketched in Fig.\ref{sketch2}, it can distinguish between localized and 
superfluid phases. Several key features are to be noted. First, the presence of two peaks in each phase is a direct consequence of 
the strong coupling -- through Bose-Fermi interactions -- between both componnents of the gas, even in the fully localized phase. Second, the width of 
the peaks in the localized phases is controlled by the inverse of the
localization lengths. For the parameters of  
Fig. \ref{Fig:diagramme_exp} where fermions are the fast component,
$L_b \gg L_f$ in the fully localzed phase, and the bosonic peak is
much sharper than its fermionic counterpart. In the intermediate
phase, where bosons are superfluid it becomes a Dirac delta. Note that
according to our analysis of the dynamical functions $I_f$ and $I_b$
 -- see equations \eqref{eq:alphaf} and \eqref{eq:alphab} as well as
\eqref{sf_3} and \eqref{sb_3}-- in both localized phases the
structure factor grows linearly at small frequencies. In the Luttinger
liquid phase one should 
be able to  retrieve the two sound modes from the peak positions,
 $\omega_+ = v_+q$ and $\omega_- = v_- q$. 
One should also bear in mind that for non-zero
$q$, deviations from the linear 
dispersion (assumed in a Luttinger liquid  description) 
will lead to a broadening  of these peaks. 
It then might be difficult to
distinguish peaks from the Luttinger liquid phase and peaks from the
localized phase in the case of very large localization lengths. The
comparison will be easier for short localization lengths, a regime
likely to be attained for either strong bosonic repulsions
(Tonks-Girardeau regime) or small Bose-Fermi interactions.  

Finally, we would like to point out that dynamical quantities are 
key observables to investigate the effects of the various 
levels of replica symmetry breaking.  
Here we computed the structure factors by making simple approximations
for  the dynamical parts of the self-energies.
It remains to solve completely the system of self-consistent
equations to obtain a definitive view on the structure factor, 
to go beyond the sketches presented in Fig.~\ref{sketch2}. 

\begin{acknowledgments}
We would like to acknowledge fruitful discussions with K. Damle,
T. Giamarchi, L. Glazman, N. Laflorencie, and L. Sanchez-Palencia. 
Part of this work has been carried on thanks to the support of the Institut
Universitaire de France, and  Hungarian research funds 
OTKA and NKTH under Grant Nos.~K73361 and CNK80991. 
G.Z. acknowledges support from the Humboldt Foundation and the 
DFG.
\end{acknowledgments}

\appendix
\section{Normal modes of the homogeneous Bose-Fermi mixture}
\label{app:coeff}
Starting from the Hamiltonian of Eq.~(\ref{eq:H0}), one uses
standard diagonalization methods to find: 
\beq
H = \sum_{\alpha=\pm} \frac{  v_\alpha}{2\pi}\int dx
\left[K_\alpha \left( \nabla\theta_\alpha\right)^2 +
  \frac{1}{K_\alpha} \left( \nabla \phi_\alpha\right)^2 \right], 
\eeq
with $K_\pm = 1$ and,
\beq
v_\pm^2 = \frac{1}{2}(v_f^2 + v_b^2) \pm \frac{1}{2}\sqrt{(v_f^2 - v_b^2)^2 + 4g^2v_f^2 v_b^2},
\eeq
where $g=\frac{U_{bf}}{  \pi}\sqrt{\frac{K_f K_b}{v_f v_b}}$ is a
dimensionless parameter. Transformation rules betweenhe fields
$(\phi_f,\phi_b)$ and $(\phi_+,\phi_-)$ are related as $\phi_f =
f_+ \phi_+ + f_- \phi_-$, $\phi_b = b_+ \phi_+ + b_-
\phi_-$, with the  coefficients are defined as follows: 
\begin{eqnarray}
f_+=\sqrt{\frac{K_f v_f}{v_+}}\sin(\theta), && \; \;
f_-=\sqrt{\frac{K_f v_f}{v_-}} \cos(\theta), \nn \\ 
b_+=\sqrt{\frac{K_b v_b}{v_+}}\cos(\theta), && \; \; b_-=
-\sqrt{\frac{K_b v_b}{v_-}} \sin(\theta). \nn   
\end{eqnarray}
A similar transformation relates $(\theta_f,\theta_b)$ to
$(\theta_+,\theta_-)$, that is,  $\theta_f = \bar{f}_+ \theta_+ +
\bar{f}_- \theta_-$ and $\theta_b = \bar{b}_+ \theta_+ + \bar{b}_-
\theta_-$. Coefficients for this transformation are: 
\begin{eqnarray}
\bar{f}_+=\sqrt{\frac{v_+}{K_f v_f}} \sin(\theta), && \; \; \bar{f}_-=\sqrt{\frac{v_-}{K_f v_f}} \cos(\theta), \nn \\
\bar{b}_+=\sqrt{\frac{v_+}{K_b v_b}} \cos(\theta), && \; \; \bar{b}_-= -\sqrt{\frac{v_-}{K_b v_b}} \sin(\theta).\nn 
\end{eqnarray}
The rotation angle $\theta$ is defined by:
\beq
\cos (2\theta) = \frac{v_b^2-v_f^2}{v_+^2-v_-^2}, \  \ \
\sin (2\theta) = \frac{2g v_f v_b}{v_+^2-v_-^2}
\eeq
One can check that for $U_{bf}=g=0$ 
\beq 
\left.
\begin{array}{ll}
f_+ = \sqrt{K_f}, & f_-=0,\;\\
b_-= \sqrt{K_b}, & b_+ = 0, \;\\
v_+=v_f, & v_-=v_b \;
\end{array}
\right\}
\; \; \textrm{if} \ v_f > v_b \;,
\eeq 
\beq
\left.
\begin{array}{lll}
f_+ = 0 , & f_-=\sqrt{K_f},\;\\
b_- = 0, & b_+ = \sqrt{K_b},\;\\
v_+=v_f, & v_-=v_b \;
\end{array}
\right\}
\; \; \textrm{if} \ v_b > v_f \;.
\eeq 

\section{RG calculation}\label{sec:app_RG}
\label{app:RG}
The renormalization group (RG) relies upon  the assumption that
all important phenomena occur over length scales
much larger than a microscopic length $\Lambda^{-1}$. In our case, we
use a hydrodynamic theory, and $\Lambda^{-1}$ can be identified as 
the mean inter-particle distance, i.e., $\Lambda \sim
\rho$, the mean density. At each RG step we 
integrate out high momentum excitations by reducing 
the cutoff $\Lambda\to \Lambda'$, while 
renormalizing the parameters of the Hamiltonian --
and possibly generating new couplings, and thereby generate 
an  RG trajectory. 

In a system with quenched disorder the thermodynamic quantity of
interest is the average free energy $F$, 
\beq
-\beta F =  \overline{\log Z}\;.
\label{eq:F_aver}
\eeq
Here $Z$ is the partition function for a given realization of the
random potential, and $\overline{\phantom i \ldots \phantom i}$
denotes averaging over disorder. To compute \eqref{eq:F_aver}
we use the so-called replica trick \cite{MezardBook}: 
we introduce $n$ identical copies of the system, 
average over the disorder and then take the limit $n\rightarrow 0$,
\beq
\lim\limits_{n\rightarrow 0} \frac{1}{n}\log \overline{Z^n} =
\overline{\log Z}\;. 
\eeq
In pratice, we work with $\overline{Z^n}$ to perform
the RG. Using a path integral formulation, the partition
function $Z$ reads: 
\beq 
Z= \int D\phi_f D\phi_b \ e^{-S[\phi_f,\phi_b]} \;.
\eeq
Here the action  $S = S_0 + S_{\text{dis}}$ is given by Eqs.~\eqref{S_0} 
and \eqref{S_dis}.
Assuming that 
$\xi_\alpha$ are  random functions with a Gaussian distribution,
 $P(\xi_\alpha,\xi_\alpha^*) = \exp[-D_\alpha^{-1}\int dx
  \ \xi_\alpha(x)\xi_\alpha(x)^*]$, we can  compute $\overline{Z^n}$, 
and arrive at 
\begin{widetext}
\bea
\overline{Z^n} &=& \int \prod_{a=1}^n D\phi_f^a D\phi_b^a
\ e^{-\sum\limits_{a=1}^n S_0[\phi_f^a, \phi_b^a]/\hbar}
\ \overline{e^{-\sum\limits_{a=1}^n S_{dis}[\phi_f^a,
      \phi_b^a]/\hbar}} \nn \\ 
&=&  \int \prod_{a=1}^n D\phi_f^a D\phi_b^a \ e^{-\sum\limits_{a=1}^n
  S_0[\phi_f^a, \phi_b^a]} \exp\left[\sum\limits_{\alpha=f,b}
  {D_\alpha \rho_\alpha^2} \sum\limits_{a,b=1}^n \int
  dxd\tau d\tau '
  e^{i2\phi_\alpha^a(x,\tau)}e^{-i2\phi_\alpha^b(x,\tau')}\right]\nn
\\ 
&=& \int \prod_{a=1}^n D\phi_f^a D\phi_b^a \  e^{-S_{\text{rep}} },
\eea
\end{widetext}
with  the {\it replicated} action, $S_{\text{rep}} = S_0^{\text{rep}}
+ S_{\text{dis}}^{\text{rep}}$ defined 
through Eqs.~\eqref{S_0^rep} and \eqref{S_dis^rep}.

To perform the RG calculation we introduce a UV cutoff $\Lambda$ on
the momenta only, and write  the fields $\phi_f,\phi_b$ as well $\phi_+$
and $\phi_-$ as
\beq
\phi_\alpha(x,\tau) = \frac{1}{\beta L}
\sum_{\underset{|q|<\Lambda}{q,\omega_n}} \phi_\alpha(q,\omega_n)
e^{iqx-i\omega_n \tau}\;. 
\eeq
Next, we introduce the  slow and fast fields, 
\bea
\phi_\alpha^<(x,\tau) &=& \frac{1}{\beta L} \sum_{\underset{|q|<\Lambda'}{q,\omega_n}} \phi_\alpha(q,\omega_n) e^{iqx-i\omega_n \tau},\\
\phi_\alpha^>(x,\tau) &=& \frac{1}{\beta L} \sum_{\underset{\Lambda'<|q|<\Lambda}{q,\omega_n}} \phi_\alpha(q,\omega_n) e^{iqx-i\omega_n \tau}\;,
\eea
and integrate out the fast fields to obtain
\beq
\overline{Z_\Lambda^n} = Z_0^> \int \prod_{a=1}^n D\phi_f^{a,<}D\phi_b^{a,<} e^{-S_0^<} \langle e^{-S_{\text{dis}}^{\text{rep}} }\rangle_>\;.
\eeq
Performing then a  cumulant expansion to first order in $D_f$ and $D_b$ we get
\begin{widetext}
\bea
\langle e^{-S_{\text{dis}}^{\text{rep}} }\rangle_> &\simeq&
\exp\left[\sum\limits_{\alpha=f,b} {D_\alpha
    \rho_\alpha^2} \sum\limits_{a=1}^n \int  dxd\tau d\tau '
  e^{i2\phi_\alpha^{a,<}(x,\tau)}e^{-i2\phi_\alpha^{a,<}(x,\tau')}\langle
  e^{i2\phi_\alpha^{a,>}(x,\tau)}e^{-i2\phi_\alpha^{a,>}(x,\tau')}\rangle_>\right]
\nn \\ 
&=&
\exp\left[\sum\limits_{\alpha=f,b} {D_\alpha
    \rho_\alpha^2} \sum\limits_{a=1}^n \int  dxd\tau d\tau '
  e^{i2\phi_\alpha^{a,<}(x,\tau)}e^{-i2\phi_\alpha^{a,<}(x,\tau')}\langle
  e^{i2\phi_\alpha^{a,>}(x,\tau)}\rangle_>^2\right. \nn \\ 
 &+& \left. \sum\limits_{\alpha=f,b} {D_\alpha
    \rho_\alpha^2} \sum\limits_{a=1}^n \int  dxd\tau d\tau '
  e^{i2\phi_\alpha^{a,<}(x,\tau)}e^{-i2\phi_\alpha^{a,<}(x,\tau')}\left[\langle
    e^{i2\phi_\alpha^{a,>}(x,\tau)}e^{-i2\phi_\alpha^{a,>}(x,\tau')}\rangle_>
    - \langle e^{i2\phi_\alpha^{a,>}(x,\tau)}\rangle_>^2\right]
  \right].\label{RG_bracket}
\nn \\
\label{eq:cumulant} 
\eea
\end{widetext}

Note that, to first order, every correlation functions appearing 
in \eqref{eq:cumulant} is
diagonal in replica space. Let us therefore drop the replica 
indices for the rest of this Section. Also, let us 
keep  only $D_f$ for convenience. Using the
normal modes $\phi_+$ and $\phi_-$, we find that: $\langle
e^{i2\phi_\alpha^{>}(x,\tau)}\rangle_>^2 = (\Lambda'/\Lambda)^{2f_+^2
  + 2f_-^2}$. Therefore the first term in the bracket of equation
\eqref{RG_bracket} reads 
\begin{widetext}
\bea
\frac{D_f \rho_f^2}{ v_f^2 \Lambda^3} \left(\frac{\Lambda'}{\Lambda}\right)^{2f_+^2 + 2f_-^2}\Lambda^3\int  dxd(v_f\tau) d(v_f\tau ')  e^{i2\phi_f^{<}(x,\tau)}e^{-i2\phi_f^{<}(x,\tau')} \\
= \tilde{D}_f \left(\frac{\Lambda'}{\Lambda}\right)^{2f_+^2 + 2f_-^2-3} \Lambda'^3 \int  dxd(v_f\tau) d(v_f\tau ')  e^{i2\phi_f^{<}(x,\tau)}e^{-i2\phi_f^{<}(x,\tau')}\;,
\eea
\end{widetext}
where we have defined the dimensionless coupling $\tilde{D}_f =
\frac{D_f \rho_f^2}{v_f^2 \Lambda^3} $. To preserve the low
energy form of the action for a rescaled cutoff, $\Lambda'$, one
should rescale $\tilde{D}_f $ so that: 
\beq
\tilde{D}_f(\Lambda') = \tilde{D}_f (\Lambda) \left(\frac{\Lambda'}{\Lambda}\right)^{2f_+^2 + 2f_-^2-3}.
\eeq
Assuming an infinitesimal change of the cutoff, 
$\Lambda' = \Lambda(1-dl)$, we obtain Eq.~\eqref{eq:Df}. 
Eq.~\eqref{eq:Db} can be obtained i na similar way.

Let us now take care of the second bracket in equation
\eqref{RG_bracket}. It contributes mainly when $\tau$ and $\tau'$ are
close, and will essentially renormalize the coefficient of 
 $(\partial_\tau \phi_f)^2$ in
the quadratic action. First let us deal with: 
\bea
\mathcal{A} = \langle
e^{i2\phi_f^{>}(x,\tau)}e^{-i2\phi_f^{>}(x,\tau')}\rangle_> - \langle
e^{i2\phi_f^{>}(x,\tau)}\rangle_>^2. 
\eea
We have:
\vspace{0.5cm}
\bea
&&\hspace{-0.3cm}\langle
e^{i2\phi_f^{>}(x,\tau)}e^{-i2\phi_f^{>}(x,\tau')}\rangle_> =
e^{-\hspace{-0.1cm}\sum\limits_{\alpha=\pm}
  2f_\alpha^2\langle(\phi_\alpha(x,\tau)-\phi_\alpha(x,\tau'))^2\rangle_>}\nn
\\ 
&&\hspace{-0.3cm}= \exp\Biggl[-\sum\limits_{\alpha =\pm}
  2f_\alpha^2\hspace{-0.3cm}
  \sum\limits_{\underset{\Lambda'<|q|<\Lambda}{q,\omega_n}}  
 \hspace{-0.2cm}[2-2\cos(\omega_n\bar{\tau})]   
\frac{\pi      v_\alpha}{\omega_n^2+v_\alpha^2q^2}\Biggr]    
\eea
with $\bar{\tau}=\tau-\tau'$ and:
\bea
&&\hspace{-1.2cm}\langle e^{i2\phi_f^{>}(x,\tau)}\rangle_>^2
=\exp\left[-\sum\limits_{\alpha=\pm}
  4f_\alpha^2\hspace{-0.2cm}\sum\limits_{\underset{\Lambda'<|q|<\Lambda}{q,\omega_n}}\hspace{-0.2cm}\frac{\pi  
     v_\alpha}{\omega_n^2+v_\alpha^2q^2}\right] .
\eea
Then factorizing $\langle
e^{i2\phi_f^{>}(x,\tau)}e^{-i2\phi_f^{>}(x,\tau')}\rangle_> $ in
$\mathcal{A}$ and using the fact that $\Lambda' = \Lambda(1-dl)$, an
expansion to first order in $dl$ leads to. 
\beq
\mathcal{A} = \langle e^{i2\phi_f^{>}(x,\tau)}e^{-i2\phi_f^{>}(x,\tau')}\rangle_>\sum_{\alpha=\pm}2f_\alpha^2 e^{-v_\alpha|\bar{\tau}|\Lambda} dl.
\eeq
We are left with:
\begin{widetext}
\bea
\mathcal{B} &=& {D_\alpha \rho_\alpha^2} \int  dxd\tau d\tau ' e^{i2\phi_f^{<}(x,\tau)}e^{-i2\phi_f^{<}(x,\tau')}\mathcal{A} \nn \\
&=& {D_\alpha \rho_\alpha^2} \int  dxd\tau d\tau ' :e^{i2\phi_f^{<}(x,\tau)}e^{-i2\phi_f^{<}(x,\tau')}:e^{-\hspace{-0.1cm}\sum\limits_{\alpha=\pm} 2f_\alpha^2\int_0^\Lambda dq (1-e^{-v_\alpha|\bar{\tau}|q})/q}\sum_{\alpha=\pm}2f_\alpha^2 e^{-v_\alpha|\bar{\tau}|\Lambda} dl.
\eea
\end{widetext}
:\ldots: stands for normal-ordering. The function
$\mathcal{G}(\bar{\tau}) =
\exp\left[-\hspace{-0.1cm}\sum\limits_{\alpha=\pm}
  2f_\alpha^2\int_0^\Lambda dq
  (1-e^{-v_\alpha|\bar{\tau}|q})/q\right]$ is obtained after taking
the normal order and combining the extra factor $\langle
e^{i2\phi_f^{<}(x,\tau)}e^{-i2\phi_f^{<}(x,\tau')}\rangle_<$ with
$\mathcal{A}$. Finally we expand the exponential in powers of
$\bar{\tau}$: 
\bea
\mathcal{B} \simeq -dl {D_\alpha
  \rho_\alpha^2}\left[\int dx dT(\partial_T \phi_f)^2
  \right]\int d\bar{\tau} \bar{\tau}^2 \mathcal{G}(\bar{\tau})\nn
\\ \times \sum_{\alpha=\pm}4f_\alpha^2
e^{-v_\alpha|\bar{\tau}|\Lambda} 
\eea 
$\mathcal{G}(\bar{\tau})$ is easily evaluated to be:
\beq
\mathcal{G}(\bar{\tau}) =
\prod_{\alpha=\pm}(v_\alpha\Lambda|\bar{\tau}|)^{-2f_\alpha^2}e^{-2f_\alpha^2(\gamma
  + \Gamma(0,v_\alpha\Lambda|\bar{\tau}|) } 
\eeq
with $\gamma$ Euler's constant and $\Gamma(0,z)$ the incomplete Gamma
function. In the end we find that $\mathcal{B}$ can be cast into: 
\bea
\mathcal{B} \simeq -dl \tilde{D}_f \left[ f_+^2
  \mathcal{C}_+\frac{v_f^2}{v_+^3}\left(
  \frac{v_+}{v_-}\right)^{2f_-^2} \hspace{-0.2cm}+ f_-^2
  \mathcal{C}_-\frac{v_f^2}{v_-^3}\left(
  \frac{v_-}{v_+}\right)^{2f_+^2}\right]\nn \\ 
&&\hspace{-4.5cm}\times \left[\int dx dT(\partial_T \phi_f)^2 \right].
\eea
Here we have defined  $\mathcal{C}_\pm(K_b,K_f,v_f/v_b)$ as
\bea\mathcal{C}_+ &=& 8\int_0^\infty dz z^{2-X_f}e^{-2\gamma X_f}
e^{-2f_+^2\Gamma(0,z)-2f_-^2\Gamma(0,\frac{v_-}{v_+}z)}\;,\nn 
\\ 
\mathcal{C}_- &=& 8\int_0^\infty dz z^{2-X_f}e^{-2\gamma X_f}
e^{-2f_+^2\Gamma(0,\frac{v_+}{v_-}z)-2f_-^2\Gamma(0,z)}\;. \nn 
\\  
\eea
Both $v_f$ and $K_f$ are renormalized by this term, and we find for
the flow equations: 
\begin{widetext}
\bea
\frac{dK_f}{dl} &=& -\frac{K_f^2}{2}\tilde{D}_f \left[f_+^2
  \mathcal{C}_+\frac{v_f^3}{v_+^3}\left(
  \frac{v_+}{v_-}\right)^{2f_-^2} \hspace{-0.2cm}+ f_-^2
  \mathcal{C}_-\frac{v_f^3}{v_-^3}\left(
  \frac{v_-}{v_+}\right)^{2f_+^2} \right], \\ 
\frac{dv_f}{dl} &=& -\frac{K_f^2}{2}v_f\tilde{D}_f \left[f_+^2
  \mathcal{C}_+\frac{v_f^3}{v_+^3}\left(
  \frac{v_+}{v_-}\right)^{2f_-^2} \hspace{-0.2cm}+ f_-^2
  \mathcal{C}_-\frac{v_f^3}{v_-^3}\left(
  \frac{v_-}{v_+}\right)^{2f_+^2} \right].  
\eea
\end{widetext}
These flow equations describe the phase transition 
for small but finite values of $\tilde D_f$ and $\tilde D_b$. 

\section{Self-consistent equations for the RSB solutions}
\label{app:RSB}

We start from the replicated action. The idea of the Gaussian
variational method is to replace the complicated action $S$ by its
best Gaussian approximation $S_G$: 
\begin{equation}
S_G = \frac{1}{2}\frac{1}{\beta L}
\sum_{q,i\omega_n}\phi_\alpha^a(q,i\omega_n)(G^{-1})_{\alpha
  \beta}^{ab}(q,i\omega_n)\phi_\beta^b(-q,-i\omega_n),\nn 
\end{equation}
The propagator $G$ is a $2n\times 2n$ matrix with the following structure:
\beq
G^{-1} = \left( 
\begin{array}{ll}
\left[G^{-1}\right]_{ff}^{ab} & \left[G^{-1}\right]_{fb}^{ab} \\ 
\left[G^{-1}\right]_{bf}^{ab} & \left[G^{-1}\right]_{bb}^{ab} 
\end{array}
\right)
\eeq
where $\left[G^{-1}\right]_{\alpha\beta}$, $\alpha,\beta=f,b$ is
consequently a $n\times n$ matrix. 
Using the well-known inequality, 
$F \le F_{\rm var}[G] \equiv F_G + \frac{1}{\beta}\langle
S-S_G\rangle_G$, we can obtain an estimate for 
$F$ by minimizing the variational free energy
$F_{var}$ with respect to $G$.
Here $F_G =
-\frac{1}{\beta} \log [\textrm{Tr} e^{-S_G}]$ is the Free energy of
the Gaussian theory.  The three terms, $F_G$, 
$\langle S_0 \rangle_G$ and $\langle S_{\rm dis}\rangle_G$
can be easily computed to obtain  
\bea
F_{var} &=&  - \frac{1}{2\beta}\sum_{q,i\omega_n} \textrm{Tr} \log[G(q,i\omega_n)]\nn \\ &+&\frac{1}{2}\sum_{\alpha,\beta}\sum_{q,i\omega_n}\left(G_0^{-1}\right)_{\alpha\beta}(q,i\omega_n)\textrm{Tr}[G_{\alpha\beta}(q,i\omega_n)] 
\nn \\ &+& \frac{1}{2} \sum_{a,b}  L \int d\tau \left( V_F[F^{ab}(\tau)] + V_B[B^{ab}(\tau)] \right).
\eea

Here we have defined the two functions, $V_F$ and $V_B$, so that $V_F(x) =
-2{\rho_f^2 D_f} e^{-2x}$ and  $V_B(x) = -2{\rho_b^2D_b} e^{-2x}$, and introduced 
\bea
F^{ab}(\tau) &\equiv& G^{aa}_{ff}(0,0) +  G^{bb}_{ff}(0,0) - 2
G^{ab}_{ff}(0,\tau), \\ 
B^{ab}(\tau) &\equiv& G^{aa}_{bb}(0,0) +  G^{bb}_{bb}(0,0) - 2
G^{ab}_{bb}(0,\tau).  
\eea
Notice that only the replica-diagonal elements, 
$F^{aa}(\tau)$  and $F^{bb}(\tau)$ turn out to be time-dependent, 
and the replica-offdiagonal elements, representing disorder-generated 
correlations between replicas  are just constants in time.  
We now look for the saddle-point equations by
differentiating $F_{var}$ with respect to $G$, and 
requiring  $\delta F_{var}$=0. This yields 
\begin{widetext}
\beq
\left.
\begin{array}{ll}
(G^{-1})_{ff}^{ab}(q,i\omega_n) &= -2\beta \delta_{n,0} V'_F(F^{ab}) \\
(G^{-1})_{bb}^{ab}(q,i\omega_n) &= -2\beta \delta_{n,0} V'_B(B^{ab})
\end{array}
\right\} \;\; \mbox{ a$\neq$ b},
\eeq
with
\beq
\left.
\begin{array}{ll}
F^{ab} = \frac{1}{\beta L} \sum\limits_{q,i\omega_n} \left[ G_{ff}^{aa}(q,i\omega_n) + G_{ff}^{bb}(q,i\omega_n)\right]-\frac{2}{\beta L}\sum\limits_q G_{ff}^{ab}(q,\omega_n=0)\\\\
B^{ab} = \frac{1}{\beta L} \sum\limits_{q,i\omega_n} \left[ G_{bb}^{aa}(q,i\omega_n) + G_{bb}^{bb}(q,i\omega_n)\right]-\frac{2}{\beta L}\sum\limits_q G_{bb}^{ab}(q,\omega_n=0)
\end{array}
\right\} \;\; \mbox{ a$\neq$ b},
\eeq
and
\bea
(G^{-1})_{ff}^{aa}(q,i\omega_n) &=& (G_0^{-1})_{ff}(q,i\omega_n) +2\int_0^\beta d\tau \left(1-\cos[\omega_n\tau]\right) V'_F(F^{aa}(\tau)) + 2\int_0^\beta d\tau \sum_{b\neq a} V_F'[F^{ab}], \\
(G^{-1})_{bb}^{aa}(q,i\omega_n) &=& (G_0^{-1})_{bb}(q,i\omega_n) +2\int_0^\beta d\tau \left(1-\cos[\omega_n\tau]\right) V'_B(B^{aa}(\tau)) + 2\int_0^\beta d\tau \sum_{b\neq a} V_B'[B^{ab}].
\eea
\end{widetext}

We remark here that the matrix elements that mix species are 
unaffected by disorder:
\beq
(G^{-1})_{fb}^{ab}(q,i\omega_n)=(G^{-1})_{bf}^{ab}(q,i\omega_n) = (G_0^{-1})_{bf}(q,i\omega_n)\delta_{a,b}\;.
\eeq
We now take the limit $n \rightarrow 0$ and introduce Parisi's
parameterization for $0\times 0$ matrices.\cite{Mezard91} 
Let $A$ be  a matrix in replica space.  Taking $n=0$, 
$A$ can be  described by a
couple $(\tilde{a},a(u))$ with $\tilde{a}$ corresponding to the 
diagonal elements of $A$, and  $a(u)$ a function of $u \in [0,1]$,
parameterizing the off-diagonal elements. For multiplication and  
inversion rules of Parisi matrices, see for example Ref.~\onlinecite{Mezard91}.

With the Parisi parametrization, the previous equations read: 
\be
(G^{-1}(u))(q,\omega_n)  
= -2\beta\delta_{n,0}\;\begin{pmatrix} V_F'[F(u)]  & 0 
\\ 0 & V_B'[B(u)]  
\end{pmatrix}
\ee
and 
\be 
\widetilde {G^{-1}}_{\alpha\beta}(q,\omega_n) =   (G^{-1}_0)_{\alpha\beta}(q,\omega_n) - \delta_{\alpha\beta}\Pi_{\alpha}(q,\omega_n)\;,
\ee 
with the fermionic ''self-energy'' defined as 
\begin{widetext}
\bea 
\Pi_{f}(q,\omega_n) &= & 
- 2\int_0^\beta d\tau \left(1-\cos[\omega_n\tau]\right)
V'_F[\widetilde{F}(\tau)] + 2\beta \int_0^1 \,du\, 
V_F'[F(u)] \;,
\eea
and the bosonic self-energy given by a similar expression. 
\end{widetext}
The replica-diagonal and off-diagonal parts of $F^{ab}$ then read 
\bea 
\widetilde {F}(\tau) &\equiv& 2 \bigl(
\widetilde{G}_{ff}(0,0)  - \widetilde{G}_{ff}(0,\tau)\bigr)
\;,
\nn
\\
F (u) &=& \frac{2}{\beta L} 
\sum\limits_{q,i\omega_n} \left[
\widetilde  G_{ff}(q,i\omega_n) 
-\delta_{\omega_n,0}\, \widetilde G_{ff}(q,u)
\right]\;,
\nn
\eea
and similar expressions hold for the functions 
$B(u)$ and $\widetilde B(\tau)$.

The connected Green's function, more precisely its inverse,
$(G^{-1})^c_{\alpha \beta} 
\equiv \sum\limits_{b}(G^{-1})^{ab}_{\alpha \beta}$, was already defined
in the main text.  Let us finally express this 
in terms of the Parisi parametrization, 
\be 
(G^{-1})^c_{\alpha \beta} = \widetilde{G^{-1}}_{\alpha
    \beta} - \int_0^1 du \ G^{-1}_{\alpha \beta}(u)\;.
\ee

The integral equations above need be solved self-consistently. 
However, as we shall see, they  do not have a unique
solution. Therefore, we need to supplement the above set of integral equations by
yet another condition, which shall be the marginality condition of the 
replicon mode, discussed later.\cite{Ledoussal96} Similar to other 
quantum glass phases, this condition turns out to
yield physically meaningful solutions in all phases.\cite{Georges01}

\subsection{Level 1 RSB}
To describe the phase with localized fermions and superfluid bosons, we
assume a level one replica symmetry breaking in the fermionic sector,
while the bosonic sector remains replica symmetric. To simplify notations, we
introduce the self-energy 
\be 
\sigma_f(u) = 2\beta V_F'[F(u)]\;.
\ee
Level one
RSB implies that there exists a value $0<u_f<1$ such that $\sigma_f(u<u_f)=0$ and
$\sigma_f(u>u_f) = \sigma_f$, or equivalently $F(u<u_f)=\infty$ and
$F(u>u_f)=F$. The corresponding bosonic self-energy $\sigma_b(u)$ is
identically zero in this phase. Then the matrix
elements of  $(G^{-1})^c_{\alpha \beta}$ read:
\bea
(G^{-1})^c_{ff} &=& (G_0^{-1})_{ff}(i\omega_n,q) + I_F(\omega_n) + 
\Sigma_F(1-\delta_{n,0})\;,
\nn
\\
(G^{-1})^c_{bb}&=& (G_0^{-1})_{bb}(i\omega_n,q) + I_B(\omega_n)\;,
\nn
\\
(G^{-1})^c_{fb} &=& (G^{-1})^c_{bf} = (G_0^{-1})_{fb}(i\omega_n,q)\;,
\nn
\\
\eea
with the functions $ I_{B/F}(\omega_n)$ 
defined as 
\bea \label{eq:A_Ib}
I_{B}(\omega_n) &=& 2\int_0^\beta d\tau
\left(1-\cos[\omega_n\tau]\right) V'_B[\widetilde{B}(\tau)] \;,
\\
I_{F}(\omega_n) &=&2\int_0^\beta d\tau
\left(1-\cos[\omega_n\tau]\right)\times\nn\\ \label{eq:A_If}
&&\hskip 1cm \times (V'_F[\widetilde{F}(\tau)]- V_F'[F])\;,
\eea
and the 'mass' $\Sigma_F$ given by 
$$
\Sigma_f = 2\beta u_f V_F'[F]\;.
$$

To  obtain the 
self-consistency equations for $\sigma_f$ and $I_F, I_B$,
one needs to invert the matrix $(G^{-1})_{\alpha \beta}$. This can be 
carried out using Parisi's 
multiplication and inversion formulas\cite{Mezard91}, and 
we find: 
\begin{widetext}
\bea
F &=& \frac{2}{\beta L} \sum_{q,i\omega_n} \frac{\pi K_f}{v_f}
\frac{\frac{\omega_n^2}{v_b^2}+q^2+\hat{I}_b(\omega_n)}{\left(\frac{\omega_n^2}{v_b^2}+q^2+\hat{I}_b(\omega_n)\right)\left(\frac{\omega_n^2}{v_f^2}+q^2+\hat{I}_f(\omega_n)+\hat{\Sigma}_f\right)-g^2
  q^4}\;,   \\  
\widetilde{F}(\tau)& =& \frac{2}{\beta L} \sum_{q,i\omega_n} \frac{\pi
  K_f}{v_f}\left(1-\cos[\omega_n\tau]\right)
\frac{\frac{\omega_n^2}{v_b^2}+q^2+\hat{I}_b(\omega_n)}{\left(\frac{\omega_n^2}{v_b^2}+q^2+\hat{I}_b(\omega_n)\right)\left(\frac{\omega_n^2}{v_f^2}+q^2+\hat{I}_f(\omega_n)+\hat{\Sigma}_f\right)-g^2
  q^4} 
\;,
\\ 
\widetilde{B}(\tau) &=& \frac{2}{\beta L} \sum_{q,i\omega_n} \frac{\pi
  K_b}{v_b}\left(1-\cos[\omega_n\tau]\right)
\frac{\frac{\omega_n^2}{v_f^2}+q^2+\hat{I}_f(\omega_n)+\hat{\Sigma}_f}{\left(\frac{\omega_n^2}{v_b^2}+q^2+\hat{I}_b(\omega_n)\right)\left(\frac{\omega_n^2}{v_f^2}+q^2+\hat{I}_f(\omega_n)+\hat{\Sigma}_f\right)-g^2 
  q^4}  
\;.
\eea
\end{widetext}
Here  we have introduced
$\hat{I}_f = \frac{\pi K_f}{v_f} I_F$,  $\hat{\Sigma}_F = \frac{\pi
  K_f}{v_f} \Sigma_F$, and used similar notations for bosons.
One can check that $B$ is indeed infinity. The self-consistent 
set of equations can be cast into: 
\bea
\hat{\Sigma}_F &=& 2\beta \frac{\pi K_f}{v_f} u_f V'_F[F], \nn \\
\hat{I}_f(\omega_n) &=& 2  \frac{\pi K_f}{v_f}\int_0^\beta d\tau \left(1-\cos[\omega_n\tau]\right) (V'_F[\widetilde{F}(\tau)]- V_F'[F]), \nn \\
\hat{I}_b(\omega_n) &=& 2 \frac{\pi K_b}{v_b}\int_0^\beta d\tau \left(1-\cos[\omega_n\tau]\right) V'_B[\widetilde{B}(\tau)]. 
\eea
However, these equations do not determine the break point, 
$u_f$. The value of the latter can be  determined using the so-called
marginality condition on the replicon mode \cite{Ledoussal96}. We
expand the variational free energy to second order, around the
saddle-point. To do so, we write $G(q,i\omega_n) = G^{(0)}(q,i\omega_n)
+ g(q)$, where $G^{(0)}$ denotes the saddle-point solution, and 
and $G^{(0)}$ (as well as $g(q)$) is
a Parisi matrix with matrix elements
\beq
[G^{(0)}]^{-1}(i\omega_n=0,q)=\left(
\begin{array}{cc}
(\widetilde{\Gamma}_f,\Gamma_f(u)) & (\widetilde{\Gamma}_{fb},0) \\
(\widetilde{\Gamma}_{fb},0) & (\widetilde{\Gamma}_b,0)
\end{array}
\right)\;.
\eeq
In a similar way: 
\beq
g(q) = \left(
\begin{array}{cc}
(0,g_f(q,u)) & (0,g_{fb}(q,u)) \\
(0,g_{bf}(q,u)) & (0,g_b(q,u))
\end{array}
\right)
\eeq
Note that since the RSB only happens for the $\omega_n=0$ mode, 
we only need to perturb that particular mode. We then expand 
$F_{\rm var}$ up to second order in $g(q)$, yielding
\begin{widetext} 
\beq
\delta^2 F_{var} = \frac{1}{4\beta}\sum_q \textrm{Tr} \left[ [G^{(0)}(q)]^{-1}g(q)  \right]^2 \nn \\
- n \frac{1}{\beta L}\int_0^1 du \sum_{q,q'}\left[g_f(q,u)g_f(q',u)V_F''[F(u)] + g_b(q,u)g_b(q',u)V_B''[B(u)]  \right].
\eeq
\end{widetext}

This can be written as  $\delta^2 F_{var} = \sum\limits_{q,q'}\int\limits_0^1 du
\int\limits_0^1 du' [g]^T(q,u) M(q,q',u,u') [g](q,u)$, 
with  $[g]^T =    [g_f,g_b,g_{fb},g_{bf}]$.
The stability matrix $M$ greatly simplifies if $g(q,u)$ is a
so-called replicon mode, for which $\int_0^{u_f} du \ g(u) = 0$ and $\int_{u_f}^1 du \ g(u) = 0$. In this 
case we find a symmetrical stability matrix, which, after introducing the notation $\langle \Gamma_f \rangle = \int_0^1 du \Gamma_f(u)$, takes the form
\begin{widetext}
\beq
M^{(u<u_f)\;}(q,q',u,u') = -\frac{1}{4\beta}\delta(u-u')
\left(
\begin{array}{cccc}
(\widetilde{\Gamma}_f-\langle\Gamma_f\rangle)^2\delta_{qq'}& \widetilde{\Gamma}_{fb}^2\delta_{qq'} & (\widetilde{\Gamma}_f-\langle\Gamma_f\rangle)\widetilde{\Gamma}_{fb}\delta_{qq'} & (\widetilde{\Gamma}_f-\langle\Gamma_f\rangle)\widetilde{\Gamma}_{fb}\delta_{qq'} 
\\
\star & \widetilde{\Gamma}_b^2\delta_{qq'} & \widetilde{\Gamma}_b\widetilde{\Gamma}_{fb}\delta_{qq'} & \widetilde{\Gamma}_b\widetilde{\Gamma}_{fb}\delta_{qq'}\\
\star & \star & \widetilde{\Gamma}_{fb}^2\delta_{qq'} & (\widetilde{\Gamma}_f-\langle\Gamma_f\rangle)\widetilde{\Gamma}_{b}\delta_{qq'}\\
\star & \star & \star &  \widetilde{\Gamma}_{fb}^2\delta_{qq'}
\end{array}
\right)\;,
\eeq
\beq
M^{(u>u_f)\;}(q,q',u,u') = -\frac{1}{4\beta}\delta(u-u')
\left(
\begin{array}{cccc}
(\widetilde{\Gamma}_f-\Gamma_f)^2\delta_{qq'}+\frac{4}{ L}V_F''[F]& \widetilde{\Gamma}_{fb}^2\delta_{qq'} & (\widetilde{\Gamma}_f-\Gamma_f)\widetilde{\Gamma}_{fb}\delta_{qq'} & (\widetilde{\Gamma}_f-\Gamma_f)\widetilde{\Gamma}_{fb}\delta_{qq'} 
\\
\star & \widetilde{\Gamma}_b^2\delta_{qq'} & \widetilde{\Gamma}_b\widetilde{\Gamma}_{fb}\delta_{qq'} & \widetilde{\Gamma}_b\widetilde{\Gamma}_{fb}\delta_{qq'}\\
\star & \star & \widetilde{\Gamma}_{fb}^2\delta_{qq'} & (\widetilde{\Gamma}_f-\Gamma_f)\widetilde{\Gamma}_{b}\delta_{qq'}\\
\star & \star & \star &  \widetilde{\Gamma}_{fb}^2\delta_{qq'}
\end{array}
\right)\;,
\eeq
\end{widetext}
\noindent 
 and the marginality
 condition is $\sum\limits_{q'} M(q,q',u,u) [g(q',u)] = 0$. 
For $u<u_f$ this is trivially satisfied, while for
$u>u_f$ it leads to the following condition: 
\beq
- \frac{4}{L}\left( \frac{\pi K_f}{v_f}\right)^2V_F''(F) \sum_q
\frac{1}{\left[q^2(1-g^2)+\hat{\Sigma}_f\right]^2} = 1\;, 
\eeq
which becomes for $L\rightarrow \infty$, 
\beq
\hat{\Sigma}_f^{3/2} = -\left( \frac{\pi K_f}{v_f}\right)^2\frac{V_F''(F)}{\sqrt{1-g^2}}\;,
\eeq
and closes the system of self-consistent equations.

\subsection{Level 2 RSB}
To describe the fully localized phase, we assume replica symmetry
breaking in both the fermionic and the bosonic sectors. It turns out that
one cannot reach a self-consistent set of equations with a level 1 RSB
in each sector,  however, it is sufficient to 
  assume a level 2 RSB in one of the sectors, 
 and a level 1 RSB in the other one. We proceed along the
same path as before, excepting that now there are two break
points, $u_1$ and $u_2$, such that 
\bea 
\sigma_f(u<u_1) &=& 0,\nn \\
\sigma_f(u_1<u<u_2) &=& \sigma_f^{(1)} \equiv 2\beta V_F'[F^{(1)}], \nn \\
\sigma_f(u_2<u<1) &=& \sigma_f^{(2)}\equiv 2\beta V_F'[F^{(2)}], \nn \\
\sigma_b(u<u_2) &=& 0, \nn \\
\sigma_b(u_2<u<1) &=& \sigma_b^{(2)}\equiv 2\beta V_B'[B^{(2)}].
\eea
Now we have:
\bea
\hat{I}_f(\omega_n) &=& 2  \frac{\pi K_f}{v_f}\int_0^\beta d\tau \left(1-\cos[\omega_n\tau]\right) (V'_F[\widetilde{F}(\tau)]- V_F'[F^{(2)}]), \nn \\
\hat{I}_b(\omega_n) &=& 2 \frac{\pi K_b}{v_b}\int_0^\beta d\tau \left(1-\cos[\omega_n\tau]\right) (V'_B[\widetilde{B}(\tau)]- V_B'[B^{(2)}]),\nn \\
\hat{\Sigma}_F &=& 2\beta \frac{\pi K_f}{v_f} \left[ u_1 V'_F[F^{(1)}] + u_2\left(V'_F[F^{(2)}]-V'_F[F^{(1)}]\right)\right] , \nn \\
\hat{\Sigma}_B &=& 2\beta \frac{\pi K_b}{v_b} u_2 V'_B[B^{(2)}]
\eea
The inversion of the propagator leads to:
\begin{widetext}
\bea
B^{(2)} &=& \frac{2}{\beta L} \sum_{q,i\omega_n} \frac{\pi K_b}{v_b} \frac{\frac{\omega_n^2}{v_f^2}+q^2+\hat{I}_f(\omega_n)+\hat{\Sigma}_F}{\left(\frac{\omega_n^2}{v_b^2}+q^2+\hat{I}_b(\omega_n)\right)\left(\frac{\omega_n^2}{v_f^2}+q^2+\hat{I}_f(\omega_n)+\hat{\Sigma}_f\right)-g^2 q^4} \\
F^{(2)} &=& \frac{2}{\beta L} \sum_{q,i\omega_n} \frac{\pi K_f}{v_f} \frac{\frac{\omega_n^2}{v_b^2}+q^2+\hat{I}_b(\omega_n)+\hat{\Sigma}_B}{\left(\frac{\omega_n^2}{v_b^2}+q^2+\hat{I}_b(\omega_n)\right)\left(\frac{\omega_n^2}{v_f^2}+q^2+\hat{I}_f(\omega_n)+\hat{\Sigma}_f\right)-g^2 q^4} \\
F^{(2)}-F^{(1)}&=&-2\pi\frac{K_f}{v_f}\frac{1}{u_2\beta L}\sum_q
\frac{(q^2+\hat{\Sigma}_B)\Delta\hat{\Sigma}_F^{(2)}+g^2
  q^2\Sigma_B}{\left((q^2+\hat{\Sigma}_B)(q^2+\hat{\Sigma}_F)-g^2q^4
  \right)\left( q^2(1-g^2) + \hat{\Sigma}_F^{(1)}\right)} 
\eea
\end{widetext}
where we have introduced $\Delta\hat{\Sigma}_F^{(2)}=2\beta \frac{\pi
  K_f}{v_f}  u_2\left(V'_F[F^{(2)}]-V'_F[F^{(1)}]\right)$ and
$\hat{\Sigma}_F^{(1)} = 2\beta \frac{\pi K_f}{v_f}  u_1
V'_F[F^{(1)}]$ -- so that $\hat{\Sigma}_F = \hat{\Sigma}_F^{(1)} +
\Delta\hat{\Sigma}_F^{(2)} $. As in the previous section we need two
more equations to find $u_1$ and $u_2$ and close the system. We also
look for the marginality condition of the replicon mode, which we
define as a mode satisfying 
$\int\limits_{u_1}^{u_2} du \ g(u) \equiv 0$ and
$\int\limits_{u_2}^{1} du \ g(u) \equiv 0$. 
Now the stability matrix reads: 
\begin{widetext}
\noindent $\bullet$ for $u<u_1$,
\beq
M(q,q',u,u') = -\frac{1}{4\beta}\delta(u-u')
\left(
\begin{array}{cccc}
(\widetilde{\Gamma}_f-\langle\Gamma_f\rangle)^2\delta_{qq'}& \widetilde{\Gamma}_{fb}^2\delta_{qq'} & (\widetilde{\Gamma}_f-\langle\Gamma_f\rangle)\widetilde{\Gamma}_{fb}\delta_{qq'} & (\widetilde{\Gamma}_f-\langle\Gamma_f\rangle)\widetilde{\Gamma}_{fb}\delta_{qq'} 
\\
\star & (\widetilde{\Gamma}_b-\langle \Gamma_b \rangle)^2\delta_{qq'} &  (\widetilde{\Gamma}_b-\langle \Gamma_b \rangle)\widetilde{\Gamma}_{fb}\delta_{qq'} &  (\widetilde{\Gamma}_b-\langle \Gamma_b \rangle)\widetilde{\Gamma}_{fb}\delta_{qq'}\\
\star & \star & \widetilde{\Gamma}_{fb}^2\delta_{qq'} & (\widetilde{\Gamma}_f-\langle\Gamma_f\rangle) (\widetilde{\Gamma}_b-\langle \Gamma_b \rangle)\delta_{qq'}\\
\star & \star & \star &  \widetilde{\Gamma}_{fb}^2\delta_{qq'}
\end{array}
\right)\;,
\nn
\eeq
\noindent $\bullet$ for $u_1<u<u_2$,
{\small
\beq
\left(
\begin{array}{cccc}
\left(\widetilde{\Gamma}_f-\Gamma_f^{(2)} + \Delta \Gamma_f^{(2)}\right)^2\delta_{qq'}+\frac{4}{ L}V_F''[F^{(1)}]& \widetilde{\Gamma}_{fb}^2\delta_{qq'} & \left(\widetilde{\Gamma}_f-\Gamma_f^{(2)} + \Delta \Gamma_f^{(2)}\right)\widetilde{\Gamma}_{fb}\delta_{qq'} & \left(\widetilde{\Gamma}_f-\Gamma_f^{(2)} + \Delta \Gamma_f^{(2)}\right)\widetilde{\Gamma}_{fb}\delta_{qq'} 
\\
\star & (\widetilde{\Gamma}_b-\langle\Gamma_b\rangle)^2\delta_{qq'} & (\widetilde{\Gamma}_b-\langle\Gamma_b\rangle)\widetilde{\Gamma}_{fb}\delta_{qq'} & (\widetilde{\Gamma}_b-\langle\Gamma_b\rangle)\widetilde{\Gamma}_{fb}\delta_{qq'}\\
\star & \star & \widetilde{\Gamma}_{fb}^2\delta_{qq'} & \left(\widetilde{\Gamma}_f-\Gamma_f^{(2)} + \Delta \Gamma_f^{(2)}\right)(\widetilde{\Gamma}_b-\langle\Gamma_b\rangle)\delta_{qq'}\\
\star & \star & \star &  \widetilde{\Gamma}_{fb}^2\delta_{qq'}
\end{array}
\right)\;,
\nn
\eeq}
$\bullet$ for $u_2<u<1$,
\beq
\left(
\begin{array}{cccc}
(\widetilde{\Gamma}_f-\Gamma_f^{(2)})^2\delta_{qq'}+\frac{4}{ L}V_F''[F^{(2)}]& \widetilde{\Gamma}_{fb}^2\delta_{qq'} & (\widetilde{\Gamma}_f-\Gamma_f^{(2)})\widetilde{\Gamma}_{fb}\delta_{qq'} & (\widetilde{\Gamma}_f-\Gamma_f^{(2)})\widetilde{\Gamma}_{fb}\delta_{qq'} 
\\
\star & (\widetilde{\Gamma}_b-\Gamma_b^{(2)})^2\delta_{qq'}+\frac{4}{ L}V_B''[B^{(2)}] & (\widetilde{\Gamma}_b-\Gamma_b^{(2)})\widetilde{\Gamma}_{fb}\delta_{qq'} & (\widetilde{\Gamma}_b-\Gamma_b^{(2)})\widetilde{\Gamma}_{fb}\delta_{qq'}\\
\star & \star & \widetilde{\Gamma}_{fb}^2\delta_{qq'} & (\widetilde{\Gamma}_f-\Gamma_f^{(2)})(\widetilde{\Gamma}_b-\Gamma_b^{(2)})\delta_{qq'}\\
\star & \star & \star &  \widetilde{\Gamma}_{fb}^2\delta_{qq'}
\end{array}
\right)\;.
\nn
\eeq
\end{widetext}

As before, on the first interval the marginality condition gives a trivial condition. On the second interval it gives:
\beq
- \frac{4}{L}\left( \frac{\pi K_f}{v_f}\right)^2V_F''[F^{(1)}] \sum_q \frac{1}{\left[q^2(1-g^2)+\hat{\Sigma}_f^{(1)}\right]^2} = 1,
\eeq
and on the third we obtain:

\begin{widetext}
\beq
\left[4\left( \frac{\pi K_f}{v_f}\right)^2
   V_F''[F^{(2)}]\ A_{ff}+1\right]\left[4\left( \frac{\pi
    K_f}{v_f}\right)^2 V_B''[B^{(2)}]\ A_{bb}+1 \right] = 
 16\ \left(\frac{\pi K_f}{v_f}\right)^2\left(\frac{\pi K_b}{v_b}
\right)^2V_F''[F^{(2)}]V_B''[B^{(2)}]\ A_{fb}^2\;,
\nn
\eeq
\end{widetext}

with:
\bea
A_{ff} &=&\frac{1}{L} \sum_q
\frac{\left(q^2+\hat{\Sigma}_b\right)^2}{\left[\left(q^2+\hat{\Sigma}_f\right)\left(q^2+\hat{\Sigma}_b\right)-g^2q^4\right]^2}   \;,
\nn \\ A_{bb} &=& \frac{1}{L} \sum_q
\frac{\left(q^2+\hat{\Sigma}_f\right)^2}{\left[\left(q^2+\hat{\Sigma}_f\right)
    \left(q^2+\hat{\Sigma}_b\right)-g^2q^4\right]^2}\;, 
\nn \\ A_{fb} &=& \frac{1}{L}\sum_q \frac{\pi^2g^2q^4}
    {\left[\left(q^2+\hat{\Sigma}_f\right)
        \left(q^2+\hat{\Sigma}_b\right)-g^2q^4\right]^2}    
\;.
\nn
\eea
These conditions effectively close the system of self-consistent equations.\\

\section{Computation of the structure factor from the variational solution}
\label{app:structure}

As stated in the main text -- see equation \eqref{eq:sf_1} -- the structure factor for fermions is given by  
\beq
\label{eq:sf_2}
S_{f}({q},\omega) = - \textrm{Im} \left[ q^2 \widetilde{G}_{ff}(q,i\omega_n \to \omega + i\epsilon) \right],
\eeq
and we have a similar expression for bosons. Here, $\widetilde{G}_{ff(bb)}$ is the replica-diagonal contribution for the fermion (boson) propagator. We recall their expressions in the three phases. In the Luttinger liquid phase,

\bea
\widetilde{G_{ff}}(q,\omega_n) &=& \frac{\pi K_f}{v_f} \frac{q^2 +
  b(\omega_n)}{[q^2 + b(\omega_n)][q^2 + f(\omega_n)]-g^2q^4},\nn
\\ \\ 
\widetilde{G_{bb}}(q,\omega_n) &=& \frac{\pi K_b}{v_b} \frac{q^2 +
  b(\omega_n)}{[q^2 + f(\omega_n)][q^2 + f(\omega_n)]-g^2q^4}. \nn \\ 
\eea
Remember we have introduced the following general notation 
\bea
b(\omega_n) &=&
\omega_n^2/v_b^2 + \hat{I}_b(\omega_n) + \hat{\Sigma}_b\;,
\\
f(\omega_n) &= &\omega_n^2/v_f^2 + \hat{I}_f(\omega_n) +
\hat{\Sigma}_f\;.
\eea
In the Luttinger liquid phase $ \hat{\Sigma}_f = \hat{\Sigma}_b = 0$. In the phase where fermions are localized and bosons
superfluid, the propagators read  
\bea
\widetilde{G_{ff}}(q,\omega_n) &=& \frac{\pi K_f}{v_f} 
\Biggl(
\frac{q^2 +
  b(\omega_n)}{[q^2 + b(\omega_n)][q^2 + f(\omega_n)]-g^2q^4} \nn \\ 
&+&  \delta_{n,0} 
\frac{1}{1-g^2}\frac{\sigma_f}{q^2[q^2(1-g^2) + \hat{\Sigma}_f]}\Biggr)\;,
\\   
\widetilde{G_{bb}}(q,\omega_n) &=& \frac{\pi K_b}{v_b}
\Biggl(
 \frac{q^2 +
  f(\omega_n)}{[q^2 + b(\omega_n)][q^2 + f(\omega_n)]-g^2q^4} \nn 
\\ 
&+&  \delta_{n,0} 
 \frac{g^2}{1-g^2}
\frac{\sigma_f}{q^2[q^2(1-g^2) + \hat{\Sigma}_f]}
\Biggr)\;
.   
\eea

Finally we add here the expressions of the propagators in the fully localized phase (for clarity they do not appear in the main text). They are
\bea
\widetilde{G_{ff}}(q,\omega_n) &=& \frac{\pi K_f}{v_f} \left.\frac{q^2 + b(\omega_n)}{[q^2 + b(\omega_n)][q^2 + f(\omega_n)]-g^2q^4}\right. \nn \\
 &+& \delta_{n,0} \frac{\pi K_f}{v_f} \left[ \frac{1}{1-g^2}\frac{\sigma_f^{(1)}}{q^2[q^2(1-g^2) + \hat{\Sigma}_f^{(1)}]} \right. \nn \\
&+& \left.  \frac{(\hat{\Sigma}_b + q^2)\Delta \hat{\sigma}_f^{(2)}+ q^2 g^2\hat{\sigma}_b }{[q^2(1-g^2) + \hat{\Sigma}_f^{(1)}][(q^2+\hat{\Sigma}_f)(q^2+\hat{\Sigma}_b)-g^2q^4]}\right], \nn \\ \\
\widetilde{G_{bb}}(q,\omega_n) &=& \frac{\pi K_b}{v_b} \frac{q^2 + f(\omega_n)}{[q^2 + b(\omega_n)][q^2 + f(\omega_n)]-g^2q^4} \nn \\
&+& \delta_{n,0} \frac{\pi K_b}{v_b} \frac{\sigma_b^{(2)}}{1-g^2}\frac{q^2 + \hat{\Sigma}_f}{q^2[(q^2+\hat{\Sigma}_f)(q^2+\hat{\Sigma}_b)-g^2q^4]} \nn \\
 &+& \ \delta_{n,0} \frac{\pi K_b}{v_b} \frac{g^2}{1-g^2}\left[\frac{\sigma_f^{(1)}}{q^2[q^2(1-g^2) + \hat{\Sigma}_f^{(1)}]} \right. \nn \\
 &+& \left. \frac{(\hat{\Sigma}_f + q^2) \hat{\sigma}_b^{(2)}+ q^2\Delta\hat{\sigma}_f^{(2)} }{[q^2(1-g^2) + \hat{\Sigma}_f^{(1)}][(q^2+\hat{\Sigma}_f)(q^2+\hat{\Sigma}_b)-g^2q^4]} \right].\nn \\
\eea

After the analytical continuation, the $\delta_{n,0}$ do not contribute and  $\widetilde{G}_{ff}(q,\omega_+)$ and $\widetilde{G}_{bb}(q,\omega_+)$, with $\omega_+ = \omega + i\epsilon$, are of the general form
\bea
\widetilde{G_{ff}}(q,-i\omega_+) &=& \frac{\pi K_f}{v_f} \frac{q^2 +
  b(-i\omega_+)}{[q^2 + b(-i\omega_+)][q^2 + f(-i\omega_+)]-g^2q^4},\nn
\\ \\ 
\widetilde{G_{bb}}(q,-i\omega_+) &=& \frac{\pi K_b}{v_b} \frac{q^2 +
  b(-i\omega_+)}{[q^2 + f(-i\omega_+)][q^2 + f(-i\omega_+)]-g^2q^4}. \nn \\ 
\eea
To get a useful form we introduce real and imaginary parts of $\hat{I}_f(-i\omega_+)$ and $\hat{I}_b(-i\omega_+)$ as $\hat{I}_f(-i\omega_+) = \hat{I}_f'(\omega) + i \hat{I}_f''(\omega)$ and $\hat{I}_b(-i\omega_+) = \hat{I}_b'(\omega) + i \hat{I}_b''(\omega)$. Finally we find for the structure factors

\begin{widetext}
\bea
\label{sf_3}
S_f(q,\omega) &=& -\frac{K_f}{\pi  v_f}q^2 \frac{\hat{I}_f''(\omega)\mathcal{P}_b(q,\omega)^2+\hat{I}_b''(\omega)\mathcal{P}_{fb}(q,\omega)}{\left[\hat{I}_f''(\omega)\mathcal{P}_{b}(q,\omega)+\hat{I}_b''(\omega)\mathcal{P}_{f}(q,\omega)\right]^2+\big[\mathcal{P}_{fb}(q,\omega)-\mathcal{P}_{f}(q,\omega)\mathcal{P}_{b}(q,\omega)\big]^2}, \\
\label{sb_3}
S_b(q,\omega) &=& -\frac{K_b}{\pi  v_b}q^2
\frac{\hat{I}_b''(\omega)\mathcal{P}_f(q,\omega)^2+\hat{I}_f''(\omega)\mathcal{P}_{fb}(q,\omega)}{\left[\hat{I}_f''(\omega)\mathcal{P}_{b}(q,\omega)+\hat{I}_b''(\omega)\mathcal{P}_{f}(q,\omega)\right]^2+\left[\mathcal{P}_{fb}(q,\omega)-\mathcal{P}_{f}(q,\omega)\mathcal{P}_{b}(q,\omega)\right]^2}, 
\eea
\end{widetext}

with 

\bea
\mathcal{P}_b(q,\omega) &=&  q^2+\hat{\Sigma}_b-\frac{\omega^2}{v_b^2}+\hat{I}_b'(\omega ), \\
\mathcal{P}_f(q,\omega) &=&  q^2+\hat{\Sigma}_f-\frac{\omega^2}{v_f^2}+\hat{I}_f'(\omega ), \\
\mathcal{P}_{fb}(q,\omega) &=&  \hat{I}_f''(\omega)\hat{I}_b''(\omega)+g^2 q^4. 
\eea

Remember that $\hat{I}_f', \hat{I}_f'', \hat{I}_b', \hat{I}_b'', \hat{\Sigma}_f, \hat{\Sigma}_b$ depend on the phase one considers. Although for weak disorder the functions $\hat{I}_f$ and $\hat{I}_b$ might alter the dynamics only weakly, probing the dynamics would be a good way to test the effect of different levels of replica symmetry breaking. Note that according to \eqref{eq:alphaf} and \eqref{eq:alphab}, the sructure factors grow linearly at small frequency in the localized phases.

\bibliography{Bib_melanges}
\end{document}